\documentclass[10pt]{IEEEtran}
\usepackage[LGR,T1]{fontenc}
\usepackage[latin9]{inputenc}
\usepackage{color}
\usepackage{array}
\usepackage{units}
\usepackage{url}
\usepackage{multirow}
\usepackage{amsmath}
\usepackage{amsthm}
\usepackage{amssymb}
\usepackage{graphicx}

\usepackage{enumitem}
\usepackage[unicode=true,
 bookmarks=true,bookmarksnumbered=true,bookmarksopen=true,bookmarksopenlevel=3,
 breaklinks=false,pdfborder={0 0 0},pdfborderstyle={},backref=false,colorlinks=true]
 {hyperref}
\hypersetup{pdftitle={Renyi},
 pdfauthor={Lei Yu},
 pdfborderstyle={},pdfborderstyle={},pdfborderstyle={},pdfborderstyle={},pdfborderstyle={},pdfborderstyle={},pdfborderstyle={},pdfborderstyle={},pdfborderstyle={},pdfborderstyle={},pdfborderstyle={},pdfborderstyle={},pdfborderstyle={},pdfborderstyle={},pdfborderstyle={},pdfborderstyle={},pdfborderstyle={},pdfborderstyle={},pdfpagelayout=OneColumn,pdfnewwindow=true,pdfstartview=XYZ,plainpages=false,linkcolor=blue,urlcolor=blue,citecolor=red,anchorcolor=blue,linkcolor=blue,urlcolor=blue,citecolor=red,anchorcolor=blue}

\makeatletter

\providecommand{\tabularnewline}{\\}

\theoremstyle{plain}
\newtheorem{thm}{\protect\theoremname}
\theoremstyle{plain}
\newtheorem{lem}[]{\protect\lemmaname}
\theoremstyle{remark}
\newtheorem{rem}[]{\protect\remarkname}
\theoremstyle{plain}
\newtheorem{cor}[]{\protect\corollaryname}

\usepackage{cite}

\newcommand{\calM}{\mathcal{M}}

\newcommand{\calP}{\mathcal{P}}

\newcommand{\calX}{\mathcal{X}}

\newcommand{\bbN}{\mathbb{N}}

\newcommand{\bbR}{\mathbb{R}}

\DeclareMathAlphabet{\mathbsf}{OT1}{cmss}{bx}{n}
\DeclareMathAlphabet{\mathssf}{OT1}{cmss}{m}{sl}

\DeclareSymbolFont{bsfletters}{OT1}{cmss}{bx}{n}
\DeclareSymbolFont{ssfletters}{OT1}{cmss}{m}{n}
\DeclareMathSymbol{\bsfGamma}{0}{bsfletters}{'000}
\DeclareMathSymbol{\ssfGamma}{0}{ssfletters}{'000}
\DeclareMathSymbol{\bsfDelta}{0}{bsfletters}{'001}
\DeclareMathSymbol{\ssfDelta}{0}{ssfletters}{'001}
\DeclareMathSymbol{\bsfTheta}{0}{bsfletters}{'002}
\DeclareMathSymbol{\ssfTheta}{0}{ssfletters}{'002}
\DeclareMathSymbol{\bsfLambda}{0}{bsfletters}{'003}
\DeclareMathSymbol{\ssfLambda}{0}{ssfletters}{'003}
\DeclareMathSymbol{\bsfXi}{0}{bsfletters}{'004}
\DeclareMathSymbol{\ssfXi}{0}{ssfletters}{'004}
\DeclareMathSymbol{\bsfPi}{0}{bsfletters}{'005}
\DeclareMathSymbol{\ssfPi}{0}{ssfletters}{'005}
\DeclareMathSymbol{\bsfSigma}{0}{bsfletters}{'006}
\DeclareMathSymbol{\ssfSigma}{0}{ssfletters}{'006}
\DeclareMathSymbol{\bsfUpsilon}{0}{bsfletters}{'007}
\DeclareMathSymbol{\ssfUpsilon}{0}{ssfletters}{'007}
\DeclareMathSymbol{\bsfPhi}{0}{bsfletters}{'010}
\DeclareMathSymbol{\ssfPhi}{0}{ssfletters}{'010}
\DeclareMathSymbol{\bsfPsi}{0}{bsfletters}{'011}
\DeclareMathSymbol{\ssfPsi}{0}{ssfletters}{'011}
\DeclareMathSymbol{\bsfOmega}{0}{bsfletters}{'012}
\DeclareMathSymbol{\ssfOmega}{0}{ssfletters}{'012}

\DeclareMathOperator{\supp}{supp}

\def\dotle{\mathrel{\dot{\le}}}

\def\dotge{\mathrel{\dot{\ge}}}

  \providecommand{\lemmaname}{Lemma}
  \providecommand{\remarkname}{Remark}
\providecommand{\theoremname}{Theorem}

\allowdisplaybreaks[1]
\providecommand{\corollaryname}{Corollary}
\providecommand{\lemmaname}{Lemma}
\providecommand{\remarkname}{Remark}
\providecommand{\theoremname}{Theorem}

\providecommand{\corollaryname}{Corollary}
\providecommand{\lemmaname}{Lemma}
\providecommand{\remarkname}{Remark}
\providecommand{\theoremname}{Theorem}

\providecommand{\corollaryname}{Corollary}
\providecommand{\lemmaname}{Lemma}
\providecommand{\remarkname}{Remark}
\providecommand{\theoremname}{Theorem}

\@ifundefined{showcaptionsetup}{}{%
 \PassOptionsToPackage{caption=false}{subfig}}
\usepackage{subfig}
\makeatother

\providecommand{\corollaryname}{Corollary}
\providecommand{\lemmaname}{Lemma}
\providecommand{\remarkname}{Remark}
\providecommand{\theoremname}{Theorem}

\begin{document}

\title{Simulation of Random Variables under R\'enyi Divergence Measures of
All Orders}

\author{Lei Yu and Vincent Y. F. Tan, \IEEEmembership{Senior Member,~IEEE}
\thanks{ 
This work was supported by a Singapore National Research Foundation
(NRF) National Cybersecurity R\&D Grant (R-263-000-C74-281 and NRF2015NCR-NCR003-006).
The first author was also supported by a National Natural Science
Foundation of China (NSFC) under Grant (61631017). This paper was
presented in part at the 2018 IEEE Information Theory Workshop (ITW).} \thanks{ L.~Yu is with the Department of Electrical and Computer Engineering,
National University of Singapore (NUS), Singapore 117583 (e-mail:
leiyu@nus.edu.sg). V.~Y.~F.~Tan is with the Department of Electrical
and Computer Engineering and the Department of Mathematics, NUS, Singapore
119076 (e-mail: vtan@nus.edu.sg).} \thanks{ Communicated by V. Prabhakaran, Associate Editor for Shannon Theory. }
\thanks{Copyright (c) 2018 IEEE. Personal use of this material is permitted.
However, permission to use this material for any other purposes must
be obtained from the IEEE by sending a request to pubs-permissions@ieee.org.}}
\maketitle
\begin{abstract}
The random variable simulation problem consists in using a $k$-dimensional
i.i.d. random vector $X^{k}$ with distribution $P_{X}^{k}$ to simulate
an $n$-dimensional i.i.d. random vector $Y^{n}$ so that its distribution
is approximately $Q_{Y}^{n}$. In contrast to previous works, in this
paper we consider the standard R\'enyi divergence and two variants of
all orders to measure the level of approximation. These two variants
are the max-R\'enyi divergence $D_{\alpha}^{\mathsf{max}}(P,Q)$ and
the sum-R\'enyi divergence $D_{\alpha}^{+}(P,Q)$. When $\alpha=\infty$,
these two measures are strong because for any $\epsilon\ge0$, $D_{\infty}^{\mathsf{max}}(P,Q)\leq\epsilon$
or $D_{\infty}^{+}(P,Q)\leq\epsilon$ implies $e^{-\epsilon}\leq\frac{P(x)}{Q(x)}\leq e^{\epsilon}$
for all $x$. Under these R\'enyi divergence measures, we characterize
the asymptotics of normalized divergences as well as the R\'enyi conversion
rates. The latter is defined as the supremum of $\frac{n}{k}$ such
that the R\'enyi divergences vanish asymptotically. Our results show
that when the R\'enyi parameter is in the interval $(0,1)$, the R\'enyi
conversion rates equal the ratio of the Shannon entropies $\frac{H\left(P_{X}\right)}{H\left(Q_{Y}\right)}$,
which is consistent with traditional results in which the total variation
measure was adopted. When the R\'enyi parameter is in the interval $(1,\infty]$,
the R\'enyi conversion rates are, in general, smaller than $\frac{H\left(P_{X}\right)}{H\left(Q_{Y}\right)}$.
When specialized to the case in which either $P_{X}$ or $Q_{Y}$
is uniform, the simulation problem reduces to the source resolvability
and intrinsic randomness problems. The preceding results are used
to characterize the asymptotics of R\'enyi divergences and the R\'enyi
conversion rates for these two cases. 
\end{abstract}


\section{\label{sec:Introduction}Introduction}

How can we use a $k$-dimensional i.i.d. random vector $X^{k}$ with
distribution $P_{X}^{k}$ to simulate an $n$-dimensional i.i.d. random
vector $Y^{n}$ so that its distribution is approximately $Q_{Y}^{n}$?
This is so-called {\em random variable simulation problem} or {\em
distribution approximation problem} \cite{Han03}. In \cite{Han03}
and \cite{kumagai2017second}, the total variation (TV) distance and
the Bhattacharyya coefficient (the R\'enyi divergence of order $\frac{1}{2}$)
were respectively used to measure the level of approximation. In these
works, the asymptotic conversion rate was studied. This rate is defined
as the supremum of $\frac{n}{k}$ such that the employed measure vanishes
asymptotically as the dimensions $n$ and $k$ tend to infinity. For
both the TV distance and the Bhattacharyya coefficient, the asymptotic
(first-order) conversion rates are the same, and both equal to the
ratio of the Shannon entropies $\frac{H\left(P_{X}\right)}{H\left(Q_{Y}\right)}$.
Furthermore, in \cite{kumagai2017second}, Kumagai and Hayashi also
investigated the asymptotic second order conversion rate. Note that
by Pinsker's inequality \cite{Erven}, the Bhattacharyya coefficient
(the R\'enyi divergence of order $\frac{1}{2}$) is stronger than the
TV distance, i.e., if the Bhattacharyya coefficient tends to $1$
(or the R\'enyi divergence of order $\frac{1}{2}$ tends to $0$), then
the TV distance tends to $0$. In this paper, we strengthen the TV
distance and the Bhattacharyya coefficient by considering R\'enyi divergences
of orders in $[0,\infty]$.

As two important special cases of the distribution approximation problem,
the source resolvability and intrinsic randomness problems have been
extensively studied in the literature, e.g., \cite{Han,Hayashi06,Hayashi11,Liu,Yu,vembu1995generating,Han03}. 

\begin{enumerate}[leftmargin=*]
\item Resolvability: When $P_{X}$ is set to the Bernoulli distribution
$\mathsf{Bern}(\frac{1}{2})$, the distribution approximation problem
reduces to the {\em source resolvability problem}, i.e., determining
how much information is needed to simulate a random process so that
it approximates a target output distribution. If the simulation is
realized through a given channel, and we require that the channel
output approximates a target output distribution, then we obtain the
{\em channel resolvability problem}. These resolvability problems
were first studied by Han and Verd\'u \cite{Han}. In \cite{Han}, the
total variation (TV) distance and the normalized relative entropy
(Kullback-Leibler divergence) were used to measure the level of approximation.
The resolvability problems with the \emph{unnormalized} relative entropy
were studied by Hayashi \cite{Hayashi06,Hayashi11}. Recently, Liu,
Cuff, and Verd\'u \cite{Liu} and Yu and Tan \cite{Yu} extended the
theory of resolvability by respectively using the so-called $E_{\gamma}$
metric with $\gamma\geq1$ and various R\'enyi divergences of orders
in $[0,2]\cup\{\infty\}$ to measure the level of approximation. In
this paper, we extend the results in \cite{Yu} to the R\'enyi divergences
of orders in $[0,\infty]$. 
\item Intrinsic randomness: When $Q_{Y}$ is set to the Bernoulli distribution
$\mathsf{Bern}(\frac{1}{2})$, the distribution approximation problem
reduces to the \emph{intrinsic randomness}, i.e., determining the
amount of randomness contained in a source \cite{vembu1995generating}.
Given an arbitrary general source $\boldsymbol{X}=\left\{ X^{n}\right\} _{n=1}^{\infty}$,
we approximate, by using $\boldsymbol{X}$, a uniform random number
with as large a rate as possible. Vembu and Verd\'u \cite{vembu1995generating}
and Han \cite{Han03} determined the supremum of achievable uniform
random number generation rates by invoking the information spectrum
method. In this paper, we extend the results in \cite{vembu1995generating}
to the family of R\'enyi divergence measures. 
\end{enumerate}

\subsection{Main Contributions }

Our main contributions are as follows: 

\begin{enumerate}[leftmargin=*]
\item For the distribution approximation problem, we use the standard R\'enyi
divergences $D_{\alpha}(P_{Y^{n}}\|Q_{Y}^{n})$ and $D_{\alpha}(Q_{Y}^{n}\|P_{Y^{n}})$,
as well as two variants, namely the max-R\'enyi divergence $D_{\alpha}^{\mathsf{max}}(P,Q)$
and the sum-R\'enyi divergence $D_{\alpha}^{+}(P,Q)$, to measure the
distance between the simulated and target output distributions. For
these measures, we consider all orders in $\alpha\in[0,\infty]$.
We characterize the asymptotics of these R\'enyi divergences, as well
as the R\'enyi conversion rates, which are defined as the supremum of
$\frac{n}{k}$ to guarantee that the R\'enyi divergences vanish asymptotically.
Interestingly, when the R\'enyi parameter is in the interval $(0,1]$
for the measure $D_{\alpha}(P_{Y^{n}}\|Q_{Y}^{n})$ and in $(0,1)$
for the measures $D_{\alpha}(Q_{Y}^{n}\|P_{Y^{n}})$ and $D_{\alpha}^{\mathsf{max}}(P_{Y^{n}},Q_{Y}^{n})$
(or $D_{\alpha}^{+}(P_{Y^{n}},Q_{Y}^{n})$), the R\'enyi conversion
rates are simply equal to the ratio of the Shannon entropies $\frac{H\left(P_{X}\right)}{H\left(Q_{Y}\right)}$.
This is consistent with the existing results in \cite{kumagai2017second}
where the R\'enyi parameter is $\frac{1}{2}$. In contrast if the R\'enyi
parameter is in $(1,\infty]$ for the measure $D_{\alpha}(P_{Y^{n}}\|Q_{Y}^{n})$
and $\in[1,\infty]$ for the measures $D_{\alpha}(Q_{Y}^{n}\|P_{Y^{n}})$
and $D_{\alpha}^{\mathsf{max}}(P_{Y^{n}},Q_{Y}^{n})$ (or $D_{\alpha}^{+}(P_{Y^{n}},Q_{Y}^{n})$),
the R\'enyi conversion rates are, in general, larger than $\frac{H\left(P_{X}\right)}{H\left(Q_{Y}\right)}$.
It is worth noting that the obtained expressions for the asymptotics
of R\'enyi divergences and the R\'enyi conversion rates involve R\'enyi
entropies of all real orders, even including negative orders. To the
best of our knowledge, this is the first time that an explicit operational
interpretation of the R\'enyi entropies of negative orders is provided. 
\item When specialized to the cases in which either $P_{X}$ or $Q_{Y}$
is uniform, the preceding results are used to derive results for the
source resolvability and intrinsic randomness problems. These results
extend the existing results in \cite{Han,vembu1995generating,Han03,Yu},
where the TV distance, the relative entropy, and the R\'enyi divergences
of orders in $[0,2]$ were used to measure the level of approximation. 
\end{enumerate}

\subsection{Paper Outline }

The rest of this paper is organized as follows. In Subsections \ref{subsec:Information-Distance-Measures}
and \ref{subsec:Problem-Formulation}, we introduce several R\'enyi
information quantities and use them to formulate the random variable
simulation problem. In Section \ref{sec:Renyi-Distribution-Approximation},
we present our main results on characterizing asymptotics of R\'enyi
divergences and R\'enyi conversion rates. As consequences, in Sections
\ref{sec:Special-Case-1:} and \ref{sec:Special-Case-2:}, we apply
our main results to the problems of R\'enyi source resolvability and
R\'enyi intrinsic randomness. Finally, we conclude the paper in Section
\ref{sec:Concluding-Remarks}. For seamless presentation of results,
the proofs of all theorems and the notations involved in these proofs
are deferred to the appendices.

\subsection{\label{subsec:Information-Distance-Measures}Notations and Information
Distance Measures }

The set of probability measures on $\mathcal{X}$ is denoted as $\mathcal{P}\left(\mathcal{X}\right)$,
and the set of conditional probability measures on $\mathcal{Y}$
given a variable in $\mathcal{X}$ is denoted as $\mathcal{P}\left(\mathcal{Y}|\mathcal{X}\right):=\left\{ P_{Y|X}:P_{Y|X}\left(\cdot|x\right)\in\mathcal{P}\left(\mathcal{Y}\right),x\in\mathcal{X}\right\} $.
For a distribution $P_{X}\in\mathcal{P}\left(\mathcal{X}\right)$,
the support of $P_{X}$ is defined as $\supp\left(P_{X}\right):=\left\{ x\in\mathcal{X}:\,P_{X}(x)>0\right\} $.

We use $T_{x^{n}}\left(x\right):=\frac{1}{n}\sum_{i=1}^{n}1\left\{ x_{i}=x\right\} $
to denote the type (empirical distribution) of a sequence $x^{n}$,
$T_{X}$ and $V_{Y|X}$ to respectively denote a type of sequences
in $\mathcal{X}^{n}$ and a conditional type of sequences in $\mathcal{Y}^{n}$
(given a sequence $x^{n}\in\calX^{n}$). For a type $T_{X}$, the
type class (set of sequences having the same type $T_{X}$) is denoted
by $\mathcal{T}_{T_{X}}$. For a conditional type $V_{Y|X}$ and a
sequence $x^{n}$, the \emph{V-shell of $x^{n}$} (the set of $y^{n}$
sequences having the same conditional type $V_{Y|X}$ given $x^{n}$)
is denoted by $\mathcal{T}_{V_{Y|X}}\left(x^{n}\right)$. The set
of types of sequences in $\mathcal{X}^{n}$ is denoted as 
\begin{equation}
\mathcal{P}^{\left(n\right)}\left(\mathcal{X}\right):=\left\{ T_{x^{n}}:x^{n}\in\mathcal{X}^{n}\right\} .
\end{equation}
The set of conditional types of sequences in $\mathcal{Y}^{n}$ given
a sequence in $\mathcal{X}^{n}$ with the type $T_{X}$ is denoted
as 
\begin{align}
 & \mathcal{P}^{\left(n\right)}\left(\mathcal{Y}|T_{X}\right)\nonumber \\
 & :=\{V_{Y|X}\in\mathcal{P}\left(\mathcal{Y}|\mathcal{X}\right):V_{Y|X}\times T_{X}\in\mathcal{P}^{\left(n\right)}\left(\mathcal{X}\times\mathcal{Y}\right)\}.
\end{align}
For brevity, sometimes we use $T\left(x,y\right)$ to denote the joint
distributions $T\left(x\right)V\left(y|x\right)$ or $T\left(y\right)V\left(x|y\right)$.

The $\epsilon$-typical set of $Q_{X}$ is denoted as 
\begin{align}
 & \mathcal{T}_{\epsilon}^{n}\left(Q_{X}\right)\nonumber \\
 & :=\left\{ x^{n}\in\mathcal{X}^{n}:\left|T_{x^{n}}\left(x\right)-Q_{X}\left(x\right)\right|\leq\epsilon Q_{X}\left(x\right),\forall x\in\mathcal{X}\right\} .\label{eqn:typ_set}
\end{align}
The conditionally $\epsilon$-typical set of $Q_{XY}$ is denoted
as 
\begin{equation}
\mathcal{T}_{\epsilon}^{n}\left(Q_{XY}|x^{n}\right):=\left\{ y^{n}\in\mathcal{X}^{n}:\left(x^{n},y^{n}\right)\in\mathcal{T}_{\epsilon}^{n}\left(Q_{XY}\right)\right\} .
\end{equation}
For brevity, sometimes we write $\mathcal{T}_{\epsilon}^{n}\left(Q_{X}\right)$
and $\mathcal{T}_{\epsilon}^{n}\left(Q_{XY}|x^{n}\right)$ as $\mathcal{T}_{\epsilon}^{n}$
and $\mathcal{T}_{\epsilon}^{n}\left(x^{n}\right)$ respectively.

For a distribution $P_{X}\in\calP(\calX)$, the {\em R\'enyi entropy
of order}\footnote{In the literature, the R\'enyi entropy was defined usually only for
orders $\alpha\in[0,+\infty]$ \cite{renyi1961measures}, except for
a recent work \cite{sason2018arimoto}, but here we define it for
orders $\alpha\in[-\infty,+\infty]$. This is due to the fact that
our results involve R\'enyi entropies of all real orders, even including
negative orders. Indeed, in the axiomatic definitions of R\'enyi entropy
and R\'enyi divergence, R\'enyi restricted the parameter $\alpha\in(0,1)\cup(1,+\infty)$
\cite{renyi1961measures}. However, it is easy to verify that in \cite{renyi1961measures},
the postulates 1, 2, 3, 4, and 5' in the definition of R\'enyi entropy
with $g_{\alpha}(x)=e^{(\alpha-1)x}$ and the postulates 6, 7, 8,
9, and 10 in the definition of R\'enyi divergence with the same function
$g_{\alpha}(x)$ are also satisfied when $\alpha\in(-\infty,0)$.
It is worth noting that the R\'enyi entropy for $\alpha\in(-\infty,0)$
is always non-negative, but the R\'enyi divergence for $\alpha\in(-\infty,0)$
is always non-positive. The R\'enyi divergence of negative orders was
studied in \cite{Erven}. Observe that $D_{\alpha}(P\|Q)=\frac{\alpha}{1-\alpha}D_{1-\alpha}(Q\|P)$
holds for $\alpha\in[-\infty,0)\cup(0,1)\cup(1,+\infty]$. Hence we
only need to consider the divergences $D_{\alpha}(P\|Q)$ and $D_{\alpha}(Q\|P)$
with $\alpha\in[0,+\infty]$, since these divergences completely characterize
the divergences $D_{\alpha}(P\|Q)$ and $D_{\alpha}(Q\|P)$ with $\alpha\in[-\infty,+\infty]$.
Furthermore, it is also worth noting that the R\'enyi entropy is non-increasing
and the R\'enyi divergence is non-decreasing in $\alpha$ for $\alpha\in[-\infty,\infty]$
\cite{sason2018arimoto,Erven}.}{} $\alpha\in(-\infty,1)\cup(1,+\infty),$ is defined as 
\begin{align}
H_{\alpha}(P_{X}) & :=\frac{1}{1-\alpha}\log\sum_{x\in\supp\left(P_{X}\right)}P_{X}(x)^{\alpha},
\end{align}
and the {\em R\'enyi entropy of order $\alpha=1,-\infty,+\infty$}
is defined as the limit by taking $\alpha\rightarrow1,-\infty,+\infty$,
respectively. It is known that 
\begin{align}
H_{-\infty}(P_{X}) & =-\log\inf_{x\in\supp\left(P_{X}\right)}P_{X}(x);\\
H_{1}(P_{X}) & =H(P_{X})\\
 & :=-\sum_{x\in\supp\left(P_{X}\right)}P_{X}(x)\log P_{X}(x);\\
H_{+\infty}(P_{X}) & =-\log\sup_{x\in\supp\left(P_{X}\right)}P_{X}(x).
\end{align}
 Hence  the usual Shannon entropy $H(P_{X})$ is a special (limiting)
case of the R\'enyi entropy. Some properties of R\'enyi entropies of all
real orders (including negative orders) can be found in a recent work
\cite{sason2018arimoto}, e.g., $H_{\alpha}(P_{X})$ is monotonically
decreasing in $\alpha$ throughout the real line, and $\frac{\alpha-1}{\alpha}H_{\alpha}(P_{X})$
is monotonically increasing in $\alpha$ on $(0,+\infty)$ and $(-\infty,0)$.

For a distribution $P_{X}\in\calP(\calX)$, the {\em mode entropy}\footnote{Here the concept of ``mode entropy'' is consistent with the concept
of ``mode'' in statistics. This is because, in statistics, the mode
of a set of data values is the value that appears most often. On the
other hand, for a product set $\supp\left(P_{X}\right)^{n}$, the
type class $\mathcal{T}_{T_{X}}$ with type $T_{X}\approx\mathrm{Unif}\left(\supp\left(P_{X}\right)\right)$
has more elements than any other type class, and under the product
distribution $P_{X}^{n}$, the probability values of sequences in
the type class $\mathcal{T}_{T_{X}}$ is $e^{-nH^{\mathrm{u}}(P_{X})}$.
Hence, under the product distribution $P_{X}^{n}$, the probability
value $e^{-nH^{\mathrm{u}}(P_{X})}$ is the mode of the data values
$\left(P_{X}^{n}\left(x^{n}\right)>0:\:x^{n}\in\mathcal{X}^{n}\right)$.} is defined as 
\begin{align}
H^{\mathrm{u}}(P_{X}) & :=-\sum_{x\in\supp\left(P_{X}\right)}\frac{1}{\left|\supp\left(P_{X}\right)\right|}\log P_{X}(x).\label{eq:modeentropy}
\end{align}
The mode entropy is also known as the cross (Shannon) entropy between
$\mathrm{Unif}\left(\supp\left(P_{X}\right)\right)$ and $P_{X}$.
For a distribution $P_{X}\in\calP(\calX)$ and $\alpha\in[-\infty,\infty]$,
the {\em $\alpha$-tilted distribution } is defined as 
\begin{align}
P_{X}^{(\alpha)}(\cdot) & :=\frac{P_{X}^{\alpha}(\cdot)}{\sum_{x'\in\supp\left(P_{X}\right)}P_{X}^{\alpha}(x')},\label{eq:modeentropy-1-1}
\end{align}
and the {\em $\alpha$-tilted cross entropy} is defined as 
\begin{align}
H_{\alpha}^{\mathrm{u}}(P_{X}) & :=-\sum_{x\in\supp\left(P_{X}\right)}P_{X}^{(\alpha)}(x)\log P_{X}(x).\label{eq:modeentropy-1}
\end{align}
Obviously, $H_{0}^{\mathrm{u}}(P_{X})=H^{\mathrm{u}}(P_{X})$, and
$H_{\alpha}^{\mathrm{u}}(P_{X})=H_{\alpha}(P_{X})$ for $\alpha\in\left\{ -\infty,1,\infty\right\} $.

Fix distributions $P_{X},Q_{X}\in\calP(\calX)$. Then the {\em R\'enyi
divergence of order $(0,1)\cup(1,+\infty)$} is defined as 
\begin{align}
D_{\alpha}(P_{X}\|Q_{X}) & :=\frac{1}{\alpha-1}\log\sum_{x\in\supp\left(P_{X}\right)}P_{X}(x)^{\alpha}Q_{X}(x)^{1-\alpha},
\end{align}
and the {\em R\'enyi divergence of order $\alpha=0,1,+\infty$} is
defined as the limit by taking $\alpha\rightarrow0,1,+\infty$, respectively.
It is known that 
\begin{align}
D_{0}(P_{X}\|Q_{X}) & =-\log\{Q_{X}(\supp\left(P_{X}\right))\};\label{eq:-29}\\
D_{1}(P_{X}\|Q_{X}) & =D(P_{X}\|Q_{X})\\
 & :=\sum_{x\in\supp\left(P_{X}\right)}P_{X}(x)\log\frac{P_{X}(x)}{Q_{X}(x)};\\
D_{\infty}(P_{X}\|Q_{X}) & =\log\sup_{x\in\supp\left(P_{X}\right)}\frac{P_{X}(x)}{Q_{X}(x)}.
\end{align}
 Hence the usual relative entropy is a special case of the R\'enyi divergence.

We define the max-R\'enyi divergence as 
\begin{equation}
D_{\alpha}^{\mathsf{max}}(P,Q)=\max\left\{ D_{\alpha}(P\|Q),D_{\alpha}(Q\|P)\right\} ,
\end{equation}
and the sum-R\'enyi divergence as 
\begin{equation}
D_{\alpha}^{+}(P,Q)=D_{\alpha}(P\|Q)+D_{\alpha}(Q\|P).
\end{equation}
The sum-R\'enyi divergence reduces to Jeffrey's divergence $D(P\|Q)+D(Q\|P)$
\cite{jeffreys1946invariant} when the parameter $\alpha$ is set
to $1$. Observe that $D_{\alpha}^{\mathsf{max}}(P,Q)\leq D_{\alpha}^{+}(P,Q)\leq2D_{\alpha}^{\mathsf{max}}(P,Q)$.
Hence $D_{\alpha}^{\mathsf{max}}(P,Q)$ is ``equivalent'' to $D_{\alpha}^{+}(P,Q)$
in the sense that for any sequences of distribution pairs $\left\{ (P^{(n)},Q^{(n)})\right\} _{n=1}^{\infty}$,
$D_{\alpha}^{\mathsf{max}}(P^{(n)},Q^{(n)})\to0$ if and only if $D_{\alpha}^{+}(P^{(n)},Q^{(n)})\to0$.
Hence in this paper, we only consider the max-R\'enyi divergence. For
$\alpha=\infty$, 
\begin{align}
D_{\infty}^{\mathsf{max}}(P,Q) & =\sup_{x\in\calX}|\log P(x)-\log Q(x)|\\
 & =\sup_{\mathcal{A}\subseteq\calX}\left|\log P(\mathcal{A})-\log Q(\mathcal{A})\right|.
\end{align}
This expression is similar to the definition of TV distance, hence
we term $D_{\infty}^{\mathsf{max}}$ as the \emph{logarithmic variation
distance.}\footnote{In \cite{cuff2016differential}, $D_{\infty}^{\mathsf{max}}(P,Q)\leq\epsilon$
is termed the $\left(\epsilon,0\right)$-closeness.} 
\begin{lem}
\label{lem:The-following-properties}The following properties hold. 
\begin{enumerate}
\item $D_{\infty}^{\mathsf{max}}$ is a metric. Similarly, $D_{\infty}^{+}$
is also a metric. 
\item $D_{\infty}^{\mathsf{max}}(P,Q)\leq\epsilon\Longleftrightarrow e^{-\epsilon}\leq\frac{P(x)}{Q(x)}\leq e^{\epsilon},\forall x.$ 
\item For any $f$, $-D_{\infty}(Q\|P)\leq\log\frac{\mathbb{E}_{P}f(X)}{\mathbb{E}_{Q}f(X)}\leq D_{\infty}(P\|Q)$,
hence $D_{\infty}^{\mathsf{max}}(P,Q)\leq\epsilon\Longrightarrow e^{-\epsilon}\leq\frac{\mathbb{E}_{P}f(X)}{\mathbb{E}_{Q}f(X)}\leq e^{\epsilon}.$ 
\item $D_{\infty}^{\mathsf{max}}(P_{X}P_{Y|X},Q_{X}P_{Y|X})=D_{\infty}^{\mathsf{max}}(P_{X},Q_{X})$. 
\end{enumerate}
\end{lem}
The proof of this lemma is omitted.

\subsection{\label{subsec:Problem-Formulation}Problem Formulation and Result
Summary}

We consider the \emph{distribution approximation problem}, which can
be described as follows. We are given a target ``output'' distribution
$Q_{Y}$ that we would like to simulate. At the same time, we are
given a $k$-length sequence of a memoryless source $X^{k}\sim P_{X}^{k}$.
We would like to design a function $f:\mathcal{X}^{k}\to\mathcal{Y}^{n}$
such that the distance, according to some divergence measure, of the
simulated distribution $P_{Y^{n}}$ with $Y^{n}:=f(X^{k})$ and $n$
independent copies of the target distribution $Q_{Y}^{n}$ is minimized.
Here we let $n=\left\lceil kR\right\rceil $, where $R$ is a fixed
positive number known as the {\em rate}. We assume the alphabets
$\mathcal{X}$ and $\mathcal{Y}$ are finite. We also assume $P_{X}(x)>0,\forall x\in\mathcal{X}$
and $Q_{Y}(y)>0,\forall y\in\mathcal{Y}$, i.e., $\mathcal{X}$ and
$\mathcal{Y}$ are the supports of $P_{X}$ and $Q_{Y}$, respectively.
There are now two fundamental questions associated to this simulation
task: (i) As $k\to\infty$, what is the asymptotic level of approximation
as a function of $(R,P_{X},Q_{Y})$? (ii) As $k\to\infty$, what is
the maximum rate $R$ such that the discrepancy between the distribution
$P_{Y^{n}}$ and $Q_{Y}^{n}$ tends to zero? In contrast to previous
works on this problem~\cite{Han03,kumagai2017second}, here we employ
R\'enyi divergences $D_{\alpha}(P_{Y^{n}}\|Q_{Y}^{n}),D_{\alpha}(Q_{Y}^{n}\|P_{Y^{n}})$,
and $D_{\alpha}^{\mathsf{max}}(P_{Y^{n}},Q_{Y}^{n})$ of all orders
$\alpha\in[0,\infty]$ to measure the discrepancy between $P_{Y^{n}}$
and $Q_{Y}^{n}$.

Furthermore, our results are summarized in Table \ref{tab:Summary-of-our}.
\begin{table*}
\caption{\label{tab:Summary-of-our}Summary of results on asymptotics of R\'enyi
divergences. Here $a(t')$ and $b(t')$ are defined in \eqref{eq:-48}
and \eqref{eq:-49} respectively, and $c(\alpha):=\left|\frac{\alpha-1}{\alpha}\right|$
for $\alpha\protect\neq0$. For $\alpha\in[0,1]\cup\{\infty\}$, R\'enyi
conversion rates for unnormalized R\'enyi divergences are the same to
those for normalized R\'enyi divergences. Furthermore, for $\alpha\in(1,\infty)$,
an achievability result on the R\'enyi conversion rate for unnormalized
R\'enyi divergence $D_{\alpha}(P_{Y^{n}}\|Q_{Y}^{n})$ is given in \eqref{eq:-57}.
All of our results summarized here are new, except that the R\'enyi
conversion rates for the unnormalized R\'enyi divergence $D_{\alpha}(P_{Y^{n}}\|Q_{Y}^{n})$
with $\alpha\in(0,\frac{1}{2}]$ are implied by Kumagai and Hayashi
\cite{kumagai2017second} and Han \cite{Han03}. }
\centering{}\centering %
\begin{tabular}{|>{\centering}m{2.2cm}|>{\centering}m{1.7cm}|c|}
\hline 
R\'enyi Divergences & Cases & Asymptotics of R\'enyi Divergences\tabularnewline
\hline 
\hline 
$\frac{1}{n}D_{\alpha}(P_{Y^{n}}\|Q_{Y}^{n})$ & $\alpha\in[0,\infty]$ & $\sup_{t\in[0,1)}\left\{ tH_{\frac{1}{1-t}}(Q_{Y})-\frac{t}{R}H_{\frac{1}{1-c(\alpha)t}}(P_{X})\right\} $ \tabularnewline
\hline 
\multirow{4}{2.2cm}{$\frac{1}{n}D_{\alpha}(Q_{Y}^{n}\|P_{Y^{n}})$} & $\alpha=0$  & $0$\tabularnewline
\cline{2-3} 
 & $\alpha\in(0,1)$ & $\frac{1}{c(\alpha)}\max_{t\in[0,1]}\left\{ tH_{\frac{1}{1-t}}(Q_{Y})-\frac{t}{R}H_{\frac{1}{1+\frac{t}{c(\alpha)}}}(P_{X})\right\} $\tabularnewline
\cline{2-3} 
 & $\alpha\in[1,\infty]$

$R<\frac{H_{0}(P_{X})}{H_{0}(Q_{Y})}$ & $\sup_{t\in(0,\infty)}\left\{ tH_{\frac{1}{1+c(\alpha)t}}(Q_{Y})-\frac{t}{R}H_{\frac{1}{1+t}}(P_{X})\right\} $\tabularnewline
\cline{2-3} 
 & $\alpha\in[1,\infty]$

$R>\frac{H_{0}(P_{X})}{H_{0}(Q_{Y})}$ & $\infty$ \tabularnewline
\hline 
\multirow{5}{2.2cm}{$\frac{1}{n}D_{\alpha}^{\mathsf{max}}(P_{Y^{n}},Q_{Y}^{n})$} & $\alpha=0$  & $\sup_{t\in[0,1)}\left\{ tH_{\frac{1}{1-t}}(Q_{Y})-\frac{t}{R}H_{0}(P_{X})\right\} $\tabularnewline
\cline{2-3} 
 & $\alpha\in(0,1)$ & $\sup_{t\in[0,1)}\max_{t'\in[0,1]}\left\{ tb(t')H_{\frac{1}{1-t}}(Q_{Y})-\frac{tb(t')}{R}H_{\frac{1}{1+\frac{b(t')}{a(t')}t}}(P_{X})\right\} $\tabularnewline
\cline{2-3} 
 & \multirow{2}{1.7cm}{$\,\,\,\, \alpha\in[1,\infty]$ $R<\frac{H_{0}(P_{X})}{H_{0}(Q_{Y})}$} & $\max\biggl\{\sup_{t\in[0,1)\cup(\frac{1}{c(\alpha)},\infty)}\left\{ tH_{\frac{1}{1-t}}(Q_{Y})-\frac{t}{R}H_{\frac{1}{1-c(\alpha)t}}(P_{X})\right\} ,$ \tabularnewline
 &  & $\qquad\sup_{t\in(0,\infty)}\left\{ tH_{\frac{1}{1+c(\alpha)t}}(Q_{Y})-\frac{t}{R}H_{\frac{1}{1+t}}(P_{X})\right\} \biggr\}$
\tabularnewline
\cline{2-3} 
 & $\alpha\in[1,\infty]$

$R>\frac{H_{0}(P_{X})}{H_{0}(Q_{Y})}$ & $\infty$ \tabularnewline
\hline 
\multicolumn{3}{c}{}\tabularnewline
\hline 
R\'enyi Divergences & Cases & R\'enyi Conversion Rates\tabularnewline
\hline 
\hline 
\multirow{3}{2.2cm}{$\frac{1}{n}D_{\alpha}(P_{Y^{n}}\|Q_{Y}^{n})$} & $\alpha=0$  & $\frac{H_{0}(P_{X})}{H(Q_{Y})}$\tabularnewline
\cline{2-3} 
 & $\alpha\in(0,1)$ & $\frac{H(P_{X})}{H(Q_{Y})}$\tabularnewline
\cline{2-3} 
 & $\alpha\in[1,\infty]$ & $\inf_{t\in(0,1)}\frac{H_{\frac{1}{1-c(\alpha)t}}(P_{X})}{H_{\frac{1}{1-t}}(Q_{Y})}$ \tabularnewline
\hline 
\multirow{4}{2.2cm}{$\frac{1}{n}D_{\alpha}(Q_{Y}^{n}\|P_{Y^{n}})$} & $\alpha=0$  & $\infty$ \tabularnewline
\cline{2-3} 
 & $\alpha\in(0,1)$ & $\frac{H(P_{X})}{H(Q_{Y})}$\tabularnewline
\cline{2-3} 
 & $\alpha=1$  & $\min\left\{ \frac{H(P_{X})}{H(Q_{Y})},\frac{H_{0}(P_{X})}{H_{0}(Q_{Y})}\right\} $ \tabularnewline
\cline{2-3} 
 & $\alpha\in(1,\infty]$ & $\inf_{t\in(0,\infty)}\frac{H_{\frac{1}{1+t}}(P_{X})}{H_{\frac{1}{1+c(\alpha)t}}(Q_{Y})}$ \tabularnewline
\hline 
\multirow{4}{2.2cm}{$\frac{1}{n}D_{\alpha}^{\mathsf{max}}(P_{Y^{n}},Q_{Y}^{n})$} & $\alpha=0$  & $\frac{H_{0}(P_{X})}{H(Q_{Y})}$\tabularnewline
\cline{2-3} 
 & $\alpha\in(0,1)$ & $\frac{H(P_{X})}{H(Q_{Y})}$\tabularnewline
\cline{2-3} 
 & $\alpha=1$  & $\min\left\{ \frac{H(P_{X})}{H(Q_{Y})},\frac{H_{0}(P_{X})}{H_{0}(Q_{Y})}\right\} $\tabularnewline
\cline{2-3} 
 & $\alpha\in(1,\infty]$ & $\min\left\{ \inf_{t\in[0,1)\cup(\frac{1}{c(\alpha)},\infty)}\frac{H_{\frac{1}{1-c(\alpha)t}}(P_{X})}{H_{\frac{1}{1-t}}(Q_{Y})},\inf_{t\in(0,\infty)}\frac{H_{\frac{1}{1+t}}(P_{X})}{H_{\frac{1}{1+c(\alpha)t}}(Q_{Y})}\right\} $ \tabularnewline
\hline 
\end{tabular}
\end{table*}

\subsection{\label{subsec:Mappings}Mappings}

\begin{figure*}[t]
\centering \subfloat[\label{fig:Mapping-1}Mapping 1]{\includegraphics[width=0.5\textwidth]{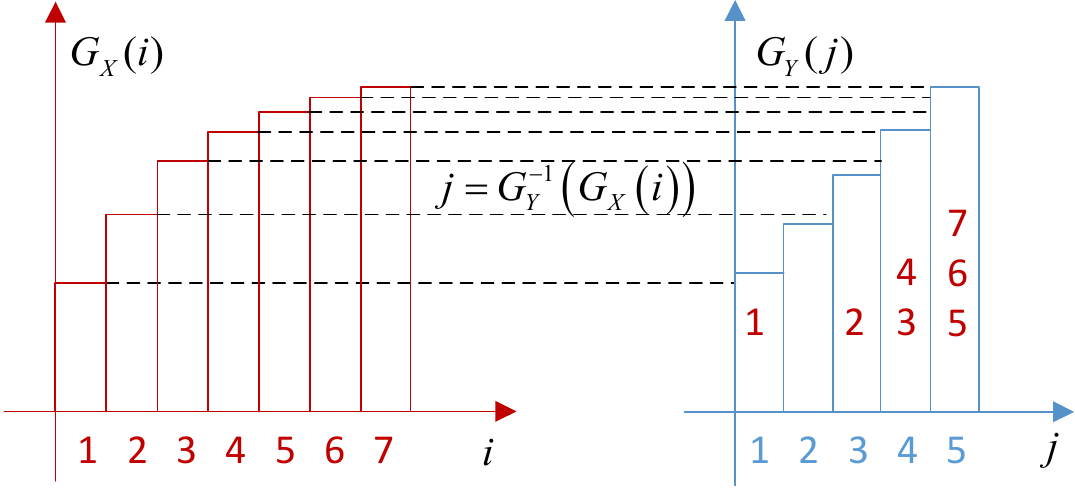}

}

\subfloat[\label{fig:Mapping-2}Mapping 2]{\includegraphics[width=0.8\textwidth]{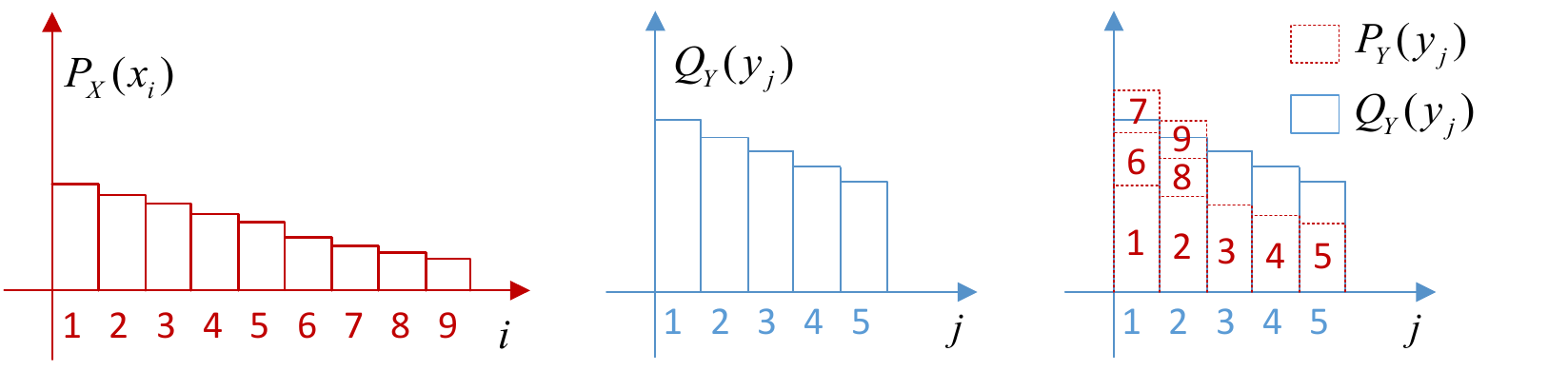}

}

\caption{Illustrations of Mappings 1 and 2.}
\label{fig:mappings} 
\end{figure*}

The following two fundamental mappings, illustrated in Fig.~\ref{fig:mappings},
will be used in our constructions of the functions $f:\mathcal{X}^{k}\to\mathcal{Y}^{n}$
described in Subsection \ref{subsec:Problem-Formulation}.

Consider two (possibly unnormalized) nonnegative measures $P_{X}$
and $Q_{Y}$. Sort the elements in $\mathcal{X}$ as $x_{1},x_{2},...,x_{|\mathcal{X}|}$
such that $P_{X}(x_{1})\geq P_{X}(x_{2})\geq...\geq P_{X}(x_{|\mathcal{X}|})$.
Similarly, sort the elements in $\mathcal{Y}$ as $y_{1},y_{2},...,y_{|\mathcal{Y}|}$
such that $Q_{Y}(y_{1})\geq Q_{Y}(y_{2})\geq...\geq Q_{Y}(y_{|\mathcal{Y}|})$.
Consider two mappings from $\mathcal{X}$ to~$\mathcal{Y}$ as follows: 
\begin{itemize}
\item Mapping 1 (Inverse-Transform): If $P_{X}$ and/or $Q_{Y}$ are unnormalized,
then normalize them first. Define $G_{X}(i):=P_{X}\left(x_{l}:l\leq i\right)$
and $G_{X}^{-1}(\theta):=\max\left\{ i\in\mathbb{N}:G_{X}(i)\leq\theta\right\} $.
Similarly, for $Q_{Y}$, we define $G_{Y}(j):=Q_{Y}\left(y_{l}:l\leq j\right)$
and $G_{Y}^{-1}(\theta):=\min\left\{ j\in\mathbb{N}:G_{Y}(j)\geq\theta\right\} $.
Consider the following mapping. For each $i\in[1:|\mathcal{X}|]$,
$x_{i}$ is mapped to $y_{j}$ where $j=G_{Y}^{-1}(G_{X}(i))$. The
resulting distribution is denoted as $P_{Y}$. This mapping is illustrated
in Fig. \ref{fig:Mapping-1}. For such a mapping, the following properties
hold: 
\begin{enumerate}
\item If $P_{X}(x_{i})\geq Q_{Y}(y_{j})$ where $i:=G_{X}^{-1}(G_{Y}(j))$,
then $|\left\{ i:G_{Y}^{-1}(G_{X}(i))=j\right\} |\leq1$. Hence, $P_{Y}(y_{j})\le P_{X}(x_{i})$.
\item If $P_{X}(x_{i})<Q_{Y}(y_{j})$ where $i:=G_{X}^{-1}(G_{Y}(j))$,
then $|\left\{ i:G_{Y}^{-1}(G_{X}(i))=j\right\} |\geq1$ and 
\begin{align}
 & \max\left\{ \frac{1}{2}Q_{Y}(y_{j}),Q_{Y}(y_{j})-P_{X}(x_{i})\right\} \nonumber \\
 & \leq P_{Y}(y_{j})\le Q_{Y}(y_{j})+P_{X}(x_{i}).
\end{align}
\end{enumerate}
\item Mapping 2: Denote $k_{m},m\in[1:L]$ with $k_{L}:=|\mathcal{X}|$
as a sequence of integers such that for $m\in[1:L-1]$, $\sum_{i=k_{m-1}+1}^{k_{m}-1}P_{X}(x_{i})<Q_{Y}(y_{m})\le\sum_{i=k_{m-1}+1}^{k_{m}}P_{X}(x_{i})$,
and $\sum_{i=k_{L-1}+1}^{k_{L}}P_{X}(x_{i})\leq Q_{Y}(y_{L})$ or
$\sum_{i=k_{L-1}+1}^{k_{L}-1}P_{X}(x_{i})<Q_{Y}(y_{L})\le\sum_{i=k_{L-1}+1}^{k_{L}}P_{X}(x_{i})$.
Obviously $L\leq|\mathcal{Y}|$. For each $m\in[1:L]$, map $x_{k_{m-1}+1},...,x_{k_{m}}$
to $y_{m}$. The resulting distribution is denoted as $P_{Y}$. This
mapping is illustrated in Fig. \ref{fig:Mapping-2}. For such a mapping,
we have 
\begin{equation}
Q_{Y}(y_{m})\leq P_{Y}(y_{m})<Q_{Y}(y_{m})+P_{X}(x_{k_{m}})
\end{equation}
for $m\in[1:L-1]$, 
\begin{equation}
P_{Y}(y_{m})<Q_{Y}(y_{m})+P_{X}(x_{k_{m}})
\end{equation}
for $m=L$, and $P_{Y}(y_{m})=0$ for $m>L$. 
\end{itemize}

\section{\label{sec:Renyi-Distribution-Approximation}R\'enyi Distribution Approximation }

\subsection{Asymptotics of R\'enyi Divergences}

We first characterize the asymptotics of R\'enyi divergences $D_{\alpha}(P_{Y^{n}}\|Q_{Y}^{n})$,
$D_{\alpha}(Q_{Y}^{n}\|P_{Y^{n}})$, and $D_{\alpha}^{\mathsf{max}}(P_{Y^{n}},Q_{Y}^{n})$,
as shown by the following theorems. 
\begin{thm}[Asymptotics of $\frac{1}{n}D_{\alpha}(P_{Y^{n}}\|Q_{Y}^{n})$]
\label{thm:RenyiPQ} For any $\alpha\in[0,\infty]$, we have 
\begin{align}
 & \lim_{n\to\infty}\frac{1}{n}\inf_{f}D_{\alpha}(P_{Y^{n}}\|Q_{Y}^{n})\nonumber \\
 & =\sup_{t\in[0,1)}\left\{ tH_{\frac{1}{1-t}}(Q_{Y})-\frac{t}{R}H_{\frac{1}{1-\frac{\alpha-1}{\alpha}t}}(P_{X})\right\} .
\end{align}
\end{thm}
\begin{thm}[Asymptotics of $\frac{1}{n}D_{\alpha}(Q_{Y}^{n}\|P_{Y^{n}})$]
\label{thm:RenyiQP} For any $\alpha\in[0,\infty]$, we have 
\begin{align}
 & \lim_{n\to\infty}\frac{1}{n}\inf_{f}D_{\alpha}(Q_{Y}^{n}\|P_{Y^{n}})\nonumber \\
 & =\begin{cases}
\infty,\quad\alpha\in[1,\infty]\textrm{ and }R>\frac{H_{0}(P_{X})}{H_{0}(Q_{Y})};\\
\sup_{t\in(0,\infty)}\left\{ tH_{\frac{1}{1+\frac{\alpha-1}{\alpha}t}}(Q_{Y})-\frac{t}{R}H_{\frac{1}{1+t}}(P_{X})\right\} ,\\
\qquad\;\alpha\in[1,\infty]\textrm{ and }R<\frac{H_{0}(P_{X})}{H_{0}(Q_{Y})};\\
\frac{\alpha}{1-\alpha}\max_{t\in[0,1]}\left\{ tH_{\frac{1}{1-t}}(Q_{Y})-\frac{t}{R}H_{\frac{1}{1+\frac{\alpha}{1-\alpha}t}}(P_{X})\right\} ,\\
\qquad\;\alpha\in(0,1);\\
0,\quad\;\alpha=0.
\end{cases}\label{eq:-4}
\end{align}
\end{thm}
\begin{thm}[Asymptotics of $\frac{1}{n}D_{\alpha}^{\mathsf{max}}(P_{Y^{n}},Q_{Y}^{n})$]
\label{thm:Renyimax} For any $\alpha\in[0,\infty]$, we have \eqref{eq:-12}
(given on page \pageref{eq:-12}), 
\begin{figure*}
\begin{align}
 & \lim_{n\to\infty}\frac{1}{n}\inf_{f}D_{\alpha}^{\mathsf{max}}(P_{Y^{n}},Q_{Y}^{n})\nonumber \\
 & =\begin{cases}
\infty, & \alpha\in[1,\infty]\textrm{ and }R>\frac{H_{0}(P_{X})}{H_{0}(Q_{Y})}\\
\max\biggl\{\sup_{t\in[0,1)\cup(\frac{\alpha}{\alpha-1},\infty)}\left\{ tH_{\frac{1}{1-t}}(Q_{Y})-\frac{t}{R}H_{\frac{1}{1-\frac{\alpha-1}{\alpha}t}}(P_{X})\right\} ,\\
\qquad\sup_{t\in(0,\infty)}\left\{ tH_{\frac{1}{1+\frac{\alpha-1}{\alpha}t}}(Q_{Y})-\frac{t}{R}H_{\frac{1}{1+t}}(P_{X})\right\} \biggr\}, & \alpha\in(1,\infty]\textrm{ and }R<\frac{H_{0}(P_{X})}{H_{0}(Q_{Y})}\\
\max\biggl\{\sup_{t\in[0,1)}\left\{ tH_{\frac{1}{1-t}}(Q_{Y})-\frac{t}{R}H(P_{X})\right\} ,\\
\qquad\sup_{t\in(0,\infty)}\left\{ tH(Q_{Y})-\frac{t}{R}H_{\frac{1}{1+t}}(P_{X})\right\} \biggr\}, & \alpha=1\textrm{ and }R<\frac{H_{0}(P_{X})}{H_{0}(Q_{Y})}\\
\sup_{t\in[0,1)}\max_{t'\in[0,1]}\left\{ tb(t')H_{\frac{1}{1-t}}(Q_{Y})-\frac{tb(t')}{R}H_{\frac{1}{1+\frac{b(t')}{a(t')}t}}(P_{X})\right\} , & \alpha\in(0,1)\\
\sup_{t\in[0,1)}\left\{ tH_{\frac{1}{1-t}}(Q_{Y})-\frac{t}{R}H_{0}(P_{X})\right\} , & \alpha=0
\end{cases}\label{eq:-12}
\end{align}

\hrulefill{} 
\end{figure*}

where 
\begin{align}
a(t') & =\left(\frac{\alpha}{1-\alpha}-1\right)t'+1\label{eq:-48}\\
b(t') & =\left(1-\frac{\alpha}{1-\alpha}\right)t'+\frac{\alpha}{1-\alpha}.\label{eq:-49}
\end{align}
\end{thm}
\begin{rem}
For $\alpha\in[1,\infty]$ and $R=\frac{H_{0}(P_{X})}{H_{0}(Q_{Y})}$,
the asymptotic behavior of $\frac{1}{n}\inf_{f}D_{\alpha}(Q_{Y}^{n}\|P_{Y^{n}})$
and $\frac{1}{n}\inf_{f}D_{\alpha}^{\mathsf{max}}(P_{Y^{n}},Q_{Y}^{n})$
depends on how fast $\frac{n}{k}$ converges to $R$. In this paper,
we set $n=\left\lceil kR\right\rceil $, i.e., the fastest case. For
this case, $\frac{1}{n}\inf_{f}D_{\alpha}(Q_{Y}^{n}\|P_{Y^{n}})=\frac{1}{n}\inf_{f}D_{\alpha}^{\mathsf{max}}(P_{Y^{n}},Q_{Y}^{n})=\infty$,
if $kR\notin\mathbb{N}$; and $\frac{1}{n}\inf_{f}D_{\alpha}(Q_{Y}^{n}\|P_{Y^{n}})=\frac{1}{n}D_{\alpha}(\{Q_{i}\}\|\{P_{i}\})$
and $\frac{1}{n}\inf_{f}D_{\alpha}^{\mathsf{max}}(P_{Y^{n}},Q_{Y}^{n})=\frac{1}{n}\max\left\{ D_{\alpha}(\{P_{i}\}\|\{Q_{i}\}),D_{\alpha}(\{Q_{i}\}\|\{P_{i}\})\right\} $,
if $kR\in\mathbb{N}$, where $\{P_{i}\}$ and $\{Q_{i}\}$ respectively
denote the resulting sequences after sorting the elements of $P_{X}^{k}$
and $Q_{Y}^{n}$ in descending order. 
\end{rem}
The proofs of Theorems \ref{thm:RenyiPQ}, \ref{thm:RenyiQP}, and
\ref{thm:Renyimax} are provided in Appendices \ref{sec:Proof-of-Theorem-RenyiPQ},
\ref{sec:Proof-of-Theorem-RenyiQP}, and \ref{sec:Proof-of-Theorem-Renyimax},
respectively. For the achievability parts, we partition the sequences
in $\mathcal{X}^{k}$ and $\mathcal{Y}^{n}$ into type classes, and
design codes on the level of type classes. More specifically, for
Theorem \ref{thm:RenyiPQ}, we first design a function $g:\mathcal{P}^{\left(k\right)}\left(\mathcal{X}\right)\to\mathcal{P}^{\left(n\right)}\left(\mathcal{Y}\right)$
that maps $k$-types on $\mathcal{X}$ to $n$-types on $\mathcal{Y}$;
and then a code $f$ induced by $g$ is obtained by mapping the sequences
in $\mathcal{T}_{T_{X}}$ to the sequences in $\mathcal{T}_{g(T_{X})}$
as uniformly as possible for all $T_{X}\in\mathcal{P}^{\left(k\right)}\left(\mathcal{X}\right)$,
i.e., $f$ maps approximately $\nicefrac{\left|\mathcal{T}_{T_{X}}\right|}{\left|\mathcal{T}_{g(T_{X})}\right|}$
sequences in $\mathcal{T}_{T_{X}}$ to each distinct sequence in $\mathcal{T}_{g(T_{X})}$.
Here the optimal selection of the function $g$ depends on $s$ and
requires careful analysis (the detail can be found in the proof).
The intuition of designing such a code is given in the following.
On one hand, observe that 
\begin{align}
 & \frac{1}{n}D_{1+s}(P_{Y^{n}}\|Q_{Y}^{n})\nonumber \\
 & =\frac{1}{ns}\log\biggl\{\sum_{T_{Y}}\sum_{y^{n}\in\mathcal{T}_{T_{Y}}}\nonumber \\
 & \quad\Bigl(\sum_{T_{X}}\sum_{x^{k}\in\mathcal{T}_{T_{X}}}P_{X}^{k}(x^{k})1\left\{ y^{n}=f(x^{k})\right\} \Bigr)^{1+s}Q_{Y}^{n}(y^{n})^{-s}\biggr\}\\
 & =\frac{1}{ns}\log\biggl\{\max_{T_{X},T_{Y}}\sum_{y^{n}\in\mathcal{T}_{T_{Y}}}\nonumber \\
 & \quad\Bigl(\sum_{x^{k}\in\mathcal{T}_{T_{X}}}P_{X}^{k}(x^{k})1\left\{ y^{n}=f(x^{k})\right\} \Bigr)^{1+s}Q_{Y}^{n}(y^{n})^{-s}\biggr\}+o(1)\label{eq:-3}
\end{align}
where \eqref{eq:-3} follows since the number of $n$-types (or $k$-types)
is only polynomial in $n$ (or $k$). This means that for any code
$f$, the asymptotics of $\frac{1}{n}D_{1+s}(P_{Y^{n}}\|Q_{Y}^{n})$
induced by $f$ is only determined by restrictions of $f$ on $\mathcal{A}\left(T_{X},T_{Y}\right):=\left\{ x^{n}\in\mathcal{T}_{T_{X}}:\:f(x^{n})\in\mathcal{T}_{T_{Y}}\right\} $
for different $\left(T_{X},T_{Y}\right)$. In other words, the performance
of a code $f$ only depends on its restrictions to those maps from
$\mathcal{A}\left(T_{X},T_{Y}\right)$ to $\mathcal{T}_{T_{Y}}$.
On the other hand, $P_{X}^{k}(x^{k})$ and $Q_{Y}^{n}(y^{n})$ are
uniform on $\mathcal{T}_{T_{X}}$ and $\mathcal{T}_{T_{Y}}$, respectively.
Hence for different $\left(T_{X},T_{Y}\right)$, to make the objective
function of \eqref{eq:-3} as small as possible, we need to map the
sequences in $\mathcal{A}\left(T_{X},T_{Y}\right)$ to the sequences
in $\mathcal{T}_{T_{Y}}$ as uniformly as possible. Since $\bigcup_{T_{Y}}\mathcal{A}\left(T_{X},T_{Y}\right)=\mathcal{T}_{T_{X}}$
and the number of types $T_{Y}$ is polynomial in $n$, for each $T_{X}$,
there is a dominant type $T_{Y}=g(T_{X})$ such that redefining $f$
to satisfy $\left\{ f(x^{n}),x^{n}\in\mathcal{T}_{T_{X}}\right\} \subseteq\mathcal{T}_{T_{Y}}$
with $T_{Y}=g(T_{X})$ does not affect the asymptotics of $\frac{1}{n}D_{1+s}(P_{Y^{n}}\|Q_{Y}^{n})$.
Therefore, we only need to consider the codes consisting of a function
$g$ that maps $k$-types on $\mathcal{X}$ to $n$-types on $\mathcal{Y}$,
and mappings that map sequences in $\mathcal{T}_{T_{X}}$ to sequences
in $\mathcal{T}_{g(T_{X})}$ as uniformly as possible.

The achievability proof for Theorem \ref{thm:RenyiQP} follows similar
ideas. However, in contrast, to ensure that $\frac{1}{n}\inf_{f}D_{\alpha}(Q_{Y}^{n}\|P_{Y^{n}})$
is finite and also as small as possible, it is required that $\supp\left(P_{Y^{n}}\right)\supseteq\supp\left(Q_{Y}^{n}\right)$
and $P_{Y^{n}}(y^{n})$ should be as large as possible for all $y^{n}$.
On the other hand, observe that $\left|\mathcal{P}^{\left(n\right)}\left(\mathcal{Y}\right)\right|$
is polynomial in $n$. Hence for each $T_{X}$, we should partition
$\mathcal{T}_{T_{X}}$ into $\left|\mathcal{P}^{\left(n\right)}\left(\mathcal{Y}\right)\right|$
subsets with equal size, and for each $T_{Y}$, map the sequences
in each subset to the sequences in the set $\mathcal{T}_{T_{Y}}$
as uniformly as possible. Observe that for each $T_{Y}$, there must
exist a type $T_{X}$ such that $H(T_{X})\ge H(T_{Y})+o(1)$ (otherwise
$\frac{1}{n}\inf_{f}D_{\alpha}(Q_{Y}^{n}\|P_{Y^{n}})=\infty$) and
moreover, similar to \eqref{eq:-3}, the summation term is dominated
by some type $T_{X}$ such that $H(T_{X})\ge H(T_{Y})+o(1)$. Hence
without loss of any optimality, it suffices to consider the following
mapping. For each $T_{X}$ and $\delta>0$, partition $\mathcal{T}_{T_{X}}$
into $\left|\left\{ T_{Y}:H(T_{X})\ge H(T_{Y})+\delta\right\} \right|$
subsets with approximately same size. For each $T_{Y}$ such that
$H(T_{X})\ge H(T_{Y})+\delta$, map the sequences in each subset to
the sequences in the set $\mathcal{T}_{T_{Y}}$ as uniformly as possible.

The code used to prove the achievability part of Theorem \ref{thm:Renyimax}
is a combination of the two codes above.

\subsection{R\'enyi Conversion Rates}

As shown in the theorems above, when the code rate is large, the normalized
R\'enyi divergences $\frac{1}{n}D_{\alpha}(P_{Y^{n}}\|Q_{Y}^{n})$,
$\frac{1}{n}D_{\alpha}(Q_{Y}^{n}\|P_{Y^{n}})$, and $\frac{1}{n}D_{\alpha}^{\mathsf{max}}(P_{Y^{n}},Q_{Y}^{n})$
converge to a positive number; however when the code rate is small
enough, the normalized R\'enyi divergences converge to zero. This threshold
rate, termed the \emph{R\'enyi conversion rate}, is important, since
it represents the maximum possible rate under the condition that the
distribution induced by the code approximates the target distribution
arbitrarily well as $n\to\infty$. We characterize the R\'enyi conversion
rates for normalized and unnormalized $D_{\alpha}(P_{Y^{n}}\|Q_{Y}^{n})$,
$D_{\alpha}(Q_{Y}^{n}\|P_{Y^{n}})$, and $D_{\alpha}^{\mathsf{max}}(P_{Y^{n}},Q_{Y}^{n})$
in the following theorems. 
\begin{thm}[R\'enyi Conversion Rate for $D_{\alpha}(P_{Y^{n}}\|Q_{Y}^{n})$]
\label{thm:Renyi1Con} For any $\alpha\in[0,\infty]$, 
\begin{align}
 & \sup\left\{ R:\frac{1}{n}D_{\alpha}(P_{Y^{n}}\|Q_{Y}^{n})\rightarrow0\right\} \nonumber \\
 & =\begin{cases}
\inf_{t\in(0,1)}\frac{H_{\frac{1}{1-\frac{\alpha-1}{\alpha}t}}(P_{X})}{H_{\frac{1}{1-t}}(Q_{Y})}, & \alpha\in[1,\infty]\\
\frac{H(P_{X})}{H(Q_{Y})}, & \alpha\in(0,1)\\
\frac{H_{0}(P_{X})}{H(Q_{Y})}, & \alpha=0
\end{cases}.\label{eq:}
\end{align}
For $\alpha\in[0,1]\cup\left\{ \infty\right\} $, we have 
\begin{align}
 & \sup\left\{ R:D_{\alpha}(P_{Y^{n}}\|Q_{Y}^{n})\rightarrow0\right\} \nonumber \\
 & =\sup\left\{ R:\frac{1}{n}D_{\alpha}(P_{Y^{n}}\|Q_{Y}^{n})\rightarrow0\right\} .\label{eq:-58}
\end{align}
For $\alpha\in[1,\infty]$, we have 
\begin{align}
 & \sup\left\{ R:\frac{1}{n}D_{\alpha}(P_{Y^{n}}\|Q_{Y}^{n})\rightarrow0\right\} \nonumber \\
 & \geq\sup\left\{ R:D_{\alpha}(P_{Y^{n}}\|Q_{Y}^{n})\rightarrow0\right\} \\
 & \geq\inf_{t\in(0,1)}\frac{H_{\frac{\alpha-1+t}{\alpha-1+t-\left(\alpha-1\right)t}}(P_{X})}{H_{\frac{1}{1-t}}(Q_{Y})}.\label{eq:-57}
\end{align}
\end{thm}
\begin{rem}
The analogous result under the TV distance measure was first shown
by Han \cite{Han03}. Theorem \ref{thm:Renyi1Con} is an extension
of \cite{Han03} to the R\'enyi divergence of all orders $\alpha\in[0,\infty]$.
Besides, the first-order and second-order rates, as well as the conversion
rates of the quantum version, for the unnormalized R\'enyi divergence
$D_{\alpha}(P_{Y^{n}}\|Q_{Y}^{n})$ with $\alpha=\frac{1}{2}$ were
given by Kumagai and Hayashi \cite{kumagai2017second}; and the corresponding
moderate deviation of the quantum R\'enyi conversion rates with the
same order was studied by Chubb, Tomamichel, and Korzekwa1 \cite{chubb2018moderate}.
The result for the unnormalized R\'enyi divergence with $\alpha\in(0,\frac{1}{2})$
can be obtained by combining two observations: 1) the achievability
for $D_{\frac{1}{2}}(P_{Y^{n}}\|Q_{Y}^{n})$ implies the achievability
for $\alpha\in(0,\frac{1}{2})$; 2) by Pinsker's inequality for R\'enyi
divergence \cite{Erven}, the converse result for the TV distance
measure \cite{Han03} implies the converse for $\alpha\in(0,\frac{1}{2})$.
Our results for orders $\alpha\in\left\{ 0\right\} \cup(\frac{1}{2},\infty]$
are new. 
\end{rem}
\begin{rem}
$D_{\alpha}(P_{Y|X=x}\|P_{Y|X=x'})\leq\epsilon$ for all neighboring
databases $x,x'$ is known as the \emph{$\epsilon$-R\'enyi differential
privacy of order $\alpha$} \cite{mironov2017renyi}, and the special
case with $\alpha=\infty$ is known as the \emph{$\epsilon$-differential
privacy} \cite{dwork2008differential}. Here, $X$ represents public
data and $Y$ represents private data. In the theorem above, this
measure is applied to the random variable simulation problem, and
we provide a ``necessary and sufficient condition'' for $\lim_{n\to\infty}\frac{1}{n}D_{\alpha}\leq\epsilon$
for any $\epsilon>0$. 
\end{rem}
\begin{thm}[R\'enyi Conversion Rate for $D_{\alpha}(Q_{Y}^{n}\|P_{Y^{n}})$]
\label{thm:Renyi2Con} For any $\alpha\in[0,\infty]$, 
\begin{align}
 & \sup\left\{ R:\frac{1}{n}D_{\alpha}(Q_{Y}^{n}\|P_{Y^{n}})\rightarrow0\right\} \nonumber \\
 & =\begin{cases}
\inf_{t\in(0,\infty)}\frac{H_{\frac{1}{1+t}}(P_{X})}{H_{\frac{1}{1+\frac{\alpha-1}{\alpha}t}}(Q_{Y})}, & \alpha\in(1,\infty]\\
\min\left\{ \frac{H(P_{X})}{H(Q_{Y})},\frac{H_{0}(P_{X})}{H_{0}(Q_{Y})}\right\} , & \alpha=1\\
\frac{H(P_{X})}{H(Q_{Y})}, & \alpha\in(0,1)\\
\infty, & \alpha=0
\end{cases}.\label{eq:-1}
\end{align}
For $\alpha\in[0,1]\cup\left\{ \infty\right\} $, we have 
\begin{align}
 & \sup\left\{ R:D_{\alpha}(Q_{Y}^{n}\|P_{Y^{n}})\rightarrow0\right\} \nonumber \\
 & =\sup\left\{ R:\frac{1}{n}D_{\alpha}(Q_{Y}^{n}\|P_{Y^{n}})\rightarrow0\right\} .\label{eq:-59}
\end{align}
\end{thm}
\begin{rem}
\label{rmk:unnormRenyi} Our results for all orders $\alpha\in[0,\infty]$
are new. 
\end{rem}
\begin{thm}[R\'enyi Conversion Rate for $D_{\alpha}^{\mathsf{max}}(P_{Y^{n}},Q_{Y}^{n})$]
\label{thm:Renyi3Con} For $\alpha\in[0,\infty]$, we have 
\begin{align}
 & \sup\left\{ R:\frac{1}{n}D_{\alpha}^{\mathsf{max}}(P_{Y^{n}},Q_{Y}^{n})\rightarrow0\right\} \nonumber \\
 & =\begin{cases}
\min\Biggl\{\inf_{t\in[0,1)\cup(\frac{\alpha}{\alpha-1},\infty)}\frac{H_{\frac{1}{1-\frac{\alpha-1}{\alpha}t}}(P_{X})}{H_{\frac{1}{1-t}}(Q_{Y})},\\
\qquad\inf_{t\in(0,\infty)}\frac{H_{\frac{1}{1+t}}(P_{X})}{H_{\frac{1}{1+\frac{\alpha-1}{\alpha}t}}(Q_{Y})}\Biggr\}, & \alpha\in(1,\infty]\\
\min\left\{ \frac{H(P_{X})}{H(Q_{Y})},\frac{H_{0}(P_{X})}{H_{0}(Q_{Y})}\right\} , & \alpha=1\\
\frac{H(P_{X})}{H(Q_{Y})}, & \alpha\in(0,1)\\
\frac{H_{0}(P_{X})}{H(Q_{Y})}, & \alpha=0
\end{cases}.\label{eq:-18}
\end{align}
For $\alpha\in[0,1]\cup\left\{ \infty\right\} $, we have 
\begin{align}
 & \sup\left\{ R:D_{\alpha}^{\mathsf{max}}(P_{Y^{n}},Q_{Y}^{n})\rightarrow0\right\} \nonumber \\
 & =\sup\left\{ R:\frac{1}{n}D_{\alpha}^{\mathsf{max}}(P_{Y^{n}},Q_{Y}^{n})\rightarrow0\right\} .
\end{align}
\end{thm}
\begin{rem}
Note that for $\alpha\in(1,\infty]$, \eqref{eq:-18} involves an
infimum taken over $(\frac{\alpha}{\alpha-1},\infty)$, and hence
it is in general smaller than the minimum of \eqref{eq:} and \eqref{eq:-1}. 
\end{rem}
\begin{rem}
For $\alpha=\infty$, the R\'enyi conversion rate in \eqref{eq:-18}
is $\min_{\beta\in[-\infty,\infty]}\frac{H_{\beta}(P_{X})}{H_{\beta}(Q_{Y})}$.
Consider $R=1$. Then this theorem implies that $P_{X}^{n}$ can approximate
$Q_{Y}^{n}$ in the sense that $\frac{1}{n}D_{\infty}^{\mathsf{max}}(P_{Y^{n}},Q_{Y}^{n})\rightarrow0$
or $D_{\infty}^{\mathsf{max}}(P_{Y^{n}},Q_{Y}^{n})\rightarrow0$,
if $H_{\beta}(P_{X})>H_{\beta}(Q_{Y})$ for all $\beta\in[-\infty,\infty]$,
and only if $H_{\beta}(P_{X})\geq H_{\beta}(Q_{Y})$ for all $\beta\in[-\infty,\infty]$.
This also implies the statement 1) of \cite[Proposition III.3]{yu2018asymptotic},
since if $H_{\beta}(P_{X})<H_{\beta}(Q_{Y})$ for some $\beta\in[-\infty,\infty]$,
then approximate simulation (under the measure $D_{\infty}^{\mathsf{max}}$)
is impossible, and hence exact simulation is also impossible. 
\end{rem}
\begin{rem}
Note that $D_{\infty}^{\mathsf{max}}$ is an extremely strong distance
measure. Theorem \ref{thm:Renyi3Con} states that the R\'enyi conversion
rate (the maximum possible rate under the condition $D_{\infty}^{\mathsf{max}}(P_{Y^{n}},Q_{Y}^{n})\rightarrow0$)
is finite. That is to say, as the dimension tends to infinity, it
is always possible to achieve $D_{\infty}^{\mathsf{max}}(P_{Y^{n}},Q_{Y}^{n})\rightarrow0$,
even though $D_{\infty}^{\mathsf{max}}$ is extremely strong. However,
in our recent work \cite[Proposition III.4]{yu2018asymptotic}, we
showed that for some special pairs of distributions, it is impossible
to achieve $P_{Y^{n}}=Q_{Y}^{n}$ (or $D_{\infty}^{\mathsf{max}}(P_{Y^{n}},Q_{Y}^{n})=0$)
for finite $n$, i.e, the exact simulation cannot be obtained for
finite-dimensional product of distributions. Hence there exists a
big ``gap'' between approximate simulation and exact simulation
(for fixed blocklength cases), even when the approximate simulation
is realized under the measure $D_{\infty}^{\mathsf{max}}$. 
\end{rem}
\begin{rem}
The condition $D_{\infty}^{\mathsf{max}}(P,Q)\leq\epsilon$ is called
$\left(\epsilon,0\right)$-closeness, and was used to measure privacy
in \cite{cuff2016differential}. In Theorem \ref{thm:Renyi3Con},
we provide a ``necessary and sufficient condition'' for $\lim_{n\to\infty}D_{\infty}^{\mathsf{max}}(P_{Y^{n}},Q_{Y}^{n})\leq\epsilon$
or $\lim_{n\to\infty}\frac{1}{n}D_{\infty}^{\mathsf{max}}(P_{Y^{n}},Q_{Y}^{n})\leq\epsilon$
for any $\epsilon>0$. $D_{\infty}^{\mathsf{max}}(P,Q)$ is a very
strong measure, hence it can be taken as a secrecy measure for a secrecy
system when secrecy stronger than the usual notion of strong secrecy
is required. Our result can be applied to this case. Furthermore,
$D_{\infty}^{\mathsf{max}}(P,Q)$ is also related to \emph{$\epsilon$-information
privacy}, which is defined as $D_{\infty}^{\mathsf{max}}(P_{XY},P_{X}P_{Y})\leq\epsilon$
where $X$ and $Y$ represent public and private datum respectively
\cite{du2012privacy}. 
\end{rem}
The proofs of Theorems \ref{thm:Renyi1Con}, \ref{thm:Renyi2Con},
and \ref{thm:Renyi3Con} are provided in Appendices \ref{sec:Proof-of-Theorem-Renyi1Con},
\ref{sec:Proof-of-Theorem-Renyi2Con}, and \ref{sec:Proof-of-Theorem-Renyi3Con},
respectively. The R\'enyi conversion rates for normalized $D_{\alpha}(P_{Y^{n}}\|Q_{Y}^{n})$,
$D_{\alpha}(Q_{Y}^{n}\|P_{Y^{n}})$, and $D_{\alpha}^{\mathsf{max}}(P_{Y^{n}},Q_{Y}^{n})$
respectively follow from Theorems \ref{thm:RenyiPQ}, \ref{thm:RenyiQP},
and \ref{thm:Renyimax}. Obviously, the unnormalized R\'enyi conversion
rates are lower bounded by the normalized ones. We believe such lower
bounds are tight. However, we do not know how to construct an efficient
coding scheme for the case $\alpha\in(1,\infty)$. Hence for the measure
$D_{\alpha}(P_{Y^{n}}\|Q_{Y}^{n})$, we consider a relatively simple
scheme \textemdash{} the inverse-transform scheme, which is described
in Subsection \ref{subsec:Mappings} and illustrated in Fig. \ref{fig:Mapping-1}.
Another reason for using the inverse-transform scheme is that such
a scheme is optimal (which results in zero divergences) when the source
distribution $P_{X}$ is continuous \cite[Proposition 1]{yu2018beyond}.
Hence we believe it should work also well for discrete source distributions.
The specific code used to prove the achievability part for this case
is illustrated in Fig. \ref{fig:DPQ}. For $\delta>0$, define $\mathcal{B}_{1}:=\left\{ y^{n}:Q_{Y}^{n}(y^{n})\geq e^{-n\left(H(Q_{Y})+\delta\right)}\right\} $.
To ensure $D_{\alpha}(P_{Y^{n}}\|Q_{Y}^{n})\to0$, we only need to
simulate a truncated version $\widetilde{Q}_{Y^{n}}(y^{n}):=\frac{Q_{Y}^{n}(y^{n})}{Q_{Y}^{n}(\mathcal{B}_{1})}1\left\{ y^{n}\in\mathcal{B}_{1}\right\} $
of $Q_{Y}^{n}$. This is because, on one hand, for any function $f:\mathcal{X}^{k}\to\mathcal{B}_{1}$
with output $Y^{n}=f(X^{k})$, 
\begin{align}
 & D_{\alpha}(P_{Y^{n}}\|Q_{Y}^{n})\nonumber \\
 & =\frac{1}{\alpha-1}\log\sum_{y^{n}\in\mathcal{A}}P_{Y^{n}}(y^{n})\left(\frac{P_{Y^{n}}(y^{n})}{\widetilde{Q}_{Y^{n}}(y^{n})}\frac{\widetilde{Q}_{Y^{n}}(y^{n})}{Q_{Y}^{n}(y^{n})}\right)^{\alpha-1}\\
 & =\frac{1}{\alpha-1}\log\sum_{y^{n}\in\mathcal{A}}P_{Y^{n}}(y^{n})\left(\frac{P_{Y^{n}}(y^{n})}{\widetilde{Q}_{Y^{n}}(y^{n})}\frac{1}{Q_{Y}^{n}(\mathcal{B}_{1})}\right)^{\alpha-1}\\
 & =D_{\alpha}(P_{Y^{n}}\|\widetilde{Q}_{Y^{n}})-\log Q_{Y}^{n}(\mathcal{B}_{1}),
\end{align}
and on the other hand, observe that $Q_{Y}^{n}(\mathcal{B}_{1})\to1$
as $n\to\infty$. That is to say, if a function $f$ is a ``good''
simulator for $\widetilde{Q}_{Y^{n}}$ in the sense that $D_{\alpha}(P_{Y^{n}}\|\widetilde{Q}_{Y^{n}})\to0$,
then it must be also ``good'' for $Q_{Y}^{n}$ in the same sense.
The reason why we consider simulating $\widetilde{Q}_{Y^{n}}$ rather
than simulating $Q_{Y}^{n}$ directly, is that by doing this, the
influence of the behavior of $\left\{ Q_{Y}^{n}(y^{n}):y^{n}\in\mathcal{Y}^{n}\backslash\mathcal{B}_{1}\right\} $
on the value of $D_{\alpha}(P_{Y^{n}}\|Q_{Y}^{n})$ is removed, since
for such a simulation, all sequences $x^{n}$ are mapped to the sequences
$y^{n}$ in $\mathcal{B}_{1}$. Hence in general, a code $f:\mathcal{X}^{k}\to\mathcal{B}_{1}$
induces a smaller $D_{\alpha}(P_{Y^{n}}\|Q_{Y}^{n})$ than a code
$f:\mathcal{X}^{k}\to\mathcal{Y}^{n}$. By using the inverse-transform
scheme, we derive an upper bound for $\alpha\in[1,\infty]$, which
is tight for $\alpha=1$ or $\infty$. This is because that to ensure
$D_{\alpha}(P_{Y^{n}}\|Q_{Y}^{n})\to0$, it is required that $\frac{P_{Y^{n}}(y^{n})}{Q_{Y}^{n}(y^{n})}\leq1+o(1)$
for all $y^{n}\in\mathcal{Y}^{n}$ when $\alpha=\infty$, and $\frac{P_{Y^{n}}(y^{n})}{Q_{Y}^{n}(y^{n})}=1+o(1)$
for all $y^{n}$ in a high probability set of $Q_{Y}^{n}$ when $\alpha=1$.

Similar ideas also apply to the cases with measures $D_{\alpha}(Q_{Y}^{n}\|P_{Y^{n}})$
and $D_{\alpha}^{\mathsf{max}}(P_{Y^{n}},Q_{Y}^{n})$. However, for
$\alpha=1$, differently from the case $D_{\alpha}(P_{Y^{n}}\|Q_{Y}^{n})$,
to ensure $D_{\alpha}(Q_{Y}^{n}\|P_{Y^{n}})\to0$ or $D_{\alpha}^{\mathsf{max}}(P_{Y^{n}},Q_{Y}^{n})\to0$,
it is required not only that $\frac{Q_{Y}^{n}(y^{n})}{P_{Y^{n}}(y^{n})}=1+o(1)$
for all $y^{n}$ in a high probability set of $Q_{Y}^{n}$, but also
that $P_{Y^{n}}(y^{n})>0$ for all $y^{n}\in\mathcal{Y}^{n}$ (otherwise,
$D_{\alpha}(Q_{Y}^{n}\|P_{Y^{n}})=D_{\alpha}^{\mathsf{max}}(P_{Y^{n}},Q_{Y}^{n})=\infty$).
Observe that there exists a code such that $P_{Y^{n}}(y^{n})>0$ for
all $y^{n}\in\mathcal{Y}^{n}$ if and only if $|\mathcal{X}|^{k}\ge|\mathcal{Y}|^{n}$,
i.e., $\frac{n}{k}\le\frac{H_{0}(P_{X})}{H_{0}(Q_{Y})}$. Hence the
term $\frac{H_{0}(P_{X})}{H_{0}(Q_{Y})}$ appears in \eqref{eq:-1}
and \eqref{eq:-18} for $\alpha=1$.

For $\alpha=\infty$ and for the measure $D_{\alpha}(Q_{Y}^{n}\|P_{Y^{n}})$,
the code used to prove the achievability part is illustrated in Fig.
\ref{fig:DPQ-1}. In contrast to the case $D_{\alpha}(P_{Y^{n}}\|Q_{Y}^{n})$,
here the sequences in $\mathcal{B}_{2}:=\left\{ y^{n}:e^{-nH^{\mathrm{u}}(Q_{Y})}\leq Q_{Y}^{n}(y^{n})\leq e^{-n\left(H(Q_{Y})-\delta\right)}\right\} $,
instead of those in $\mathcal{B}_{1}$, are dominant. That is to say,
the influence of $\{Q_{Y}^{n}(y^{n}):y^{n}\in\mathcal{Y}^{n}\backslash\mathcal{B}_{2}\}$
on the value of $D_{\alpha}(Q_{Y}^{n}\|P_{Y^{n}})$ can be removed.
However, for the measure $D_{\alpha}^{\mathsf{max}}(P_{Y^{n}},Q_{Y}^{n})$,
the influence of $Q_{Y}^{n}(y^{n}),y^{n}\in\mathcal{Y}^{n}$ cannot
be removed anymore. That is, all the sequences in $\mathcal{Y}^{n}$
are dominant. See the code illustrated in Fig. \ref{fig:DPQ-1-1},
which is used to prove the achievability part for this case.

In summary, for $\alpha=\infty$, the conversion rates are determined
by the (part of or all of) information spectrum exponents of $P_{X}^{k}$
and $Q_{Y}^{n}$, and on the other hand, the information spectrum
exponents are determined by the R\'enyi entropies (see Lemmas \ref{lem:Exponents}
and \ref{lem:ExponentsComparison}; more specifically, the infinity
order cases in Theorems \ref{thm:Renyi1Con}, \ref{thm:Renyi2Con},
and \ref{thm:Renyi3Con} respectively correspond to \eqref{eq:-82},
\eqref{eq:-86}, as well as, \eqref{eq:-82} and \eqref{eq:-85}).
Hence the conversion rates are determined by R\'enyi entropies. This
is the reason why the conversion rates are expressed as functions
of R\'enyi entropies. However, for $\alpha=1$, the conversion rates
are related to the limits of information spectrums of $P_{X}^{k}$
and $Q_{Y}^{n}$, and do not depend on how fast the information spectrums
converge. Hence they are only functions of R\'enyi entropies with orders
1 and 0.

\begin{figure}[t]
\centering \includegraphics[width=1\columnwidth]{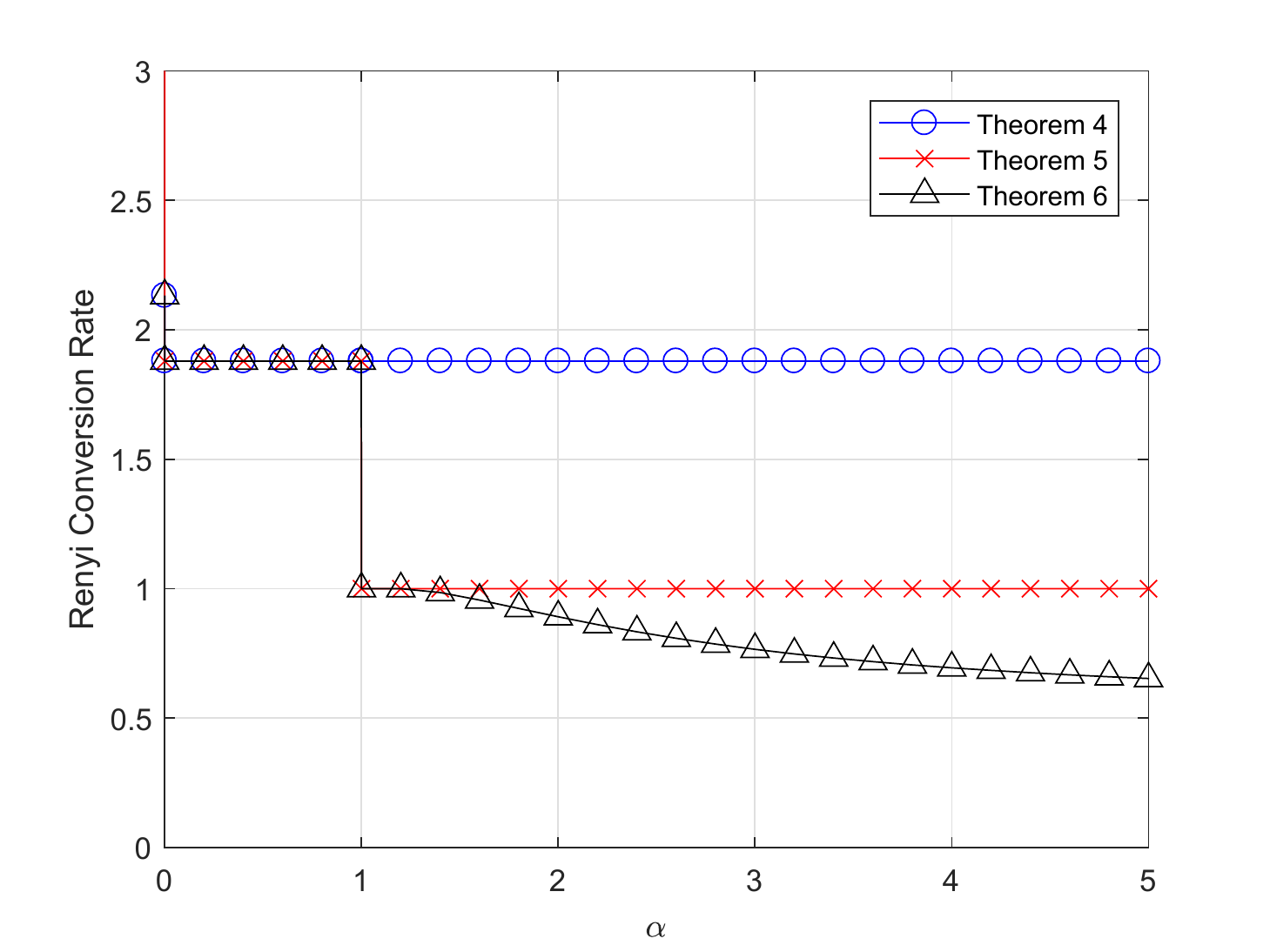} \includegraphics[width=1\columnwidth]{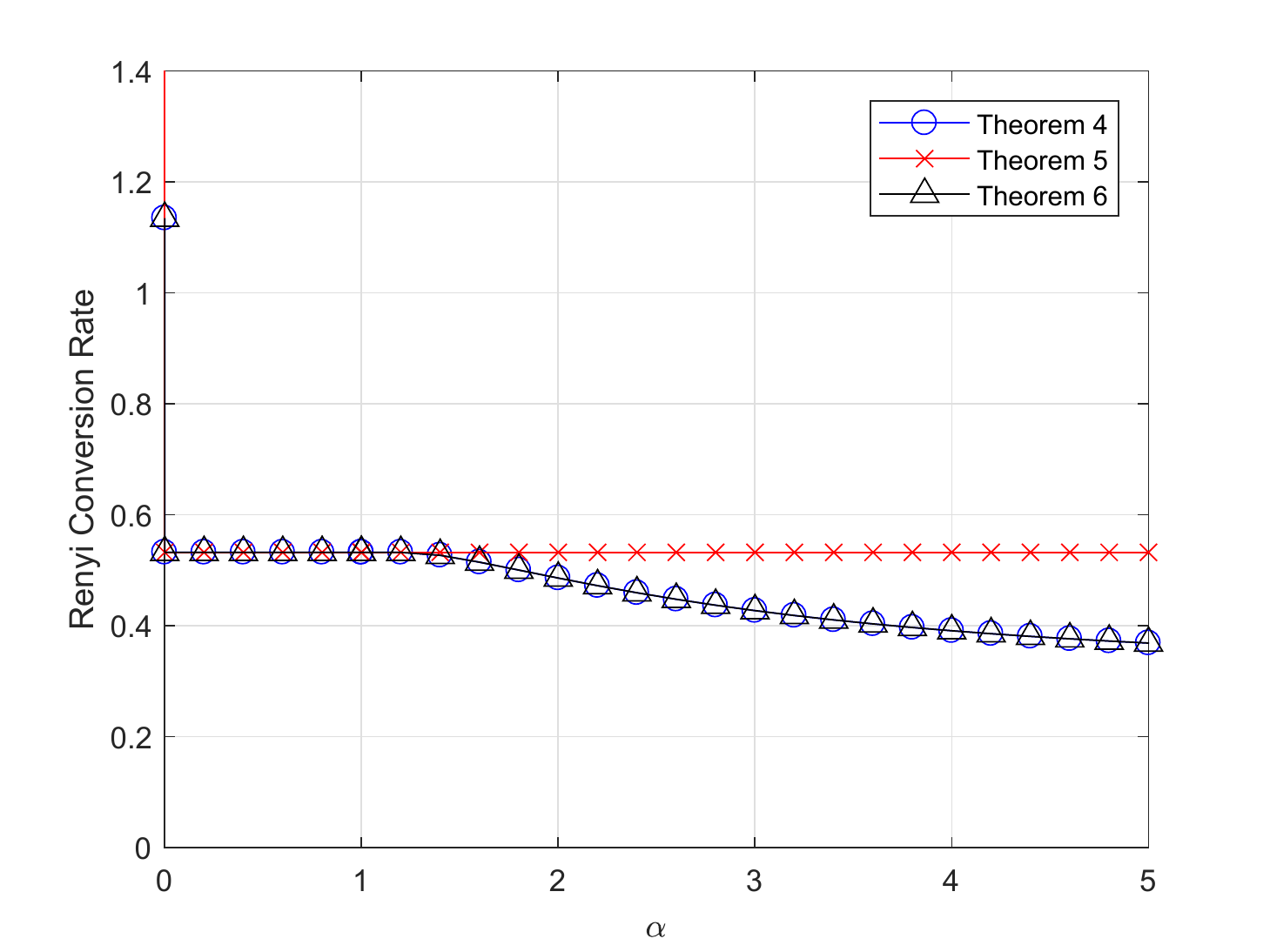}
\caption{Illustration of the R\'enyi conversion rates under normalized divergences
in Theorems \ref{thm:Renyi1Con}, \ref{thm:Renyi2Con}, and \ref{thm:Renyi3Con}
for $P_{X}=\mathsf{Bern}\left(0.3\right)$ and $Q_{Y}=\mathsf{Bern}\left(0.1\right)$
(top) and for $P_{X}=\mathsf{Bern}\left(0.1\right)$ and $Q_{Y}=\mathsf{Bern}\left(0.3\right)$
(bottom). }
\label{fig:Resolvability-1-1} 
\end{figure}

Theorems \ref{thm:Renyi1Con}, \ref{thm:Renyi2Con}, and \ref{thm:Renyi3Con}
are illustrated in Fig.~\ref{fig:Resolvability-1-1}.

\section{\label{sec:Special-Case-1:}Special Case 1: R\'enyi Source Resolvability }

If we set $P_{X}$ to the Bernoulli distribution $\mathsf{Bern}(\frac{1}{2})$,
then the distribution approximation problem reduces to the source
resolvability problem, i.e., simulating a memoryless source whose
distribution is approximately subject to a target distribution $Q_{Y}$,
using a uniform random variable $M_{n}$ that is uniformly distributed
over $\calM_{n}:=[1:\mathsf{M}]$ with $\mathsf{M}:=\lfloor e^{n\widetilde{R}}\rfloor$.
The rate $\widetilde{R}$ here is different from the $R$ defined
in Section \ref{sec:Renyi-Distribution-Approximation}, and indeed
it is approximately equal to the ratio of $\log2$ and the $R$ in
Section \ref{sec:Renyi-Distribution-Approximation} with $P_{X}$
set to $\mathsf{Bern}(\frac{1}{2})$. Given the target distribution
$Q_{Y}$, we wish to minimize the rate $\widetilde{R}$ such that
the distribution of $Y^{n}:=f(M_{n})$ forms a good approximation
to the product distribution $Q_{Y}^{n}$. In contrast to previous
works on the resolvability problem~\cite{Han,Yu}, here we employ
the R\'enyi divergences $D_{\alpha}(P_{Y^{n}}\|Q_{Y}^{n}),D_{\alpha}(Q_{Y}^{n}\|P_{Y^{n}})$,
and $D_{\alpha}^{\mathsf{max}}(P_{Y^{n}},Q_{Y}^{n})$ of all orders
$\alpha\in[0,\infty]$ to measure the discrepancy between $P_{Y^{n}}$
and $Q_{Y}^{n}$.

\subsection{Asymptotics of R\'enyi Divergences }

We consider the R\'enyi divergences $D_{\alpha}(P_{Y^{n}}\|Q_{Y}^{n}),D_{\alpha}(Q_{Y}^{n}\|P_{Y^{n}})$,
and $D_{\alpha}^{\mathsf{max}}(P_{Y^{n}},Q_{Y}^{n})$. The asymptotic
behaviors of these measures are respectively characterized in the
following corollaries. These results follow from Theorems \ref{thm:RenyiPQ},
\ref{thm:RenyiQP}, and \ref{thm:Renyimax} by setting $P_{X}=\mathsf{Bern}(\frac{1}{2})$. 
\begin{cor}[Asymptotics of $\frac{1}{n}D_{\alpha}(P_{Y^{n}}\|Q_{Y}^{n})$]
\label{thm:Renyi1} For any $\alpha\in[0,\infty]$, we have 
\begin{align}
 & \lim_{n\to\infty}\frac{1}{n}\inf_{f}D_{\alpha}(P_{Y^{n}}\|Q_{Y}^{n})\nonumber \\
 & =\sup_{t\in[0,1)}\left\{ tH_{\frac{1}{1-t}}(Q_{Y})-t\widetilde{R}\right\} .
\end{align}
\end{cor}
\begin{rem}
This result for $\alpha\in[0,2]$ was shown by our previous work \cite{Yu}.
Hence our results here for $\alpha\in(2,\infty]$ are new. 
\end{rem}
\begin{rem}
This result for $\alpha=0$ is related to the error exponent of lossless
source coding. Define 
\begin{equation}
\mathsf{P}\left(\widetilde{R}\right):=\sup_{\mathcal{A}\subseteq\mathcal{Y}:|\mathcal{A}|\le e^{n\widetilde{R}}}Q_{Y}^{n}\left(\mathcal{A}\right). \label{eq:-P}
\end{equation}
Then according to \eqref{eq:-29}, for $\alpha=0$, the asymptotics
of the normalized R\'enyi divergence 
\begin{align}
 & \lim_{n\to\infty}\frac{1}{n}\inf_{f}D_{0}(P_{Y^{n}}\|Q_{Y}^{n})\nonumber \\
 & =\lim_{n\to\infty}-\frac{1}{n}\log\mathsf{P}\left(\widetilde{R}\right)\\
 & =\min_{\widetilde{P}_{Y}:H(\widetilde{P}_{Y})\le\widetilde{R}}D(\widetilde{P}_{Y}\|Q_{Y})\\
 & =\sup_{t\in[0,1)}\left\{ tH_{\frac{1}{1-t}}(Q_{Y})-t\widetilde{R}\right\} .
\end{align}
On the other hand, the error exponent of lossless source coding with
code rate $\widetilde{R}$ for memoryless source $Q_{Y}^{n}$ is 
\begin{align}
 & \lim_{n\to\infty}-\frac{1}{n}\log\left(1-\mathsf{P}\left(\widetilde{R}\right)\right)\nonumber \\
 & =\min_{\widetilde{P}_{Y}:H(\widetilde{P}_{Y})\geq\widetilde{R}}D(\widetilde{P}_{Y}\|Q_{Y})\\
 & =\sup_{t\in[0,\infty)}\left\{ -tH_{\frac{1}{1+t}}(Q_{Y})+t\widetilde{R}\right\}. \label{eq:-P2}
\end{align}
Hence the asymptotics of the normalized R\'enyi divergence $D_{0}(P_{Y^{n}}\|Q_{Y}^{n})$
and the error exponent of lossless source coding are respectively
the exponents of $\mathsf{P}\left(\widetilde{R}\right)$ for different
regimes ($\widetilde{R}\le H(Q_{Y})$ and $\widetilde{R}\geq H(Q_{Y})$). Furthermore, by large deviation theory \cite{Dembo}, \eqref{eq:-P}-\eqref{eq:-P2} hold not only for finite alphabets, but also for countably
infinite or continuous alphabets (with the counting measure replaced by the  Lebesgue  measure,  the probability mass function 
$Q_{Y}$ replaced by the corresponding probability density function or the Radon-Nikodym
derivative, and  the summation replaced by the corresponding integration).
\end{rem}
\begin{cor}[Asymptotics of $\frac{1}{n}D_{\alpha}(Q_{Y}^{n}\|P_{Y^{n}})$]
\label{thm:RenyiResQP} For any $\alpha\in[0,\infty]$, we have 
\begin{align}
 & \lim_{n\to\infty}\frac{1}{n}\inf_{f}D_{\alpha}(Q_{Y}^{n}\|P_{Y^{n}})\nonumber \\
 & =\begin{cases}
\infty,\quad\alpha\in[1,\infty]\textrm{ and }\widetilde{R}<H_{0}(Q_{Y});\\
0,\quad\;\alpha\in[1,\infty]\textrm{ and }\widetilde{R}>H_{0}(Q_{Y});\\
\frac{\alpha}{1-\alpha}\sup_{t\in[0,1)}\left\{ tH_{\frac{1}{1-t}}(Q_{Y})-t\widetilde{R}\right\} ,\quad\alpha\in(0,1);\\
0,\quad\;\alpha=0.
\end{cases}
\end{align}
\end{cor}
\begin{cor}[Asymptotics of $\frac{1}{n}D_{\alpha}^{\mathsf{max}}(P_{Y^{n}},Q_{Y}^{n})$]
\label{thm:RenyiQP-1-1} For any $\alpha\in[0,\infty]$, we have
\begin{align}
 & \lim_{n\to\infty}\frac{1}{n}\inf_{f}D_{\alpha}^{\mathsf{max}}(P_{Y^{n}},Q_{Y}^{n})\nonumber \\
 & =\begin{cases}
\infty,\quad\alpha\in[1,\infty]\textrm{ and }\widetilde{R}<H_{0}(Q_{Y});\\
\sup_{t\in(\frac{\alpha}{\alpha-1},\infty)}\left\{ tH_{\frac{1}{1-t}}(Q_{Y})-t\widetilde{R}\right\} ,\\
\qquad\;\alpha\in(1,\infty]\textrm{ and }\widetilde{R}>H_{0}(Q_{Y});\\
0,\quad\;\alpha=1\textrm{ and }\widetilde{R}>H_{0}(Q_{Y});\\
\max\left\{ \frac{\alpha}{1-\alpha},1\right\} \sup_{t\in[0,1)}\left\{ tH_{\frac{1}{1-t}}(Q_{Y})-t\widetilde{R}\right\} ,\\
\qquad\;\alpha\in(0,1);\\
\sup_{t\in[0,1)}\left\{ tH_{\frac{1}{1-t}}(Q_{Y})-t\widetilde{R}\right\} ,\quad\alpha=0.
\end{cases}
\end{align}
\end{cor}

\subsection{R\'enyi Source Resolvability }

\begin{figure}[t]
\centering \includegraphics[width=1\columnwidth]{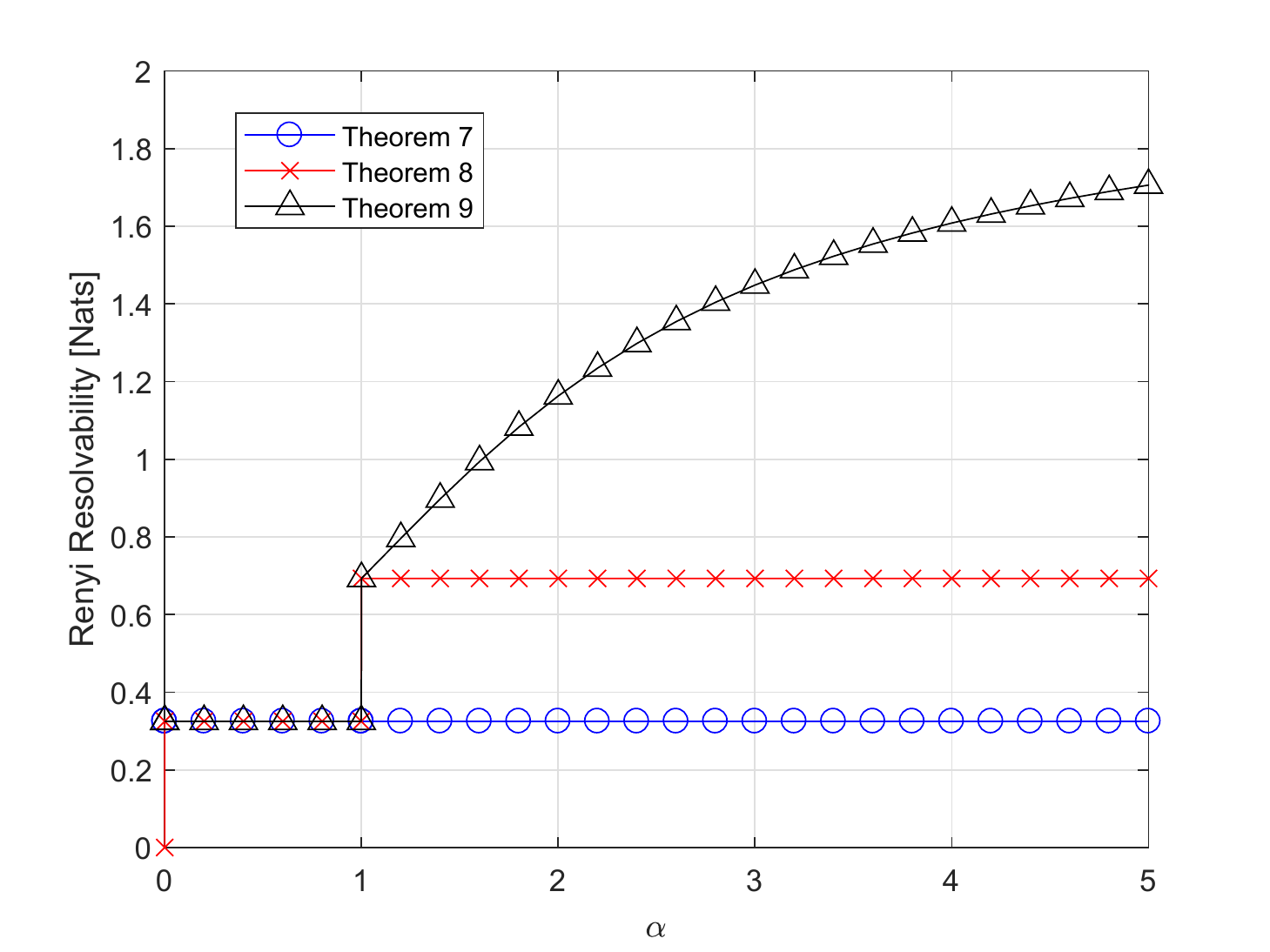} \caption{Illustration of the R\'enyi resolvabilities in Theorems \ref{thm:Renyi1rate},
\ref{thm:Renyi2rate}, and \ref{thm:Renyi3rate} for $Q_{Y}=\mathsf{Bern}\left(0.1\right)$. }
\label{fig:Resolvability-1-1-1} 
\end{figure}

As shown in the theorems above, when the code rate is small, the normalized
R\'enyi divergences $\frac{1}{n}D_{\alpha}(P_{Y^{n}}\|Q_{Y}^{n})$,
$\frac{1}{n}D_{\alpha}(Q_{Y}^{n}\|P_{Y^{n}})$, and $\frac{1}{n}D_{\alpha}^{\mathsf{max}}(P_{Y^{n}},Q_{Y}^{n})$
converge to a positive number; however when the code rate is large
enough, the normalized R\'enyi divergences converge to zero. The threshold
rate, named R\'enyi resolvability, represents the minimum rate needed
to ensure the distribution induced by the code well approximates the
target distribution. We characterize the R\'enyi resolvabilities in
the following theorems. The R\'enyi resolvabilities for normalized divergences
of all orders and the R\'enyi resolvabilities for unnormalized divergences
of orders in $[0,1]\cup\left\{ \infty\right\} $ are direct consequences
of Theorems \ref{thm:Renyi1Con}, \ref{thm:Renyi2Con}, and \ref{thm:Renyi3Con}.
Hence we only need focus on the cases for unnormalized divergences
of orders in $(1,\infty)$. Furthermore, the converse parts for these
cases follow from the fact the unnormalized divergences are stronger
than the normalized versions. Hence we only prove the achievability
parts for unnormalized divergences of orders in $(1,\infty)$. These
proofs are provided in Appendices \ref{sec:Proof-of-Theorem-Renyi1rate},
\ref{sec:Proof-of-Theorem-Renyi2rate}, and \ref{sec:Proof-of-Theorem-Renyi3rate},
respectively. 
\begin{thm}[R\'enyi Resolvability]
\label{thm:Renyi1rate} For any $\alpha\in[0,\infty]$, we have 
\begin{align}
 & \inf\left\{ \widetilde{R}:\frac{1}{n}D_{\alpha}(P_{Y^{n}}\|Q_{Y}^{n})\rightarrow0\right\} \nonumber \\
 & =\inf\left\{ \widetilde{R}:D_{\alpha}(P_{Y^{n}}\|Q_{Y}^{n})\rightarrow0\right\} \nonumber \\
 & =H(Q_{Y}).
\end{align}
\end{thm}
\begin{rem}
\label{rmk:unnormRenyi-1} The case $\alpha=1$ and the normalized
divergence (i.e., the normalized relative entropy case) was first
shown by Han and Verd\'u \cite{Han}. The case $\alpha=1$ and the unnormalized
divergence (i.e., the unnormalized relative entropy case) has been
shown in other works, such as those by Hayashi~\cite{Hayashi06,Hayashi11}
and Han, Endo, and Sasaki~\cite{Han14}. In fact, Theorem \ref{thm:Renyi1rate}
is implied by our previous work on R\'enyi channel resolvability \cite{Yu}
by setting the channel to be the identity channel. 
\end{rem}
\begin{thm}[R\'enyi Resolvability]
\label{thm:Renyi2rate} For any $\alpha\in[0,\infty]$, we have 
\begin{align}
 & \inf\left\{ \widetilde{R}:\frac{1}{n}D_{\alpha}(Q_{Y}^{n}\|P_{Y^{n}})\rightarrow0\right\} \nonumber \\
 & =\inf\left\{ \widetilde{R}:D_{\alpha}(Q_{Y}^{n}\|P_{Y^{n}})\rightarrow0\right\} \nonumber \\
 & =\begin{cases}
H_{0}(Q_{Y}), & \alpha\in[1,\infty]\\
H(Q_{Y}), & \alpha\in(0,1)\\
0, & \alpha=0
\end{cases}
\end{align}
\end{thm}
\begin{rem}
\label{rmk:unnormRenyi-2} The results in Theorem \ref{thm:Renyi2rate}
for all orders $\alpha\in[0,\infty]$ are new. 
\end{rem}
\begin{thm}[R\'enyi Resolvability]
\label{thm:Renyi3rate} For any $\alpha\in[0,\infty]$, we have 
\begin{align}
 & \inf\left\{ \widetilde{R}:\frac{1}{n}D_{\alpha}^{\mathsf{max}}(P_{Y^{n}},Q_{Y}^{n})\rightarrow0\right\} \nonumber \\
 & =\inf\left\{ \widetilde{R}:D_{\alpha}^{\mathsf{max}}(P_{Y^{n}},Q_{Y}^{n})\rightarrow0\right\} \nonumber \\
 & =\begin{cases}
H_{1-\alpha}(Q_{Y}), & \alpha\in[1,\infty]\\
H(Q_{Y}), & \alpha\in[0,1)
\end{cases}
\end{align}
\end{thm}
\begin{rem}
For special cases $\alpha=1,\infty$, the R\'enyi resolvabilities are
respectively equal to $H_{-\infty}(Q_{Y})=-\log\min_{y}Q_{Y}(y)$
and $H_{0}(Q_{Y})=\log\left|\mathrm{supp}(Q_{Y})\right|$. 
\end{rem}
\begin{rem}
To the best of our knowledge, we are the first to give an explicit
operational interpretation of R\'enyi entropies of negative orders as
R\'enyi resolvabilities. In \cite{routtenberg2008general,sason2018arimoto},
R\'enyi entropies of negative orders were used to lower bound the probability
of error for hypothesis testing. 
\end{rem}
Theorems \ref{thm:Renyi1rate}, \ref{thm:Renyi2rate}, and \ref{thm:Renyi3rate}
are illustrated in Fig.~\ref{fig:Resolvability-1-1-1}.

\section{\label{sec:Special-Case-2:}Special Case 2: R\'enyi Intrinsic Randomness}

If we set $Q_{Y}$ to the Bernoulli distribution $\mathsf{Bern}(\frac{1}{2})$,
then the distribution approximation problem reduces to the intrinsic
randomness problem, which can be seen as a ``dual'' problem of the
source resolvability problem. Consider simulating a uniform random
variable $M_{n}$ that is uniformly distributed over $\calM_{n}:=[1:\mathsf{M}]$
with $\mathsf{M}:=\lceil e^{n\ensuremath{\widehat{R}}}\rceil$ using
a memoryless source $X^{n}\sim P_{X}^{n}$. The rate $\ensuremath{\widehat{R}}$
here is approximately equal to $\log2$ times the rate $R$ in Section
\ref{sec:Renyi-Distribution-Approximation} with $Q_{Y}$ set to $\mathsf{Bern}(\frac{1}{2})$.
Given the distribution $P_{X}$, we wish to maximize the rate $\ensuremath{\widehat{R}}$
such that the distribution of $M_{n}:=f(X^{n})$ forms a good approximation
to the target distribution $Q_{M_{n}}:=\mathrm{Unif}[1:\mathsf{M}]$.

\subsection{Asymptotics of R\'enyi Divergences }

We consider the R\'enyi divergences $D_{\alpha}(P_{M_{n}}\|Q_{M_{n}}),D_{\alpha}(Q_{M_{n}}\|P_{M_{n}})$,
and $D_{\alpha}^{\mathsf{max}}(P_{M_{n}},Q_{M_{n}})$. The asymptotics
of these measures are respectively characterized in the following
corollaries. These results respectively follow from Theorems \ref{thm:RenyiPQ},
\ref{thm:RenyiQP}, and \ref{thm:Renyimax} by setting $Q_{Y}=\mathsf{Bern}(\frac{1}{2})$. 
\begin{cor}[Asymptotics of $\frac{1}{n}D_{\alpha}(P_{M_{n}}\|Q_{M_{n}})$]
\label{thm:Renyi1-1} For any $\alpha\in[0,\infty]$, we have 
\begin{align}
 & \lim_{n\to\infty}\frac{1}{n}\inf_{f}D_{\alpha}(P_{M_{n}}\|Q_{M_{n}})\nonumber \\
 & =\begin{cases}
\left[\ensuremath{\widehat{R}}-H_{\alpha}(P_{X})\right]^{+} & \alpha\in\{0\}\cup[1,\infty]\\
\max_{t\in[0,1]}\left\{ t\ensuremath{\widehat{R}}-tH_{\frac{1}{1-\frac{\alpha-1}{\alpha}t}}(P_{X})\right\}  & \alpha\in(0,1)
\end{cases}.
\end{align}
\end{cor}
\begin{rem}
The case $\alpha\in[0,2]$ was shown by Hayashi and Tan \cite{Hayashi17}.
Hence our results for $\alpha\in(2,\infty]$ are new. 
\end{rem}
\begin{cor}[Asymptotics of $\frac{1}{n}D_{\alpha}(Q_{M_{n}}\|P_{M_{n}})$]
\label{thm:RenyiIRQP} For any $\alpha\in[0,\infty]$, we have 
\begin{align}
 & \lim_{n\to\infty}\frac{1}{n}\inf_{f}D_{\alpha}(Q_{M_{n}}\|P_{M_{n}})\nonumber \\
 & =\begin{cases}
\sup_{t\in[0,\infty)}\left\{ t\ensuremath{\widehat{R}}-tH_{\frac{1}{1+t}}(P_{X})\right\} , & \alpha\in[1,\infty]\\
\frac{\alpha}{1-\alpha}\max_{t\in[0,1]}\left\{ t\ensuremath{\widehat{R}}-tH_{\frac{1}{1+\frac{\alpha}{1-\alpha}t}}(P_{X})\right\} , & \alpha\in(0,1)\\
0, & \alpha=0
\end{cases}
\end{align}
\end{cor}
\begin{rem}
If $\ensuremath{\widehat{R}}>H_{0}(P_{X})$, then $\lim_{n\to\infty}\frac{1}{n}\inf_{f}D_{\alpha}(Q_{M_{n}}\|P_{M_{n}})=\infty,\alpha\in[1,\infty]$. 
\end{rem}
\begin{cor}[Asymptotics of $\frac{1}{n}D_{\alpha}^{\mathsf{max}}(P_{Y^{n}},Q_{Y}^{n})$]
\label{thm:Renyimax-1} For any $\alpha\in[0,\infty]$, we have 
\begin{align}
 & \lim_{n\to\infty}\frac{1}{n}\inf_{f}D_{\alpha}^{\mathsf{max}}(P_{Y^{n}},Q_{Y}^{n})\nonumber \\
 & =\begin{cases}
\max\biggl\{\left[\ensuremath{\widehat{R}}-H_{\alpha}(P_{X})\right]^{+},\\
\quad\sup_{t\in[0,\infty)}\left\{ t\ensuremath{\widehat{R}}-tH_{\frac{1}{1+t}}(P_{X})\right\} \biggr\}, & \alpha\in[1,\infty]\\
\max_{t\in[0,1]}\max_{t'\in[0,1]}\\
\quad\left\{ tb(t')\ensuremath{\widehat{R}}-tb(t')H_{\frac{a(t')}{a(t')+tb(t')}}(P_{X})\right\} , & \alpha\in(0,1)\\
\left[\ensuremath{\widehat{R}}-H_{0}(P_{X})\right]^{+}, & \alpha=0
\end{cases}
\end{align}
where $a(t')$ and $b(t')$ are defined in \eqref{eq:-48} and \eqref{eq:-49}. 
\end{cor}

\subsection{R\'enyi Intrinsic Randomness}

\begin{figure}[t]
\centering \includegraphics[width=1\columnwidth]{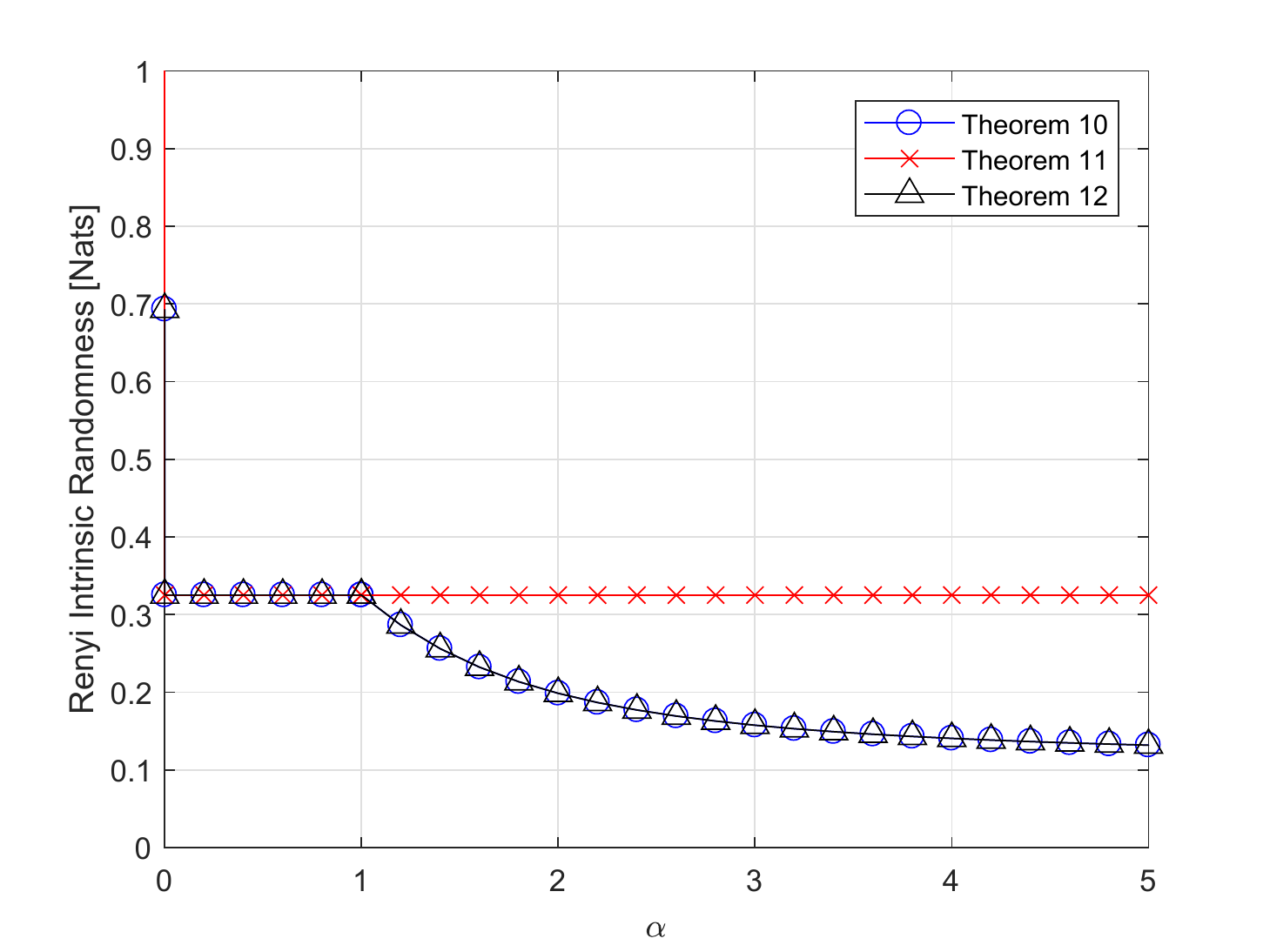} \caption{Illustration of the R\'enyi intrinsic randomness in Theorems \ref{thm:RenyiIR1rate},
\ref{thm:RenyiIR2rate}, and \ref{thm:RenyiIRmax} for $P_{X}=\mathsf{Bern}\left(0.1\right)$. }
\label{fig:Resolvability-1-1-1-1} 
\end{figure}

As shown in the theorems above, when the rate is large, the normalized
R\'enyi divergences $\frac{1}{n}D_{\alpha}(P_{M_{n}}\|Q_{M_{n}}),\frac{1}{n}D_{\alpha}(Q_{M_{n}}\|P_{M_{n}})$,
and $\frac{1}{n}D_{\alpha}^{\mathsf{max}}(P_{M_{n}},Q_{M_{n}})$ converge
to a positive number; however when the rate is small enough, the normalized
R\'enyi divergences converge to zero. The threshold rate, named R\'enyi
intrinsic randomness, represents the maximum possible rate to satisfy
that the distribution induced by a code well approximates the target
uniform distribution. We characterize the R\'enyi intrinsic randomness
in the following theorems. The R\'enyi intrinsic randomness for normalized
divergences of all orders and the R\'enyi intrinsic randomness for unnormalized
divergences of orders in $[0,1]\cup\left\{ \infty\right\} $ are direct
consequences of Theorems \ref{thm:Renyi1Con}, \ref{thm:Renyi2Con},
and \ref{thm:Renyi3Con}. Hence we only need focus on the cases for
unnormalized divergences of orders in $(1,\infty)$. Furthermore,
the converse parts for these cases follow from the fact the unnormalized
divergences are stronger than the normalized versions. Hence we only
prove the achievability parts. The proofs are provided in Appendices
\ref{sec:Proof-of-Theorem-Renyi1rate-1}, \ref{sec:Proof-of-Theorem-Renyi1rate-1-1},
and \ref{sec:Proof-of-Theorem-Renyi1rate-1-2}, respectively. 
\begin{thm}[R\'enyi Intrinsic Randomness]
\label{thm:RenyiIR1rate} For any $\alpha\in[0,\infty]$, we have
\begin{align}
 & \sup\left\{ \ensuremath{\widehat{R}}:\frac{1}{n}D_{\alpha}(P_{M_{n}}\|Q_{M_{n}})\rightarrow0\right\} \nonumber \\
 & =\sup\left\{ \ensuremath{\widehat{R}}:D_{\alpha}(P_{M_{n}}\|Q_{M_{n}})\rightarrow0\right\} \nonumber \\
 & =\begin{cases}
H_{\alpha}(P_{X}) & \alpha\in\{0\}\cup[1,\infty]\\
H(P_{X}) & \alpha\in(0,1)
\end{cases}.
\end{align}
\end{thm}
\begin{rem}
\label{rmk:unnormRenyi-1-1} The case $\alpha=1$ and the normalized
divergence (i.e., the normalized relative entropy case) was shown
in \cite{Han03}. The case $\alpha=1$ and the unnormalized divergence
(i.e., the unnormalized relative entropy case) was shown by Hayashi
\cite{hayashi2008second}. The result for the unnormalized R\'enyi divergence
with $\alpha\in(0,1)$ can be obtained by combining two observations:
1) the achievability for $D(P_{Y^{n}}\|Q_{Y}^{n})$ implies the achievability
for this case; 2) by Pinsker's inequality \cite{Erven}, the result
under the TV distance measure \cite{Han03} implies the converse for
$\alpha\in(0,1)$. The case $\alpha\in[0,2]$ was shown by Hayashi
and Tan \cite{Hayashi17}. Hence our results for $\alpha\in(2,\infty]$
are new. 
\end{rem}
\begin{thm}[R\'enyi Intrinsic Randomness]
\label{thm:RenyiIR2rate} \label{thm:RenyiIRRateQP} For any $\alpha\in[0,\infty]$,
we have 
\begin{align}
 & \sup\left\{ \ensuremath{\widehat{R}}:\frac{1}{n}D_{\alpha}(Q_{M_{n}}\|P_{M_{n}})\rightarrow0\right\} \nonumber \\
 & =\sup\left\{ \ensuremath{\widehat{R}}:D_{\alpha}(Q_{M_{n}}\|P_{M_{n}})\rightarrow0\right\} \nonumber \\
 & =\begin{cases}
H(P_{X}), & \alpha\in(0,\infty]\\
\infty, & \alpha=0
\end{cases}
\end{align}
\end{thm}
\begin{rem}
\label{rmk:unnormRenyi-3} The case $\alpha=1$ was shown by Hayashi
\cite{hayashi2008second}. Our results for all orders $\alpha\in[0,1)\cup(1,\infty]$
are new. 
\end{rem}
\begin{thm}[R\'enyi Intrinsic Randomness]
\label{thm:RenyiIRmax} For any $\alpha\in[0,\infty]$, we have 
\begin{align}
 & \sup\left\{ \ensuremath{\widehat{R}}:\frac{1}{n}D_{\alpha}^{\mathsf{max}}(P_{M_{n}},Q_{M_{n}})\rightarrow0\right\} \nonumber \\
 & =\sup\left\{ \ensuremath{\widehat{R}}:D_{\alpha}^{\mathsf{max}}(P_{M_{n}},Q_{M_{n}})\rightarrow0\right\} \nonumber \\
 & =\begin{cases}
H_{\alpha}(P_{X}), & \alpha\in\{0\}\cup[1,\infty]\\
H(P_{X}), & \alpha\in(0,1)
\end{cases}
\end{align}
\end{thm}
Theorems \ref{thm:RenyiIR1rate}, \ref{thm:RenyiIR2rate}, and \ref{thm:RenyiIRmax}
are illustrated in Fig.~\ref{fig:Resolvability-1-1-1-1}.

\section{\label{sec:Concluding-Remarks}Concluding Remarks}

\label{sec:concl}In this paper, we studied generalized versions of
random variable simulation problem or distribution approximation problem,
in which the (normalized or unnormalized) standard R\'enyi divergence
and max- or sum-R\'enyi divergence of orders in $[0,\infty]$ are used
to measure the level of approximation. As special cases, the source
resolvability problem and the intrinsic randomness problem were studied
as well.

Our results on the distribution approximation problem extend those
by Han \cite{Han03} and by Kumagai and Hayashi \cite{kumagai2017second},
as we consider R\'enyi divergences with all orders in $[0,\infty]$
instead of the TV distance or the special case with order $\frac{1}{2}$.
Similarly, our source resolvability results extend those by Han and
Verd\'u \cite{Han}, by Hayashi \cite{Hayashi06,Hayashi11}, and by
Yu and Tan \cite{Yu} for the source resolvability case, and our intrinsic
randomness results extend those by Vembu and Verd\'u \cite{vembu1995generating},
by Han \cite{Han03}, and by Hayashi and Tan \cite{Hayashi17}.

\subsection{Open Problem }

In Theorems \ref{thm:Renyi1Con}, \ref{thm:Renyi2Con}, and \ref{thm:Renyi3Con},
we completely characterized the R\'enyi conversion rates only for $\alpha\in[0,1]\cup\left\{ \infty\right\} $.
But the cases for $\alpha\in(1,\infty)$ are still open. We believe
that analogous to the case $\alpha\in[0,1]\cup\left\{ \infty\right\} $,
the unnormalized version of R\'enyi conversion rate for $\alpha\in(1,\infty)$
is also equal to the corresponding normalized version with the same
$\alpha$.

\subsection{Applications}

Similar to other results concerning simulation of random variables,
our results can be applied to the analysis of Monte Carlo methods,
randomized algorithms (or random coding), and cryptography. In the
following we apply our results to information-theoretic security.
To illustrate this point, we consider the Shannon cipher system with
a guessing wiretapper that was studied in \cite{merhav1999shannon}.
In the Shannon cipher system, the sender and the legitimate receiver
share a secret key $K_{n}\sim\mathrm{Unif}\left[1:e^{nR}\right]$,
and they want to communicate a source $X^{n}\sim P_{X}^{n}$ with
zero-error (using a variable-length code $M_{n}=f(X^{n},K_{n})$ and
$X^{n}=f^{-1}(M_{n},K_{n})$) from the sender to the legitimate receiver
through a public noiseless channel with sufficiently large capacity.
However, the cryptogram $M_{n}$ is overheard by a wiretapper, who
has a test mechanism by which s/he can identify whether any given
candidate message $\widehat{X}^{n}$ is the true message. Upon the
code $f$ used by the sender and legitimate receiver and the received
cryptogram $M_{n}$, the wiretapper conducts an optimal sequential
guessing strategy, i.e., an ordered list of guesses $\mathcal{L}\left(m\right):=\left\{ \widehat{x}_{1}^{n}\left(m\right),\widehat{x}_{2}^{n}\left(m\right),..\right\} $
with $\widehat{x}_{i}^{n}\left(m\right)$ corresponding to the $i$-th
largest probability value of $P_{X^{n}|M_{n}}(\cdot|m)$ for any given
$M_{n}=m$. It is obvious that such a guessing scheme based on maximizing
the posterior probability minimizes the expectation or positive-order
moments of the number of guesses. Let the random variable $G(X^{n}|M_{n})$
denote the number of guesses of the wiretapper until identification
of the true message. Then for $\rho>0$, the $\rho$-th moment of
$G(X^{n}|M_{n})$ can be also expressed as 
\begin{equation}
\mathbb{E}\left[G(X^{n}|M_{n})^{\rho}\right]=\inf_{\left\{ \mathcal{L}(m)\right\} }\left[\sum_{i=1}^{\infty}i^{\rho}\cdot\mathbb{P}\left\{ \left.\mathcal{L}(M_{n})\right|_{i}=X^{n}\right\} \right],
\end{equation}
where $\left.\mathcal{L}(M_{n})\right|_{i}$ denotes the $i$-th element
of $\mathcal{L}(M_{n})$. For $\rho>0$, the guessing exponents are
defined as 
\begin{align}
E^{+}(R,\rho) & :=\limsup_{n\to\infty}\sup_{f}\frac{1}{n}\log\mathbb{E}\left[G(X^{n}|M_{n})^{\rho}\right]\label{eq:-31}\\
E^{-}(R,\rho) & :=\liminf_{n\to\infty}\sup_{f}\frac{1}{n}\log\mathbb{E}\left[G(X^{n}|M_{n})^{\rho}\right].\label{eq:-32}
\end{align}
Merhav and Arikan \cite{merhav1999shannon} showed that 
\begin{align}
 & E^{+}(R,\rho)=E^{-}(R,\rho)=E(R,\rho)\label{eq:-35}\\
 & :=\max_{Q_{X}}\left\{ \rho\min\left\{ H(Q_{X}),R\right\} -D(Q_{X}\|P_{X})\right\} .
\end{align}

Now we consider a variant of this problem. Suppose the secret key
$K_{n}$ is replaced by a memoryless source $Y^{n}\sim P_{Y}^{n}$.
Correspondingly, denote the guessing exponents for this case as $\widetilde{E}^{+}(P_{Y},\rho)$
and $\widetilde{E}^{-}(P_{Y},\rho)$. Next, we apply our results to
this new problem.

For the achievability part, we use $Y^{n}$ to simulate a key $K_{n}\sim Q_{K_{n}}:=\mathrm{Unif}\left[1:e^{nR}\right]$
by our simulation code $K_{n}=g(Y^{n})$. Assume $P_{K_{n}}$ is the
key distribution induced by a generator $K_{n}=g(Y^{n})$. Then Corollary
\ref{thm:Renyi1-1} implies that $\inf_{g}\frac{1}{n}D_{\infty}(Q_{K_{n}}\|P_{K_{n}})\leq\sup_{t\in[0,\infty)}\left\{ t\ensuremath{R}-tH_{\frac{1}{1+t}}(P_{Y})\right\} $.
Furthermore, for any $f$ and any $\left\{ \mathcal{L}(m)\right\} $,
\begin{align}
 & \frac{1}{n}\log\frac{\mathbb{E}_{P_{K_{n}}P_{X}^{n}}\left[\sum_{i=1}^{\infty}i^{\rho}\cdot1\left\{ \left.\mathcal{L}(f(X^{n},K_{n}))\right|_{i}=X^{n}\right\} \right]}{\mathbb{E}_{Q_{K_{n}}P_{X}^{n}}\left[\sum_{i=1}^{\infty}i^{\rho}\cdot1\left\{ \left.\mathcal{L}(f(X^{n},K_{n}))\right|_{i}=X^{n}\right\} \right]}\nonumber \\
 & \geq-\frac{1}{n}D_{\infty}(Q_{K_{n}}\|P_{K_{n}}).
\end{align}
On the other hand, \eqref{eq:-35} implies 
\begin{align}
 & \lim_{n\to\infty}\sup_{f}\frac{1}{n}\log\inf_{\left\{ \mathcal{L}(m)\right\} }\mathbb{E}_{Q_{K_{n}}P_{X}^{n}}\nonumber \\
 & \quad\left[\sum_{i=1}^{\infty}i^{\rho}\cdot1\left\{ \left.\mathcal{L}(f(X^{n},K_{n}))\right|_{i}=X^{n}\right\} \right]=E(R,\rho).
\end{align}
Hence the guessing exponent functions are bounded as follows. 
\begin{align}
 & \sup_{R\ge0}\left\{ E(R,\rho)-\sup_{t\in[0,\infty)}\left\{ t\ensuremath{R}-tH_{\frac{1}{1+t}}(P_{Y})\right\} \right\} \nonumber \\
 & \leq\widetilde{E}^{-}(P_{Y},\rho)\leq\widetilde{E}^{+}(P_{Y},\rho).\label{eq:-36}
\end{align}

For the converse part, we use a key $K_{n}\sim Q_{K_{n}}:=\mathrm{Unif}\left[1:e^{nR}\right]$
to simulate a memoryless source $Y^{n}\sim P_{Y}^{n}$ by our simulation
code $Y^{n}=g(K_{n})$. Similarly, by our Corollary \ref{thm:Renyi1},
we obtain the following converse result. 
\begin{equation}
\widetilde{E}^{-}(P_{Y},\rho)\leq\widetilde{E}^{+}(P_{Y},\rho)\leq E(H_{0}(P_{Y}),\rho).\label{eq:-37}
\end{equation}

When $P_{X}$ is uniform, the bounds in \eqref{eq:-36} and \eqref{eq:-37}
coincide, and they reduce to the result in \eqref{eq:-35}. However,
in general, the bounds in \eqref{eq:-36} and \eqref{eq:-37} do not
coincide. Furthermore, it is worth noting that the analysis here also
applies to variants of any information-theoretic security problem in
which a key (uniform random variable) is replaced with a memoryless
source, as long as the objective of the problem is to minimize or
maximize the some expectation.

The results derived in this paper can be also applied to the information-theoretic
security problems with the information leakage measured by R\'enyi divergences.
Recently, in \cite{yu2018exact}, Theorem \ref{thm:Renyi1rate} has
been used to establish the equivalence between the exact and $\infty$-R\'enyi
common informations by the present authors. Here the $\infty$-R\'enyi
common information is defined in a distributed source simulation problem
with the approximation between the generated distribution and the
target distribution measured by the R\'enyi divergence of order $\infty$.
In \cite{yu2018exact}, R\'enyi divergences were used to build a bridge
between Wyner's common information and the exact common information.
Therefore, in consideration of the importance of R\'enyi divergences
in connecting different simulation problems, it is significant to
consider R\'enyi divergences as performance indicators for simulation
problems, and also for information-theoretic security problems.

\appendices{ }

\section{\label{sec:Notation-and-Preliminaries}Preliminaries for the Proofs}

For a function $f:\mathcal{X}\to\mathcal{Y}$, and any subsets $\mathcal{A}\subseteq\mathcal{X}$
and $\mathcal{B}\subseteq\mathcal{Y}$, define $f\left(\mathcal{A}\right):=\left\{ f(x):x\in\mathcal{A}\right\} $,
and $f^{-1}(\mathcal{B}):=\left\{ x\in\mathcal{X}:f(x)\in\mathcal{B}\right\} $.
We write $f(n)\dotle g(n)$ if $\limsup_{n\to\infty}\frac{1}{n}\log\frac{f(n)}{g(n)}\le0$.
In addition, $f(n)\doteq g(n)$ means $f(n)\dotle g(n)$ and $g(n)\dotle f(n)$.
We use $o(1)$ to denote generic sequences tending to zero as $n\rightarrow\infty$.
For $a\in\bbR$, $[a]^{+}:=\max\{a,0\}$ denotes positive clipping.
For simplicity, in the proof part, we denote $s=\alpha-1$.

\subsection{Lemmas}

The following fundamental lemmas will be used in our proofs. 
\begin{lem}
\cite{Yu}\label{lem:typecovering} 
\begin{enumerate}
\item Assume $\mathcal{X}$ is a finite set. Then for any $P_{X}\in\mathcal{P}\left(\mathcal{X}\right)$,
one can find a sequence of types $P_{X}^{\left(n\right)}\in\mathcal{P}^{\left(n\right)}\left(\mathcal{X}\right),n\in\bbN$
such that $\big|P_{X}-P_{X}^{\left(n\right)}\big|\leq\frac{\left|\mathcal{X}\right|}{2n}$
as $n\rightarrow\infty$. 
\item Assume $\mathcal{X},\mathcal{Y}$ are finite sets. Then for any sequence
of types $P_{X}^{\left(n\right)}\in\mathcal{P}^{\left(n\right)}\left(\mathcal{X}\right),n\in\bbN$
and any $P_{Y|X}\in\mathcal{P}\left(\mathcal{Y}|\mathcal{X}\right)$,
one can find a sequence of conditional types $V_{Y|X}^{(n)}\in\mathcal{P}^{\left(n\right)}\big(\mathcal{Y}|P_{X}^{\left(n\right)}\big),n\in\bbN$
such that $\big|P_{X}^{\left(n\right)}P_{Y|X}-P_{X}^{\left(n\right)}V_{Y|X}^{(n)}\big|\leq\frac{\left|\mathcal{X}\right|\left|\mathcal{Y}\right|}{2n}$
as $n\rightarrow\infty$. 
\end{enumerate}
\end{lem}
We also need the following property concerning the optimization over
the set of types and conditional types. 
\begin{lem}
\cite{Yu} \label{lem:minequality} 
\begin{enumerate}
\item Assume $\mathcal{X}$ is a finite set. Then for any continuous (under
TV distance) function $f:\mathcal{P}\left(\mathcal{X}\right)\to\mathbb{R}$,
we have 
\begin{equation}
\lim_{n\rightarrow\infty}\min_{P_{X}\in\mathcal{P}^{\left(n\right)}\left(\mathcal{X}\right)}f\left(P_{X}\right)=\min_{P_{X}\in\mathcal{P}\left(\mathcal{X}\right)}f\left(P_{X}\right).
\end{equation}
\item Assume $\mathcal{X},\mathcal{Y}$ are finite sets. Then for any continuous
function $f:\mathcal{P}\left(\mathcal{X}\times\mathcal{Y}\right)\to\mathbb{R}$
and any sequence of types $P_{X}^{\left(n\right)}\in\mathcal{P}^{\left(n\right)}\left(\mathcal{X}\right),n\in\bbN$,
we have 
\begin{align}
 & \min_{P_{Y|X}\in\mathcal{P}^{\left(n\right)}(\mathcal{Y}|P_{X}^{\left(n\right)})}f\big(P_{X}^{\left(n\right)}P_{Y|X}\big)\nonumber \\
 & =\min_{P_{Y|X}\in\mathcal{P}\left(\mathcal{Y}|\mathcal{X}\right)}f\big(P_{X}^{\left(n\right)}P_{Y|X}\big)+o\left(1\right).
\end{align}
\end{enumerate}
\end{lem}
\begin{rem}
We have 
\begin{align}
 & \lim_{n\rightarrow\infty}\min_{P_{Y|X}\in\mathcal{P}^{\left(n\right)}(\mathcal{Y}|P_{X}^{\left(n\right)})}f\big(P_{X}^{\left(n\right)}P_{Y|X}\big)\nonumber \\
 & =\lim_{n\rightarrow\infty}\min_{P_{Y|X}\in\mathcal{P}\left(\mathcal{Y}|\mathcal{X}\right)}f\big(P_{X}^{\left(n\right)}P_{Y|X}\big)
\end{align}
if either one of the limits above exists. 
\end{rem}
We also need the following lemmas.{} Lemmas \ref{lem:continuity},
\ref{lem:1+x}, \ref{lem:convexity_concavity}, and \ref{lem:RenyiEntropyEquality}
follow from basic inequalities and basic properties (continuity, monotonicity,
and convexity) of functions. To save space, the proofs are omitted. 
\begin{lem}
\label{lem:continuity} Assume $f(z)$ and $g(z)$ are continuous
functions defined on a compact set $\mathcal{Z}\subseteq\mathbb{R}^{n}$
for some positive integer $n$. Define $h(t):=\min_{z\in\mathcal{Z}:g(z)\le t}f(z)$.
Then $h(t)$ is a also continuous function. 
\end{lem}
\begin{lem}
\label{lem:norm} \cite[Problem 4.15(f)]{Gallager} Assume $\left\{ a_{i}\right\} $
are non-negative real numbers. Then for $p\geq1$, we have 
\begin{equation}
\sum_{i}a_{i}^{p}\leq\left(\sum_{i}a_{i}\right)^{p},
\end{equation}
and for $0<p\leq1$, we have 
\begin{equation}
\sum_{i}a_{i}^{p}\geq\left(\sum_{i}a_{i}\right)^{p}.\label{eqn:p_ineq}
\end{equation}
\end{lem}
\begin{lem}
\label{lem:1+x} 
\begin{align}
\left(1+x\right)^{s}\le1+x^{s}, & \qquad x\ge0,\;0\le s\le1,\\
\left(1+x\right)^{s}\le1+sx+x^{s}, & \qquad x\ge0,\;1\le s\le2,\\
\left(1+x\right)^{s}\le1+s\left(2^{s-1}-1\right)x+x^{s}, & \qquad0\le x\le1,\;s\ge2.
\end{align}
\end{lem}
\begin{lem}
\label{lem:convexity_concavity} Assume $\sum_{i=1}^{n}b_{i}=m$.
Then we have that for $\beta\leq0$ or $\beta\geq1$, $\frac{1}{n}\sum_{i=1}^{n}b_{i}^{\beta}\geq\left(\frac{m}{n}\right)^{\beta}$;
for $0<\beta<1$, $\frac{1}{n}\sum_{i=1}^{n}b_{i}^{\beta}\leq\left(\frac{m}{n}\right)^{\beta}$.
Moreover, if $m<n$ and $b_{i}\in\{0\}\cup\mathbb{N}$, we have that
for $\beta\leq0$ or $\beta\geq1$, $\frac{1}{n}\sum_{i=1}^{n}b_{i}^{\beta}\geq\frac{m}{n}$;
for $0<\beta<1$, $\frac{1}{n}\sum_{i=1}^{n}b_{i}^{\beta}\leq\frac{m}{n}$. 
\end{lem}
\begin{lem}
\label{lem:RenyiEntropyEquality}For any $a\geq0$ and any $b$, 
\begin{align}
 & \sup_{\widetilde{P}_{X}\in\mathcal{P}\left(\mathcal{X}\right)}\left\{ aH(\widetilde{P}_{X})+b\sum_{x}\widetilde{P}_{X}(x)\log P_{X}(x)\right\} \nonumber \\
 & =\left(a-b\right)H_{\frac{b}{a}}(P_{X}).
\end{align}
For any $a\leq0$ and any $b$, 
\begin{align}
 & \inf_{\widetilde{P}_{X}\in\mathcal{P}\left(\mathcal{X}\right)}\left\{ aH(\widetilde{P}_{X})+b\sum_{x}\widetilde{P}_{X}(x)\log P_{X}(x)\right\} \nonumber \\
 & =\left(a-b\right)H_{\frac{b}{a}}(P_{X}).
\end{align}
\end{lem}

\subsection{\label{subsec:Exponents-of-Information} Information Spectrum Exponents}

Since information spectrum exponents are important in our proofs of
the results in this paper, they will be introduced in the following.
Furthermore, as fundamental information-theoretic quantities, investigating
information spectrum exponents are of independent interest.

For a general distribution $P_{X^{n}}$, define $F_{P_{X^{n}}}(\jmath):=P_{X^{n}}\left(x^{n}:-\frac{1}{n}\log P_{X^{n}}(x^{n})<\jmath\right)$
and $F_{P_{X^{n}}}^{-1}(\theta):=\sup\left\{ \jmath:F_{P_{X^{n}}}(\jmath)\leq\theta\right\} $.
Now consider a product distribution $P_{X}^{n}$ with $P_{X}$ defined
on a finite set $\mathcal{X}$. Define the\emph{ information spectrum
exponents} (or \emph{entropy spectrum exponents}) for distribution
$P_{X}$ as 
\begin{align}
E_{P_{X}}(\jmath) & :=\lim_{n\to\infty}-\frac{1}{n}\log F_{P_{X}^{n}}(\jmath)\\
\widehat{E}_{P_{X}}(\jmath) & :=\lim_{n\to\infty}-\frac{1}{n}\log\left(1-F_{P_{X}^{n}}(\jmath)\right).
\end{align}
Or simply, define the\emph{ information spectrum exponent} for distribution
$P_{X}$ as 
\begin{align}
\widetilde{E}_{P_{X}}(\jmath) & :=\max\left\{ E_{P_{X}}(\jmath),\widehat{E}_{P_{X}}(\jmath)\right\} .
\end{align}
Since for each $\jmath\ge0$, either $E_{P_{X}}(\jmath)$ or $\widehat{E}_{P_{X}}(\jmath)$
can be positive (the other one must be zero), the exponent $\widetilde{E}_{P_{X}}(\jmath)$
contains all the information about the exponent pair $\left(E_{P_{X}}(\jmath),\widehat{E}_{P_{X}}(\jmath)\right)$.
Moreover, the inverse functions of $E_{P_{X}}(\jmath)$ and $\widehat{E}_{P_{X}}(\jmath)$
are denoted as $E_{P_{X}}^{-1}(\omega)$ and $\widehat{E}_{P_{X}}^{-1}(\omega)$.
Then we have the following lemmas. Observe that if $P_{X}$ is uniform,
then $\widetilde{E}_{P_{X}}(\jmath)=+\infty$ for all $\jmath$. Hence,
in the following, we exclude this trivial case. 
\begin{lem}[Information Spectrum Exponents]
\label{lem:Exponents} Assume $P_{X}$ is not uniform. For $\jmath>H_{\infty}(P_{X})$,
\begin{align}
E_{P_{X}}(\jmath) & =\min_{\widetilde{P}_{X}:-\sum_{x}\widetilde{P}_{X}(x)\log P_{X}(x)\leq\jmath}D(\widetilde{P}_{X}\|P_{X})\label{eq:-74}\\
 & =\max_{t\in[0,\infty]}\left\{ tH_{1+t}(P_{X})-t\jmath\right\} ,\label{eq:-76}
\end{align}
and for $0\leq\jmath\le H_{-\infty}(P_{X})$, 
\begin{align}
\widehat{E}_{P_{X}}(\jmath) & =\min_{\widetilde{P}_{X}:-\sum_{x}\widetilde{P}_{X}(x)\log P_{X}(x)\geq\jmath}D(\widetilde{P}_{X}\|P_{X})\label{eq:-75}\\
 & =\max_{t\in[0,\infty]}\left\{ -tH_{1-t}(P_{X})+t\jmath\right\} .\label{eq:-77}
\end{align}
For $0\leq\omega<H_{\infty}(P_{X})$, 
\begin{align}
E_{P_{X}}^{-1}(\omega) & =\min_{\widetilde{P}_{X}:D(\widetilde{P}_{X}\|P_{X})\le\omega}-\sum_{x}\widetilde{P}_{X}(x)\log P_{X}(x)\\
 & =\max_{t\in[0,\infty]}\left\{ H_{1+t}(P_{X})-\frac{\omega}{t}\right\} ,\label{eq:-65}
\end{align}
and for $0\leq\omega\leq H_{-\infty}(P_{X})$, 
\begin{align}
\widehat{E}_{P_{X}}^{-1}(\omega) & =\max_{\widetilde{P}_{X}:D(\widetilde{P}_{X}\|P_{X})\le\omega}-\sum_{x}\widetilde{P}_{X}(x)\log P_{X}(x)\label{eq:-64}\\
 & =\min_{t\in[0,\infty]}\left\{ H_{1-t}(P_{X})+\frac{\omega}{t}\right\} .\label{eq:-63}
\end{align}
Moreover, $E_{P_{X}}(\jmath)$, $\widehat{E}_{P_{X}}(\jmath)$, $E_{P_{X}}^{-1}(\omega)$,
and $\widehat{E}_{P_{X}}^{-1}(\omega)$ are continuous on the intervals
mentioned above. 
\end{lem}
\begin{rem}
We can use $E_{P_{X}}(\jmath)$, $\widehat{E}_{P_{X}}(\jmath)$, $E_{P_{X}}^{-1}(\omega)$,
and $\widehat{E}_{P_{X}}^{-1}(\omega)$ to rewrite $F_{P_{X^{n}}}(\jmath)$,
$1-F_{P_{X^{n}}}(\jmath)$, $F_{P_{X}^{n}}^{-1}(\theta)$, and $F_{P_{X}^{n}}^{-1}(1-\theta)$
as follows: 
\begin{align}
F_{P_{X^{n}}}(\jmath) & =e^{-n\left(E_{P_{X}}(\jmath)+o(1)\right)}\label{eq:-61}\\
1-F_{P_{X^{n}}}(\jmath) & =e^{-n\left(\widehat{E}_{P_{X}}(\jmath)+o(1)\right)}\label{eq:-62}\\
F_{P_{X}^{n}}^{-1}(\theta) & =E_{P_{X}}^{-1}(-\frac{1}{n}\log\theta-o(1))\\
F_{P_{X}^{n}}^{-1}(1-\theta) & =\widehat{E}_{P_{X}}^{-1}(-\frac{1}{n}\log\theta-o(1)),
\end{align}
where the first two equalities follow from the definitions of $E_{P_{X}}(\jmath)$
and $\widehat{E}_{P_{X}}(\jmath)$, and the last two follow since
\begin{align}
F_{P_{X}^{n}}^{-1}(\theta) & =\sup\left\{ \jmath:F_{P_{X}^{n}}(\jmath)\leq\theta\right\} \\
 & =\sup\left\{ \jmath:e^{-n\left(E_{P_{X}}(\jmath)+o(1)\right)}\leq\theta\right\} \\
 & =\sup\left\{ \jmath:E_{P_{X}}(\jmath)\geq-\frac{1}{n}\log\theta-o(1)\right\} \\
 & =E_{P_{X}}^{-1}(-\frac{1}{n}\log\theta-o(1))
\end{align}
and similarly for $F_{P_{X}^{n}}^{-1}(1-\theta)$. 
\end{rem}
Lemma \ref{lem:Exponents} follows by large deviation theory \cite{Dembo},
and it holds not only for finite alphabets, but also for countably
infinite or continuous alphabets (with   the probability mass function 
$P_{X}$ replaced by the corresponding probability density function or the Radon-Nikodym
derivative and   the summation replaced by the corresponding integration). Note that $tH_{1-t}(P_{X})=\log\mathbb{E}\left[e^{-t\log P_{X}(x)}\right]$
is the \emph{logarithmic moment generating function} respect to the
self-information (or self-entropy) $-\log P_{X}(x)$, and \eqref{eq:-76}
and \eqref{eq:-77} are the \emph{Fenchel\textendash Legendre transform}
of $tH_{1-t}(P_{X})$. Furthermore, by \cite[Lemma 2.2.31]{Dembo},
$tH_{1-t}(P_{X})$ is convex in $t\in\mathbb{R}$.

\begin{figure*}[t]
\centering \subfloat[The information spectrum limit]{\includegraphics[width=1\columnwidth]{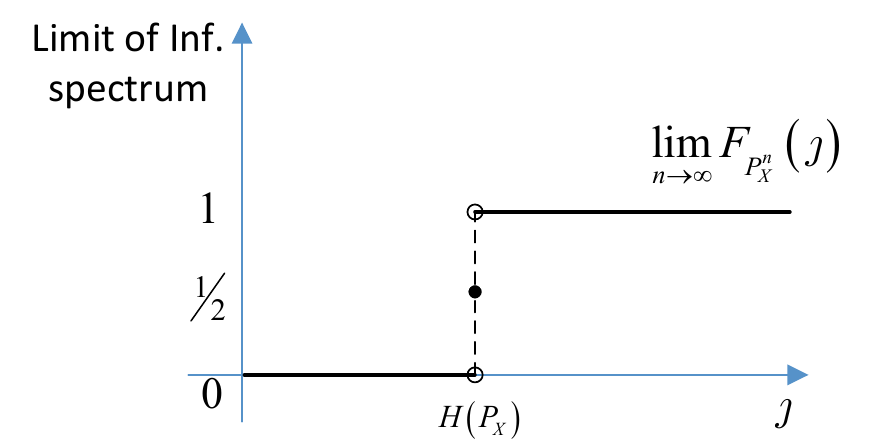}

}

\subfloat[The information spectrum exponent]{\includegraphics[width=1.4\columnwidth]{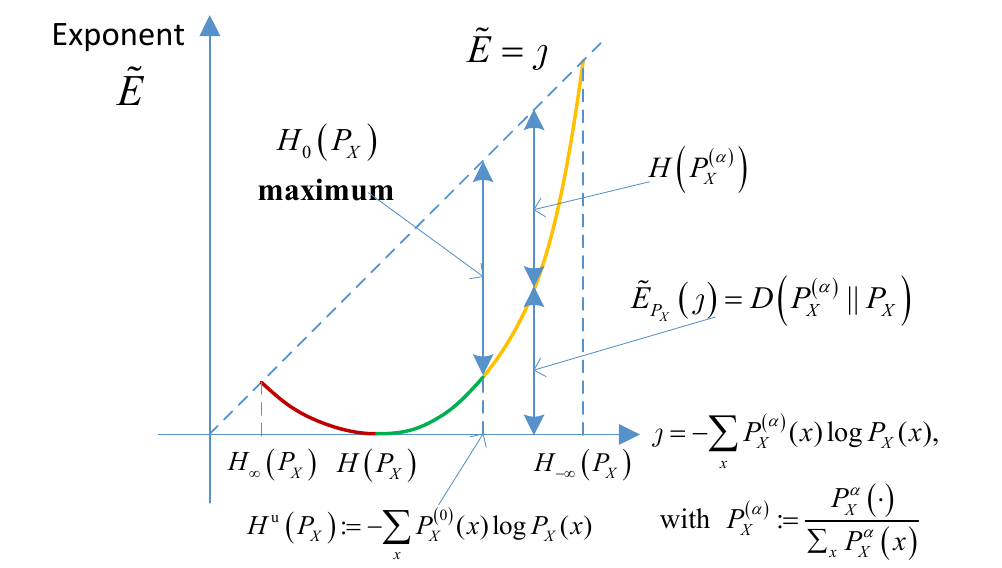}

}

\caption{Illustrations of the information spectrum limit and exponent. Note that in the bottom subfigure, the left  (resp. right)  endpoint of the information spectrum exponent $\widetilde{E}_{P_{X}}(\jmath)$ should be strictly lower than the line $\widetilde{E}=\jmath$ if there are multiple maximum (resp. minimum) probability values in $P_{X}$. } 
\label{fig:exponent} 
\end{figure*}

Note that in \eqref{eq:-74} and \eqref{eq:-75}, the minima are attained
by the $\alpha$-tilted distributions $P_{X}^{(\alpha)}(\cdot)=\frac{P_{X}^{\alpha}(\cdot)}{\sum_{x'}P_{X}^{\alpha}(x')}$
with $\alpha$ satisfying $\jmath=H_{\alpha}^{\mathrm{u}}(P_{X})$.
Hence $P_{X}^{(\alpha)}$ can be seen as a \emph{dominant} ``asymptotic
type''. We have the following lemma. 
\begin{lem}
\label{lem:Exponents-1} $\widetilde{E}_{P_{X}}(\jmath)$ can be expressed
as the following parametric representation with $\alpha\in[-\infty,\infty]$.
\[
\begin{cases}
\jmath=H_{\alpha}^{\mathrm{u}}(P_{X}),\\
\widetilde{E}_{P_{X}}=D\left(P_{X}^{(\alpha)}\|P_{X}\right).
\end{cases}
\]
Specialized to the case $\alpha=0$, it reduces to that 
\begin{align}
\widetilde{E}_{P_{X}}(H^{\mathrm{u}}(P_{X}))=\widehat{E}_{P_{X}}(H^{\mathrm{u}}(P_{X})) & =D(\mathrm{Unif}\left(\mathcal{X}\right)\|P_{X}).\label{eq:-75-2}
\end{align}
\end{lem}
\textcolor{blue}{{} }The information spectrum limit 
\begin{equation}
\lim_{n\to\infty}F_{P_{X}^{n}}(\jmath)=\begin{cases}
0 & \jmath<H(P_{X})\\
\frac{1}{2} & \jmath=H(P_{X})\\
1 & \jmath>H(P_{X})
\end{cases}
\end{equation}
and the information spectrum exponent $\widetilde{E}_{P_{X}}(\jmath)$
are illustrated in Fig. \ref{fig:exponent}.

\begin{lem}[Comparison of Exponents]
\label{lem:ExponentsComparison} Assume both $P_{X}$ and $Q_{Y}$
are not uniform. Then we have 
\begin{align}
 & \frac{1}{R}E_{P_{X}}(R\jmath)>E_{Q_{Y}}(\jmath),\:\forall\jmath\in\frac{1}{R}[H_{\infty}(P_{X}),H(P_{X})]\nonumber \\
 & \Longleftrightarrow\qquad R<\min_{t\in[1,\infty]}\frac{H_{t}(P_{X})}{H_{t}(Q_{Y})};\label{eq:-82}\\
 & \frac{1}{R}\widehat{E}_{P_{X}}(R\jmath)<\widehat{E}_{Q_{Y}}(\jmath),\:\forall\jmath\in\frac{1}{R}[H(P_{X}),H_{-\infty}(P_{X})]\nonumber \\
 & \Longleftrightarrow\qquad R<\min_{t\in[-\infty,1]}\frac{H_{t}(P_{X})}{H_{t}(Q_{Y})}.\label{eq:-85}
\end{align}
Furthermore, the equivalence in \eqref{eq:-85} can be divided into
the following two parts: 
\begin{align}
 & \begin{cases}
\frac{1}{R}\widehat{E}_{P_{X}}(R\jmath)<\widehat{E}_{Q_{Y}}(\jmath),\:\forall\jmath\in\frac{1}{R}[H(P_{X}),H^{\mathrm{u}}(P_{X})]\\
R<\frac{H_{0}(P_{X})}{H_{0}(Q_{Y})}
\end{cases}\nonumber \\
 & \Longleftrightarrow\qquad R<\min_{t\in[0,1]}\frac{H_{t}(P_{X})}{H_{t}(Q_{Y})};\label{eq:-86}\\
 & \begin{cases}
\frac{1}{R}\widehat{E}_{P_{X}}(R\jmath)<\widehat{E}_{Q_{Y}}(\jmath),\:\forall\jmath\in\frac{1}{R}[H^{\mathrm{u}}(P_{X}),H_{-\infty}(P_{X})]\\
R<\frac{H_{0}(P_{X})}{H_{0}(Q_{Y})}
\end{cases}\nonumber \\
 & \Longleftrightarrow\qquad R<\min_{t\in[-\infty,0]}\frac{H_{t}(P_{X})}{H_{t}(Q_{Y})}.\label{eq:-89}
\end{align}
In addition, the equivalences in \eqref{eq:-82}-\eqref{eq:-89} also
hold if all the ``$<$'' are replaced with \textup{``$\le$''.
} 
\end{lem}
\begin{IEEEproof}
Here we only provide a proof for the equivalence in \eqref{eq:-86}.
Other equivalences can be proven similarly.

Proof of ``$\Longleftarrow$'': Observe that the RHS of \eqref{eq:-86}
implies 
\begin{equation}
H_{t}(Q_{Y})<\frac{1}{R}H_{t}(P_{X}),\forall t\in[0,1].
\end{equation}
Hence we have 
\begin{align}
 & \max_{t\in[0,1]}\left\{ -\frac{t}{R}H_{1-t}(P_{X})+t\jmath\right\} \nonumber \\
 & <\max_{t\in[0,1]}\left\{ -tH_{1-t}(Q_{Y})+t\jmath\right\} ,\forall\jmath.
\end{align}

Observe that $-\frac{t}{R}H_{1-t}(P_{X})+t\jmath$ is concave in $t$
(which can be shown by a similar proof to that of \cite[Lemma 7]{Yu},
or directly by \cite[Lemma 2.2.31]{Dembo} since $tH_{1-t}(P_{X})=\log\mathbb{E}\left[e^{-t\log P_{X}(x)}\right]$
is the \emph{logarithmic moment generating function} respect to the
self-information $-\log P_{X}(x)$), and for $\jmath\in\frac{1}{R}[H(P_{X}),H^{\mathrm{u}}(P_{X})]$,
the extreme point of $t\mapsto-\frac{t}{R}H_{1-t}(P_{X})+t\jmath$
is in $[0,1]$. We have for $\jmath\in\frac{1}{R}[H(P_{X}),H^{\mathrm{u}}(P_{X})]$,
\begin{align}
 & \max_{t\in[0,1]}\left\{ -\frac{t}{R}H_{1-t}(P_{X})+t\jmath\right\} \nonumber \\
 & =\max_{t\in[0,\infty]}\left\{ -\frac{t}{R}H_{1-t}(P_{X})+t\jmath\right\} .
\end{align}
Hence for $\jmath\in\frac{1}{R}[H(P_{X}),H^{\mathrm{u}}(P_{X})]$,
\begin{align}
 & \max_{t\in[0,\infty]}\left\{ -\frac{t}{R}H_{1-t}(P_{X})+t\jmath\right\} \nonumber \\
 & <\max_{t\in[0,1]}\left\{ -tH_{1-t}(Q_{Y})+t\jmath\right\} \\
 & \leq\max_{t\in[0,\infty]}\left\{ -tH_{1-t}(Q_{Y})+t\jmath\right\} ,
\end{align}
which, by Lemma \ref{lem:Exponents}, implies the LHS of \eqref{eq:-86}.

Proof of ``$\Longrightarrow$'': The LHS of \eqref{eq:-86} implies
for $\jmath\in\frac{1}{R}[H(P_{X}),H^{\mathrm{u}}(P_{X})]$, 
\begin{align}
 & \max_{t\in[0,\infty]}\left\{ -\frac{t}{R}H_{1-t}(P_{X})+t\jmath\right\} \nonumber \\
 & <\max_{t\in[0,\infty]}\left\{ -tH_{1-t}(Q_{Y})+t\jmath\right\} ,\label{eq:-88}
\end{align}
By setting $\jmath=\frac{1}{R}H(P_{X})$, we have $\frac{1}{R}H(P_{X})>H(Q_{Y})$.

On the other hand, given $\jmath\in[H(Q_{Y}),H_{-\infty}(Q_{Y})]$,
the maximum in the RHS of \eqref{eq:-88} is attained at $g^{-1}(\jmath)$
which is a value $t$ satisfying $\jmath=g(t):=\frac{\partial}{\partial t}\left(tH_{1-t}(Q_{Y})\right)=-\frac{1}{\sum_{y\in\mathcal{Y}}Q_{Y}^{1-t}(y)}\sum_{y\in\mathcal{Y}}Q_{Y}^{1-t}(y)\log Q_{Y}(y)=H_{1-t}^{\mathrm{u}}(Q_{Y})$.
Here $g(t)$ is a increasing function since $tH_{1-t}(Q_{Y})$ is
convex. Hence for $\jmath$ running from $H(Q_{Y})$ to $\frac{1}{R}H^{\mathrm{u}}(P_{X})$,
$g^{-1}(\jmath)$ runs from $0$ to $t_{0}$, where $t_{0}$ is the
solution to $\frac{1}{R}H^{\mathrm{u}}(P_{X})=g(t_{0})$. Observe
$g^{-1}(\jmath)$ is continuous. Hence for each $t'\in[0,t_{0}]$,
we can find a $\jmath'\in[H(Q_{Y}),\frac{1}{R}H^{\mathrm{u}}(P_{X})]$
such that $g^{-1}(\jmath')=t'$. For such $\left(\jmath',t'\right)$,
we have 
\begin{align}
 & -t'H_{1-t'}(Q_{Y})+t'\jmath'\nonumber \\
 & =\max_{t\in[0,\infty]}\left\{ -tH_{1-t}(Q_{Y})+t\jmath'\right\} \\
 & >\max_{t\in[0,\infty]}\left\{ -\frac{t}{R}H_{1-t}(P_{X})+t\jmath'\right\} \\
 & \geq-\frac{t'}{R}H_{1-t'}(P_{X})+t'\jmath'.
\end{align}
That is, for $t'\in[0,t_{0}]$, 
\begin{equation}
RH_{1-t'}(Q_{Y})<H_{1-t'}(P_{X}).\label{eq:-90}
\end{equation}

If $t_{0}<1$, then $\frac{1}{R}H^{\mathrm{u}}(P_{X})<H^{\mathrm{u}}(Q_{Y})$.
The derivative of $\widehat{E}_{Q_{Y}}(\jmath)$ is $g^{-1}(\jmath)$
at $\jmath$, where $g(t)$ is defined above. For $\jmath\in[\frac{1}{R}H^{\mathrm{u}}(P_{X}),H^{\mathrm{u}}(Q_{Y})]$,
$g^{-1}(\jmath)\in[t_{0},1]$. Observe that $\widehat{E}_{Q_{Y}}(\jmath)$
and $\frac{1}{R}\widehat{E}_{P_{X}}(R\jmath)$ are convex, and $-H_{0}(Q_{Y})+\jmath$
and $-\frac{1}{R}H_{0}(P_{X})+\jmath$ are respectively the tangent
lines of $\widehat{E}_{Q_{Y}}(\jmath)$ at $\jmath_{0}=H^{\mathrm{u}}(Q_{Y})$
and $\frac{1}{R}\widehat{E}_{P_{X}}(R\jmath)$ at $\jmath_{0}=\frac{1}{R}H^{\mathrm{u}}(P_{X})$.
Hence combining with the assumption $R<\frac{H_{0}(P_{X})}{H_{0}(Q_{Y})}$,
we have $\widehat{E}_{Q_{Y}}(\jmath)\geq-H_{0}(Q_{Y})+\jmath>-\frac{1}{R}H_{0}(P_{X})+\jmath$.
Moreover, we also have that tangent lines of $\frac{1}{R}\widehat{E}_{P_{X}}(R\jmath)$
at $\jmath_{0}<\frac{1}{R}H^{\mathrm{u}}(P_{X})$ (with slope $t'<1$)
are below the line $-\frac{1}{R}H_{0}(P_{X})+\jmath$ for $\jmath>\frac{1}{R}H^{\mathrm{u}}(P_{X})$.

For $t'\in[t_{0},1]$, denote $\jmath'=g(t')$. Then by the analysis
above, for such $\left(\jmath',t'\right)$, we have 
\begin{align}
-t'H_{1-t'}(Q_{Y})+t'\jmath' & =\widehat{E}_{Q_{Y}}(\jmath')\\
 & >-\frac{1}{R}H_{0}(P_{X})+\jmath'\\
 & \geq-\frac{t'}{R}H_{1-t'}(P_{X})+t'\jmath'.
\end{align}
Hence for $t'\in[t_{0},1]$, \eqref{eq:-90} also holds. 
\end{IEEEproof}
For a distribution $P_{X}$, define the information spectrum exponent
for an interval $[\jmath_{1},\jmath_{2})$ as 
\begin{equation}
E_{P_{X}}(\jmath_{1},\jmath_{2}):=\lim_{n\to\infty}-\frac{1}{n}\log F_{P_{X}^{n}}(\jmath_{1},\jmath_{2}),
\end{equation}
where $F_{P_{X}^{n}}(\jmath_{1},\jmath_{2}):=P_{X}^{n}\left(x^{n}:-\frac{1}{n}\log P_{X}^{n}(x^{n})\in[\jmath_{1},\jmath_{2})\right)$. 
\begin{lem}[Information Spectrum Exponent for an Interval]
\label{lem:Exponents2} Assume $P_{X}$ is not uniform. Then for
$\jmath_{1}<\jmath_{2}$, we have 
\begin{align}
E_{P_{X}}(\jmath_{1},\jmath_{2}) & =\begin{cases}
E_{P_{X}}(\jmath_{2}), & H_{\infty}(P_{X})\leq\jmath_{1}<\jmath_{2}\leq H(P_{X})\\
\widehat{E}_{P_{X}}(\jmath_{1}), & H(P_{X})\leq\jmath_{1}<\jmath_{2}\leq H_{-\infty}(P_{X})\\
0, & H_{\infty}(P_{X})\leq\jmath_{1}\leq H(P_{X})\\
 & \qquad\leq\jmath_{2}\leq H_{-\infty}(P_{X})
\end{cases}.\label{eq:-74-1}
\end{align}
\end{lem}
Lemma \ref{lem:Exponents2} follows directly from Lemma \ref{lem:Exponents},
and hence the proof is omitted.

\section{\label{sec:Proof-of-Theorem-RenyiPQ}Proof of Theorem \ref{thm:RenyiPQ} }

In the following, we only consider the case of $R=1$. For the general
case, we can obtain the result by setting $Q_{Y}$ to the product
distribution $Q_{Y}^{R}$, if $R$ is an integer; otherwise, set $P_{X}$
to $P_{X}^{k_{0}}$ and $Q_{Y}$ to $Q_{Y}^{n_{0}}$, where $k_{0}$
and $n_{0}$ are co-prime and $R=\frac{n_{0}}{k_{0}}$.

\emph{Achievability: }Assume $g:\mathcal{P}^{\left(n\right)}\left(\mathcal{X}\right)\to\mathcal{P}^{\left(n\right)}\left(\mathcal{Y}\right)$
is a function that maps $n$-types on $\mathcal{X}$ to $n$-types
on $\mathcal{Y}$. A code $f$ induced by $g$ is obtained by mapping
the sequences in $\mathcal{T}_{T_{X}}$ to the sequences in $\mathcal{T}_{g(T_{X})}$
as uniformly as possible for all $T_{X}\in\mathcal{P}^{\left(n\right)}\left(\mathcal{X}\right)$.
That is, $f$ maps $\left\lfloor \nicefrac{\left|\mathcal{T}_{T_{X}}\right|}{\left|\mathcal{T}_{g(T_{X})}\right|}\right\rfloor $
or $\left\lceil \nicefrac{\left|\mathcal{T}_{T_{X}}\right|}{\left|\mathcal{T}_{g(T_{X})}\right|}\right\rceil $
sequences in $\mathcal{T}_{T_{X}}$ to each sequence in $\mathcal{T}_{g(T_{X})}$.
For this code $f$, and for $\alpha=1+s>1$, we have 
\begin{align}
 & \frac{1}{n}D_{1+s}(P_{Y^{n}}\|Q_{Y}^{n})\nonumber \\
 & =\frac{1}{ns}\log\sum_{y^{n}}P_{Y^{n}}(y^{n})^{1+s}Q_{Y}^{n}(y^{n})^{-s}\\
 & =\frac{1}{ns}\log\sum_{T_{Y}}\sum_{y^{n}\in\mathcal{T}_{T_{Y}}}\Bigl(\sum_{T_{X}\in g^{-1}(\left\{ T_{Y}\right\} )}\sum_{x^{n}\in\mathcal{T}_{T_{X}}}P_{X}^{n}(x^{n})\nonumber \\
 & \qquad\times1\left\{ y^{n}=f(x^{n})\right\} \Bigr)^{1+s}Q_{Y}^{n}(y^{n})^{-s}\\
 & \leq\frac{1}{ns}\log\sum_{T_{Y}}\sum_{y^{n}\in\mathcal{T}_{T_{Y}}}\Bigl(\sum_{T_{X}\in g^{-1}(\left\{ T_{Y}\right\} )}\varphi_{1}\left(T_{X},T_{Y}\right)\nonumber \\
 & \qquad+\varphi_{2}\left(y^{n},T_{X},T_{Y}\right)\Bigr)^{1+s}e^{-ns\sum_{y}T_{Y}(y)\log Q_{Y}(y)},\label{eq:-92}
\end{align}
where 
\begin{align}
\varphi_{1}\left(T_{X},T_{Y}\right) & :=e^{n\sum_{y}T_{X}(x)\log P_{X}(x)}\left(\frac{\left|\mathcal{T}_{T_{X}}\right|}{\left|\mathcal{T}_{T_{Y}}\right|}+1\right)\nonumber \\
 & \qquad\times1\left\{ \left|\mathcal{T}_{T_{X}}\right|\geq\left|\mathcal{T}_{T_{Y}}\right|\right\} \\
\varphi_{2}\left(y^{n},T_{X},T_{Y}\right) & :=e^{n\sum_{x}T_{X}(x)\log P_{X}(x)}1\left\{ y^{n}\in f\left(\mathcal{T}_{T_{X}}\right)\right\} \nonumber \\
 & \qquad\times1\left\{ \left|\mathcal{T}_{T_{X}}\right|<\left|\mathcal{T}_{T_{Y}}\right|\right\} ,
\end{align}
and \eqref{eq:-92} follows from the construction of the code $f$.

Observe that 
\begin{align}
 & \varphi_{1}\left(T_{X},T_{Y}\right)\nonumber \\
 & \leq2\widetilde{\varphi}_{1}\left(T_{X},T_{Y}\right)\\
 & :=2e^{n\sum_{y}T_{X}(x)\log P_{X}(x)}\frac{\left|\mathcal{T}_{T_{X}}\right|}{\left|\mathcal{T}_{T_{Y}}\right|}1\left\{ \left|\mathcal{T}_{T_{X}}\right|\geq\left|\mathcal{T}_{T_{Y}}\right|\right\} .
\end{align}
Hence we have \eqref{eq:-105}-\eqref{eq:-8} (given on page \pageref{eq:-105}),
\begin{figure*}
\begin{align}
 & \frac{1}{n}D_{1+s}(P_{Y^{n}}\|Q_{Y}^{n})\nonumber \\
 & \leq\frac{1}{ns}\log\sum_{T_{Y}}\sum_{y^{n}\in\mathcal{T}_{T_{Y}}}\Bigl(\sum_{T_{X}\in g^{-1}(\left\{ T_{Y}\right\} )}\widetilde{\varphi}_{1}\left(T_{X},T_{Y}\right)+\varphi_{2}\left(y^{n},T_{X},T_{Y}\right)\Bigr)^{1+s}e^{-ns\sum_{y}T_{Y}(y)\log Q_{Y}(y)}+\frac{1}{ns}\log2^{1+s}\label{eq:-105}\\
 & =\frac{1}{ns}\log\sum_{T_{Y}}\sum_{y^{n}\in\mathcal{T}_{T_{Y}}}\sum_{T_{X}\in g^{-1}(\left\{ T_{Y}\right\} )}\Bigl(\widetilde{\varphi}_{1}\left(T_{X},T_{Y}\right)+\varphi_{2}\left(y^{n},T_{X},T_{Y}\right)\Bigr)^{1+s}e^{-ns\sum_{y}T_{Y}(y)\log Q_{Y}(y)}+o(1)\label{eq:-9}\\
 & =\frac{1}{ns}\log\sum_{T_{Y}}\sum_{T_{X}\in g^{-1}(\left\{ T_{Y}\right\} )}\Bigl(e^{n\left(1+s\right)\sum_{y}T_{X}(x)\log P_{X}(x)}\frac{\left|\mathcal{T}_{T_{X}}\right|^{1+s}}{\left|\mathcal{T}_{T_{Y}}\right|^{s}}1\left\{ \left|\mathcal{T}_{T_{X}}\right|\geq\left|\mathcal{T}_{T_{Y}}\right|\right\} \nonumber \\
 & \qquad+e^{n\left(1+s\right)\sum_{x}T_{X}(x)\log P_{X}(x)}\left|\mathcal{T}_{T_{X}}\right|1\left\{ \left|\mathcal{T}_{T_{X}}\right|<\left|\mathcal{T}_{T_{Y}}\right|\right\} \Bigr)e^{-ns\sum_{y}T_{Y}(y)\log Q_{Y}(y)}+o(1)\label{eq:-10}\\
 & =\frac{1}{ns}\log\max_{T_{Y}}\max_{T_{X}\in g^{-1}(\left\{ T_{Y}\right\} )}\Bigl(e^{-n\left(1+s\right)D(T_{X}\|P_{X})-nsH(T_{Y})}1\left\{ \left|\mathcal{T}_{T_{X}}\right|\geq\left|\mathcal{T}_{T_{Y}}\right|\right\} \nonumber \\
 & \qquad+e^{n\left(1+s\right)\sum_{x}T_{X}(x)\log P_{X}(x)+nH(T_{X})}1\left\{ \left|\mathcal{T}_{T_{X}}\right|<\left|\mathcal{T}_{T_{Y}}\right|\right\} \Bigr)e^{-ns\sum_{y}T_{Y}(y)\log Q_{Y}(y)}+o(1)\label{eq:-20}\\
 & =\frac{1}{s}\max_{T_{Y}}\max_{T_{X}\in g^{-1}(\left\{ T_{Y}\right\} )}\left\{ -\left(1+s\right)D(T_{X}\|P_{X})+sD(T_{Y}\|Q_{Y})+s\left(H(T_{Y})-H(T_{X})\right)1\left\{ \left|\mathcal{T}_{T_{X}}\right|<\left|\mathcal{T}_{T_{Y}}\right|\right\} \right\} +o(1)\\
 & =\frac{1}{s}\max_{T_{Y}}\max_{T_{X}\in g^{-1}(\left\{ T_{Y}\right\} )}\left\{ -\left(1+s\right)D(T_{X}\|P_{X})+sD(T_{Y}\|Q_{Y})+s\left[H(T_{Y})-H(T_{X})\right]^{+}\right\} +o(1)\label{eq:-50}\\
 & =\frac{1}{s}\max_{T_{X}}\left.\left\{ -\left(1+s\right)D(T_{X}\|P_{X})+sD(T_{Y}\|Q_{Y})+s\left[H(T_{Y})-H(T_{X})\right]^{+}\right\} \right|_{T_{Y}=g(T_{X})}+o(1)\label{eq:-8}
\end{align}

\hrulefill{} 
\end{figure*}

where in \eqref{eq:-9}, the sum operation $\sum_{T_{X}\in g^{-1}(\left\{ T_{Y}\right\} )}$
is taken outside the $(\cdot)^{1+s}$ since by the fact that the number
of $n$-types $T_{X}$ is polynomial in $n$, we have 
\begin{align}
 & \Bigl(\sum_{T_{X}\in g^{-1}(\left\{ T_{Y}\right\} )}\widetilde{\varphi}_{1}\left(T_{X},T_{Y}\right)+\varphi_{2}\left(y^{n},T_{X},T_{Y}\right)\Bigr)^{1+s}\nonumber \\
 & \qquad\times e^{-ns\sum_{y}T_{Y}(y)\log Q_{Y}(y)}\nonumber \\
 & =\max_{T_{X}\in g^{-1}(\left\{ T_{Y}\right\} )}\Bigl(\widetilde{\varphi}_{1}\left(T_{X},T_{Y}\right)+\varphi_{2}\left(y^{n},T_{X},T_{Y}\right)\Bigr)^{1+s}\nonumber \\
 & \qquad\times e^{-ns\sum_{y}T_{Y}(y)\log Q_{Y}(y)}+o(1)\\
 & =\sum_{T_{X}\in g^{-1}(\left\{ T_{Y}\right\} )}\Bigl(\widetilde{\varphi}_{1}\left(T_{X},T_{Y}\right)+\varphi_{2}\left(y^{n},T_{X},T_{Y}\right)\Bigr)^{1+s}\nonumber \\
 & \qquad\times e^{-ns\sum_{y}T_{Y}(y)\log Q_{Y}(y)}+o(1);
\end{align}
and \eqref{eq:-20} also follows from the fact that the number of
$n$-types $T_{X}$ (or $T_{Y}$) is polynomial in $n$.

For each $T_{X}$, choose $g(T_{X})$ as the $T_{Y}$ that minimizes
the expression in \eqref{eq:-8}. Then we obtain 
\begin{align}
 & \limsup_{n\to\infty}\frac{1}{n}D_{1+s}(P_{Y^{n}}\|Q_{Y}^{n})\nonumber \\
 & \leq\limsup_{n\to\infty}\max_{T_{X}}\min_{T_{Y}}\Bigl\{-\frac{1+s}{s}D(T_{X}\|P_{X})\nonumber \\
 & \qquad+D(T_{Y}\|Q_{Y})+\left[H(T_{Y})-H(T_{X})\right]^{+}\Bigr\}\\
 & =\max_{\widetilde{P}_{X}\in\mathcal{P}\left(\mathcal{X}\right)}\min_{\widetilde{P}_{Y}\in\mathcal{P}\left(\mathcal{Y}\right)}\Bigl\{-\frac{1+s}{s}D(\widetilde{P}_{X}\|P_{X})\nonumber \\
 & \qquad+D(\widetilde{P}_{Y}\|Q_{Y})+\left[H(\widetilde{P}_{Y})-H(\widetilde{P}_{X})\right]^{+}\Bigr\}\label{eq:-22}\\
 & =\max_{\widetilde{P}_{X}\in\mathcal{P}\left(\mathcal{X}\right)}\min_{\widetilde{P}_{Y}\in\mathcal{P}\left(\mathcal{Y}\right)}\max_{t\in[0,1]}\Bigl\{-\frac{1+s}{s}D(\widetilde{P}_{X}\|P_{X})\nonumber \\
 & \qquad+D(\widetilde{P}_{Y}\|Q_{Y})+t\left(H(\widetilde{P}_{Y})-H(\widetilde{P}_{X})\right)\Bigr\}\\
 & =\max_{\widetilde{P}_{X}\in\mathcal{P}\left(\mathcal{X}\right)}\max_{t\in[0,1]}\min_{\widetilde{P}_{Y}\in\mathcal{P}\left(\mathcal{Y}\right)}\Bigl\{-\frac{1+s}{s}D(\widetilde{P}_{X}\|P_{X})\nonumber \\
 & \qquad+D(\widetilde{P}_{Y}\|Q_{Y})+t\left(H(\widetilde{P}_{Y})-H(\widetilde{P}_{X})\right)\Bigr\}\label{eq:-11}\\
 & =\max_{\widetilde{P}_{X}\in\mathcal{P}\left(\mathcal{X}\right)}\max_{t\in[0,1]}\Bigl\{ tH_{\frac{1}{1-t}}(Q_{Y})\nonumber \\
 & \qquad-\frac{1+s}{s}D(\widetilde{P}_{X}\|P_{X})-tH(\widetilde{P}_{X})\Bigr\}\label{eq:-38}\\
 & =\max_{t\in[0,1]}\left\{ tH_{\frac{1}{1-t}}(Q_{Y})-tH_{\frac{1+s}{1+s-st}}(P_{X})\right\} ,\label{eq:-39}
\end{align}
where \eqref{eq:-22} follows from Lemma \ref{lem:minequality}, the
swapping of min and max in \eqref{eq:-11} follows from the fact that
the objective function is convex and concave in $\widetilde{P}_{Y}$
and $t$ respectively, $\widetilde{P}_{Y}$ resides in a compact,
convex set (the probability simplex) and $t$ resides in a convex
set $\left[0,1\right]$ (Sion's minimax theorem \cite{Sion}); and
\eqref{eq:-38} and \eqref{eq:-39} follow from Lemma \ref{lem:RenyiEntropyEquality}.

For $\alpha=1+s\in(0,1)$, similar to \eqref{eq:-8}, we can show
that 
\begin{align}
 & \limsup_{n\to\infty}\frac{1}{n}D_{1+s}(P_{Y^{n}}||Q_{Y}^{n})\nonumber \\
 & \leq\frac{1}{s}\max_{T_{X}}\Bigl\{-\left(1+s\right)D(T_{X}\|P_{X})+sD(T_{Y}\|Q_{Y})\nonumber \\
 & \qquad\left.+s\left[H(T_{Y})-H(T_{X})\right]^{+}\Bigr\}\right|_{T_{Y}=g(T_{X})}.\label{eq:-16}
\end{align}
For each $T_{X}$, choose $g(T_{X})$ as the $T_{Y}$ that maximizes
the expression in \eqref{eq:-16}. Then similarly we obtain that 
\begin{align}
 & \limsup_{n\to\infty}\frac{1}{n}D_{1+s}(P_{Y^{n}}||Q_{Y}^{n})\nonumber \\
 & \leq\limsup_{n\to\infty}\frac{1}{s}\max_{T_{X}}\max_{T_{Y}}\Bigl\{-\left(1+s\right)D(T_{X}\|P_{X})\nonumber \\
 & \qquad+sD(T_{Y}\|Q_{Y})+s\left[H(T_{Y})-H(T_{X})\right]^{+}\Bigr\}\label{eq:-24}\\
 & =\min_{\widetilde{P}_{X}\in\mathcal{P}\left(\mathcal{X}\right)}\min_{\widetilde{P}_{Y}\in\mathcal{P}\left(\mathcal{Y}\right)}\Bigl\{-\frac{1+s}{s}D(\widetilde{P}_{X}\|P_{X})\nonumber \\
 & \qquad+D(\widetilde{P}_{Y}\|Q_{Y})+\left[H(\widetilde{P}_{Y})-H(\widetilde{P}_{X})\right]^{+}\Bigr\}\label{eq:-17}\\
 & =\min_{\widetilde{P}_{X}\in\mathcal{P}\left(\mathcal{X}\right)}\min_{\widetilde{P}_{Y}\in\mathcal{P}\left(\mathcal{Y}\right)}\max_{t\in[0,1]}\Bigl\{-\frac{1+s}{s}D(\widetilde{P}_{X}\|P_{X})\nonumber \\
 & \qquad+D(\widetilde{P}_{Y}\|Q_{Y})+t\left(H(\widetilde{P}_{Y})-H(\widetilde{P}_{X})\right)\Bigr\}\\
 & =\max_{t\in[0,1]}\min_{\widetilde{P}_{X}\in\mathcal{P}\left(\mathcal{X}\right)}\min_{\widetilde{P}_{Y}\in\mathcal{P}\left(\mathcal{Y}\right)}\Bigl\{-\frac{1+s}{s}D(\widetilde{P}_{X}\|P_{X})\nonumber \\
 & \qquad+D(\widetilde{P}_{Y}\|Q_{Y})+t\left(H(\widetilde{P}_{Y})-H(\widetilde{P}_{X})\right)\Bigr\}\\
 & =\max_{t\in[0,1]}\left\{ tH_{\frac{1}{1-t}}(Q_{Y})-tH_{\frac{1+s}{1+s-st}}(P_{X})\right\} ,\label{eq:-15}
\end{align}
where \eqref{eq:-17} follows from Lemma \ref{lem:minequality} (Note
that here $s<0$).

\emph{Converse: }Consider an optimal function $f:\mathcal{X}^{k}\to\mathcal{Y}^{n}$
attaining the minimum of $\frac{1}{n}D_{1+s}(P_{Y^{n}}\|Q_{Y}^{n})$.
Since $\left|\mathcal{P}^{\left(n\right)}\left(\mathcal{Y}\right)\right|\leq\left(n+1\right)^{\left|\mathcal{Y}\right|}$,
by the pigeonhole principle, we have that for every $T_{X}$, there
exists a type $T_{Y}=g(T_{X})$ such that at least $\frac{1}{(n+1)^{|\mathcal{Y}|}}\left|\mathcal{T}_{T_{X}}\right|$
sequences in $\mathcal{T}_{T_{X}}$ are mapped through $f$ to the
sequences in $\mathcal{T}_{T_{Y}}$. Hence for such $T_{Y}=g(T_{X})$,
we have $\sum_{y^{n}\in\mathcal{T}_{T_{Y}}}\left|f^{-1}(\left\{ y^{n}\right\} )\cap\mathcal{T}_{T_{X}}\right|=\left|f^{-1}(\mathcal{T}_{T_{Y}})\cap\mathcal{T}_{T_{X}}\right|\geq\frac{1}{(n+1)^{|\mathcal{Y}|}}\left|\mathcal{T}_{T_{X}}\right|$.

For $s>0$, we have \eqref{eq:-23}-\eqref{eq:-106} (given on page
\pageref{eq:-23}). 
\begin{figure*}
\begin{align}
 & \frac{1}{n}D_{1+s}(P_{Y^{n}}\|Q_{Y}^{n})\nonumber \\
 & =\frac{1}{ns}\log\sum_{T_{Y}}\sum_{y^{n}\in\mathcal{T}_{T_{Y}}}\left(\sum_{T_{X}}\sum_{x^{n}\in\mathcal{T}_{T_{X}}}P_{X}^{n}(x^{n})1\left\{ y^{n}=f(x^{n})\right\} \right)^{1+s}Q_{Y}^{n}(y^{n})^{-s}\label{eq:-23}\\
 & \geq\frac{1}{ns}\log\sum_{T_{Y}}\sum_{y^{n}\in\mathcal{T}_{T_{Y}}}\left(\max_{T_{X}}\sum_{x^{n}\in\mathcal{T}_{T_{X}}}P_{X}^{n}(x^{n})1\left\{ y^{n}=f(x^{n})\right\} \right)^{1+s}Q_{Y}^{n}(y^{n})^{-s}\\
 & \geq\frac{1}{ns}\log\max_{T_{X}}\sum_{T_{Y}}\sum_{y^{n}\in\mathcal{T}_{T_{Y}}}\left(\sum_{x^{n}\in\mathcal{T}_{T_{X}}}P_{X}^{n}(x^{n})1\left\{ y^{n}=f(x^{n})\right\} \right)^{1+s}Q_{Y}^{n}(y^{n})^{-s}\\
 & \geq\frac{1}{ns}\log\max_{T_{X}}\left.\left\{ \sum_{y^{n}\in\mathcal{T}_{T_{Y}}}\left(\sum_{x^{n}\in\mathcal{T}_{T_{X}}}P_{X}^{n}(x^{n})1\left\{ y^{n}=f(x^{n})\right\} \right)^{1+s}Q_{Y}^{n}(y^{n})^{-s}\right\} \right|_{T_{Y}=g(T_{X})}\\
 & =\frac{1}{ns}\log\max_{T_{X}}\left.\left\{ e^{n(1+s)\sum_{x}T_{X}(x)\log P_{X}(x)-ns\sum_{y}T_{Y}(y)\log Q_{Y}(y)}\sum_{y^{n}\in\mathcal{T}_{T_{Y}}}\left|f^{-1}(\left\{ y^{n}\right\} )\cap\mathcal{T}_{T_{X}}\right|^{1+s}\right\} \right|_{T_{Y}=g(T_{X})}\label{eq:-106}
\end{align}

\hrulefill{} 
\end{figure*}

By Lemma \ref{lem:convexity_concavity}, 
\begin{align}
 & \sum_{y^{n}\in\mathcal{T}_{T_{Y}}}\left|f^{-1}(\left\{ y^{n}\right\} )\cap\mathcal{T}_{T_{X}}\right|^{1+s}\nonumber \\
 & \geq\left|\mathcal{T}_{T_{Y}}\right|\left(\frac{\frac{1}{(n+1)^{|\mathcal{Y}|}}\left|\mathcal{T}_{T_{X}}\right|}{\left|\mathcal{T}_{T_{Y}}\right|}\right)^{1+s}1\left\{ \left|\mathcal{T}_{T_{X}}\right|\geq\left|\mathcal{T}_{T_{Y}}\right|\right\} \nonumber \\
 & \qquad+\left|\mathcal{T}_{T_{X}}\right|1\left\{ \left|\mathcal{T}_{T_{X}}\right|<\left|\mathcal{T}_{T_{Y}}\right|\right\} \\
 & \doteq e^{\left(1+s\right)nH(T_{X})-snH(T_{Y})}1\left\{ \left|\mathcal{T}_{T_{X}}\right|\geq\left|\mathcal{T}_{T_{Y}}\right|\right\} \nonumber \\
 & \qquad+e^{nH(T_{X})}1\left\{ \left|\mathcal{T}_{T_{X}}\right|<\left|\mathcal{T}_{T_{Y}}\right|\right\} 
\end{align}
Therefore, we have \eqref{eq:-2}-\eqref{eq:-21} (given on the page
\pageref{eq:-2}), 
\begin{figure*}
\begin{align}
 & \frac{1}{n}D_{1+s}(P_{Y^{n}}\|Q_{Y}^{n})\nonumber \\
 & \geq\frac{1}{ns}\log\max_{T_{X}}\biggl\{ e^{n(1+s)\sum_{x}T_{X}(x)\log P_{X}(x)-ns\sum_{y}T_{Y}(y)\log Q_{Y}(y)}\nonumber \\
 & \qquad\times\left.\left(e^{\left(1+s\right)nH(T_{X})-snH(T_{Y})}1\left\{ \left|\mathcal{T}_{T_{X}}\right|\geq\left|\mathcal{T}_{T_{Y}}\right|\right\} +e^{nH(T_{X})}1\left\{ \left|\mathcal{T}_{T_{X}}\right|<\left|\mathcal{T}_{T_{Y}}\right|\right\} \right)\biggr\}\right|_{T_{Y}=g(T_{X})}+o(1)\label{eq:-2}\\
 & =\frac{1}{s}\max_{T_{X}}\left.\left\{ -\left(1+s\right)D(T_{X}\|P_{X})+sD(T_{Y}\|Q_{Y})+s\left[H(T_{Y})-H(T_{X})\right]^{+}\right\} \right|_{T_{Y}=g(T_{X})}+o(1)\label{eq:-13}\\
 & \geq\max_{T_{X}}\min_{T_{Y}}\left\{ -\frac{1+s}{s}D(T_{X}\|P_{X})+D(T_{Y}\|Q_{Y})+\left[H(T_{Y})-H(T_{X})\right]^{+}\right\} +o(1)\\
 & =\max_{t\in[0,1]}\left\{ tH_{\frac{1}{1-t}}(Q_{Y})-tH_{\frac{1+s}{1+s-st}}(P_{X})\right\} +o(1),\label{eq:-21}
\end{align}

\hrulefill{} 
\end{figure*}

where \eqref{eq:-21} follows from the derivations in \eqref{eq:-22}-\eqref{eq:-39}.

For $s<0$, following derivations similar to \eqref{eq:-23}-\eqref{eq:-13},
we have 
\begin{align}
 & \frac{1}{n}D_{1+s}(P_{Y^{n}}\|Q_{Y}^{n})\nonumber \\
 & \geq\frac{1}{s}\max_{T_{X}}\Bigl\{-\left(1+s\right)D(T_{X}\|P_{X})+sD(T_{Y}\|Q_{Y})\nonumber \\
 & \qquad\left.+s\left[H(T_{Y})-H(T_{X})\right]^{+}\Bigr\}\right|_{T_{Y}=g(T_{X})}+o(1)\\
 & \geq\min_{T_{X}}\min_{T_{Y}}\Bigl\{-\frac{1+s}{s}D(T_{X}\|P_{X})+D(T_{Y}\|Q_{Y})\nonumber \\
 & \qquad+\left[H(T_{Y})-H(T_{X})\right]^{+}\Bigr\}+o(1)\\
 & =\max_{t\in[0,1]}\left\{ tH_{\frac{1}{1-t}}(Q_{Y})-tH_{\frac{1+s}{1+s-st}}(P_{X})\right\} +o(1),\label{eq:-25}
\end{align}
where \eqref{eq:-25} follows from the derivations in \eqref{eq:-24}-\eqref{eq:-15}.

\section{\label{sec:Proof-of-Theorem-RenyiQP}Proof of Theorem \ref{thm:RenyiQP} }

Similar to the proof in Appendix \ref{sec:Proof-of-Theorem-RenyiPQ},
we only prove the case of $R=1$.

\emph{Achievability: }By the equality $D_{\alpha}(Q\|P)=\frac{\alpha}{1-\alpha}D_{1-\alpha}(P\|Q)$
for $\alpha\in(0,1)$, the case $\alpha\in(0,1)$ has been proven
in Theorem \ref{thm:RenyiPQ}, so here we only need to consider the
case $\alpha>1$.

We consider the following mapping. For each $T_{X}$, partition $\mathcal{T}_{T_{X}}$
into $a_{T_{X}}=\left|\left\{ T_{Y}:H(T_{X})\ge H(T_{Y})+\delta\right\} \right|$
subsets with size $\left\lfloor \frac{\left|\mathcal{T}_{T_{X}}\right|}{a_{T_{X}}}\right\rfloor $
or $\left\lceil \frac{\left|\mathcal{T}_{T_{X}}\right|}{a_{T_{X}}}\right\rceil $.
For each $T_{Y}$ such that $H(T_{X})\ge H(T_{Y})+\delta$, map the
sequences in each subset to the sequences in the set $\mathcal{T}_{T_{Y}}$
as uniformly as possible, such that $\left\lfloor \nicefrac{\left\lfloor \frac{\left|\mathcal{T}_{T_{X}}\right|}{a_{T_{X}}}\right\rfloor }{\left|\mathcal{T}_{T_{Y}}\right|}\right\rfloor $
or $\left\lceil \nicefrac{\left\lfloor \frac{\left|\mathcal{T}_{T_{X}}\right|}{a_{T_{X}}}\right\rfloor }{\left|\mathcal{T}_{T_{Y}}\right|}\right\rceil $
(for subsets with size $\left\lfloor \frac{\left|\mathcal{T}_{T_{X}}\right|}{a_{T_{X}}}\right\rfloor $)
or $\left\lfloor \nicefrac{\left\lceil \frac{\left|\mathcal{T}_{T_{X}}\right|}{a_{T_{X}}}\right\rceil }{\left|\mathcal{T}_{T_{Y}}\right|}\right\rfloor $
or $\left\lceil \nicefrac{\left\lceil \frac{\left|\mathcal{T}_{T_{X}}\right|}{a_{T_{X}}}\right\rceil }{\left|\mathcal{T}_{T_{Y}}\right|}\right\rceil $
(for subsets with size $\left\lceil \frac{\left|\mathcal{T}_{T_{X}}\right|}{a_{T_{X}}}\right\rceil $)
sequences in $\mathcal{T}_{T_{X}}$ are mapped to each sequence in
$\mathcal{T}_{T_{Y}}$. If there is no such $T_{Y}$, then map the
sequences in $\mathcal{T}_{T_{X}}$ into any sequences in $\mathcal{Y}^{n}$.

For this code and for $s>0$, we have \eqref{eq:-107}-\eqref{eq:-108}
(given on page \pageref{eq:-107}), 
\begin{figure*}
\begin{align}
 & \frac{1}{n}D_{1+s}(Q_{Y}^{n}\|P_{Y^{n}})\nonumber \\
 & =\frac{1}{ns}\log\sum_{y^{n}}Q_{Y}^{n}(y^{n})^{1+s}P_{Y^{n}}(y^{n})^{-s}\label{eq:-107}\\
 & =\frac{1}{ns}\log\sum_{T_{Y}}\sum_{y^{n}\in\mathcal{T}_{T_{Y}}}\left(\sum_{T_{X}}\sum_{x^{n}\in\mathcal{T}_{T_{X}}}P_{X}^{n}(x^{n})1\left\{ y^{n}=f(x^{n})\right\} \right)^{-s}Q_{Y}^{n}(y^{n})^{1+s}\\
 & \leq\frac{1}{ns}\log\sum_{T_{Y}}\sum_{y^{n}\in\mathcal{T}_{T_{Y}}}\left(\sum_{T_{X}:H(T_{X})\ge H(T_{Y})+\delta}e^{n\sum_{x}T_{X}(x)\log P_{X}(x)}\left(\frac{\left(\frac{\left|\mathcal{T}_{T_{X}}\right|}{a_{T_{X}}}-1\right)}{\left|\mathcal{T}_{T_{Y}}\right|}-1\right)\right)^{-s}Q_{Y}^{n}(y^{n})^{1+s}\\
 & \leq\frac{1}{ns}\log\sum_{T_{Y}}\sum_{y^{n}\in\mathcal{T}_{T_{Y}}}\left(\sum_{T_{X}:H(T_{X})\ge H(T_{Y})+\delta}e^{n\sum_{x}T_{X}(x)\log P_{X}(x)}\left(e^{n\left(H(T_{X})-H(T_{Y})+o(1)\right)}-2\right)\right)^{-s}e^{\left(1+s\right)n\sum_{y}T_{Y}(y)\log Q_{Y}(y)}\\
 & \leq\frac{1}{ns}\log\sum_{T_{Y}}\sum_{y^{n}\in\mathcal{T}_{T_{Y}}}\left(\sum_{T_{X}:H(T_{X})\ge H(T_{Y})+\delta}e^{-nD(T_{X}\|P_{X})-nH(T_{Y})+no(1)}\left(1-2e^{-n\left(\delta+o(1)\right)}\right)\right)^{-s}e^{\left(1+s\right)n\sum_{y}T_{Y}(y)\log Q_{Y}(y)}\\
 & =\frac{1}{ns}\log\sum_{T_{Y}}\sum_{y^{n}\in\mathcal{T}_{T_{Y}}}\left(\sum_{T_{X}:H(T_{X})\ge H(T_{Y})+\delta}e^{-nD(T_{X}\|P_{X})-nH(T_{Y})}\right)^{-s}e^{\left(1+s\right)n\sum_{y}T_{Y}(y)\log Q_{Y}(y)}+o(1)\\
 & \leq\frac{1}{ns}\log\max_{T_{Y}}\min_{T_{X}:H(T_{X})\ge H(T_{Y})+\delta}e^{snD(T_{X}\|P_{X})+\left(1+s\right)nH(T_{Y})}e^{\left(1+s\right)n\sum_{y}T_{Y}(y)\log Q_{Y}(y)}+o(1)\label{eq:-91}\\
 & =\max_{T_{Y}}\min_{T_{X}:H(T_{X})\ge H(T_{Y})+\delta}\left\{ D(T_{X}\|P_{X})-\frac{1+s}{s}D(T_{Y}\|Q_{Y})\right\} +o(1)\label{eq:-52}\\
 & =\max_{\widetilde{P}_{Y}\in\mathcal{P}\left(\mathcal{Y}\right)}\min_{\widetilde{P}_{X}\in\mathcal{P}\left(\mathcal{X}\right):H(\widetilde{P}_{X})\ge H(\widetilde{P}_{Y})+\delta}\left\{ D(\widetilde{P}_{X}\|P_{X})-\frac{1+s}{s}D(\widetilde{P}_{Y}\|Q_{Y})\right\} +o(1),\label{eq:-108}
\end{align}

\hrulefill{} 
\end{figure*}

where \eqref{eq:-91} follows from the fact that the number of $n$-types
$T_{X}$ is polynomial in $n$. Therefore, 
\begin{align}
 & \limsup_{n\to\infty}\frac{1}{n}D_{1+s}(Q_{Y}^{n}\|P_{Y^{n}})\nonumber \\
 & \leq\max_{\widetilde{P}_{Y}\in\mathcal{P}\left(\mathcal{Y}\right)}\min_{\widetilde{P}_{X}\in\mathcal{P}\left(\mathcal{X}\right):H(\widetilde{P}_{X})\ge H(\widetilde{P}_{Y})+\delta}\nonumber \\
 & \qquad\left\{ D(\widetilde{P}_{X}\|P_{X})-\frac{1+s}{s}D(\widetilde{P}_{Y}\|Q_{Y})\right\} .\label{eq:-51}
\end{align}
Since $\delta>0$ is arbitrary, 
\begin{align}
 & \limsup_{n\to\infty}\inf_{f}\frac{1}{n}D_{1+s}(Q_{Y}^{n}\|P_{Y^{n}})\nonumber \\
 & \leq\max_{\widetilde{P}_{Y}\in\mathcal{P}\left(\mathcal{Y}\right)}\min_{\widetilde{P}_{X}\in\mathcal{P}\left(\mathcal{X}\right):H(\widetilde{P}_{X})\ge H(\widetilde{P}_{Y})}D(\widetilde{P}_{X}\|P_{X})\nonumber \\
 & \qquad-\frac{1+s}{s}D(\widetilde{P}_{Y}\|Q_{Y})\label{eq:-27}\\
 & =\max_{\widetilde{P}_{Y}\in\mathcal{P}\left(\mathcal{Y}\right)}\max_{t\in[0,\infty]}\min_{\widetilde{P}_{X}\in\mathcal{P}\left(\mathcal{X}\right)}D(\widetilde{P}_{X}\|P_{X})\nonumber \\
 & \qquad-\frac{1+s}{s}D(\widetilde{P}_{Y}\|Q_{Y})+t\left(H(\widetilde{P}_{Y})-H(\widetilde{P}_{X})\right)\\
 & =\max_{\widetilde{P}_{Y}\in\mathcal{P}\left(\mathcal{Y}\right)}\max_{t\in[0,\infty]}-\frac{1+s}{s}D(\widetilde{P}_{Y}\|Q_{Y})\nonumber \\
 & \qquad+tH(\widetilde{P}_{Y})-tH_{\frac{1}{1+t}}(P_{X})\\
 & =\max_{t\in[0,\infty]}\max_{\widetilde{P}_{Y}\in\mathcal{P}\left(\mathcal{Y}\right)}-\frac{1+s}{s}D(\widetilde{P}_{Y}\|Q_{Y})\nonumber \\
 & \qquad+tH(\widetilde{P}_{Y})-tH_{\frac{1}{1+t}}(P_{X})\\
 & =\max_{t\in[0,\infty]}tH_{\frac{1+s}{st+1+s}}(Q_{Y})-tH_{\frac{1}{1+t}}(P_{X}).\label{eq:-28}
\end{align}

\emph{Converse: } For $s>0$, we have \eqref{eq:-109}-\eqref{eq:-110}
(given on page \pageref{eq:-109}). 
\begin{figure*}
\begin{align}
 & \frac{1}{n}D_{1+s}(Q_{Y}^{n}\|P_{Y^{n}})\nonumber \\
 & =\frac{1}{ns}\log\sum_{y^{n}}Q_{Y}^{n}(y^{n})^{1+s}P_{Y^{n}}(y^{n})^{-s}\label{eq:-109}\\
 & =\frac{1}{ns}\log\sum_{T_{Y}}\sum_{y^{n}\in\mathcal{T}_{T_{Y}}}\left(\sum_{T_{X}}\sum_{x^{n}\in\mathcal{T}_{T_{X}}}P_{X}^{n}(x^{n})1\left\{ y^{n}=f(x^{n})\right\} \right)^{-s}e^{\left(1+s\right)n\sum_{y}T_{Y}(y)\log Q_{Y}(y)}\\
 & \geq\frac{1}{ns}\log\sum_{T_{Y}}\sum_{y^{n}\in\mathcal{T}_{T_{Y}}\backslash\bigcup_{T_{X}:H(T_{X})<H(T_{Y})-\delta}f\left(\mathcal{T}_{T_{X}}\right)}\left(\sum_{T_{X}}\sum_{x^{n}\in\mathcal{T}_{T_{X}}}P_{X}^{n}(x^{n})1\left\{ y^{n}=f(x^{n})\right\} \right)^{-s}\nonumber \\
 & \qquad\times e^{\left(1+s\right)n\sum_{y}T_{Y}(y)\log Q_{Y}(y)}\\
 & =\frac{1}{ns}\log\sum_{T_{Y}}\sum_{y^{n}\in\mathcal{T}_{T_{Y}}\backslash\bigcup_{T_{X}:H(T_{X})<H(T_{Y})-\delta}f\left(\mathcal{T}_{T_{X}}\right)}\left(\sum_{T_{X}:H(T_{X})\ge H(T_{Y})-\delta}\sum_{x^{n}\in\mathcal{T}_{T_{X}}}P_{X}^{n}(x^{n})1\left\{ y^{n}=f(x^{n})\right\} \right)^{-s}\nonumber \\
 & \qquad\times e^{\left(1+s\right)n\sum_{y}T_{Y}(y)\log Q_{Y}(y)}\label{eq:-110}
\end{align}

\hrulefill{} 
\end{figure*}

Observe that 
\begin{align}
A & :=\sum_{y^{n}\in\mathcal{T}_{T_{Y}}\backslash\bigcup_{T_{X}:H(T_{X})<H(T_{Y})-\delta}f\left(\mathcal{T}_{T_{X}}\right)}\sum_{T_{X}:H(T_{X})\ge H(T_{Y})-\delta}\nonumber \\
 & \qquad\sum_{x^{n}\in\mathcal{T}_{T_{X}}}P_{X}^{n}(x^{n})1\left\{ y^{n}=f(x^{n})\right\} \label{eq:-72}\\
 & \leq\sum_{y^{n}}\sum_{T_{X}:H(T_{X})\ge H(T_{Y})-\delta}\sum_{x^{n}\in\mathcal{T}_{T_{X}}}P_{X}^{n}(x^{n})1\left\{ y^{n}=f(x^{n})\right\} \\
 & =\sum_{T_{X}:H(T_{X})\ge H(T_{Y})-\delta}\sum_{x^{n}\in\mathcal{T}_{T_{X}}}P_{X}^{n}(x^{n})\\
 & \doteq\sum_{T_{X}:H(T_{X})\ge H(T_{Y})-\delta}e^{-nD(T_{X}\|P_{X})}\label{eq:-70}
\end{align}
and 
\begin{align}
N & :=\left|\mathcal{T}_{T_{Y}}\backslash\bigcup_{T_{X}:H(T_{X})<H(T_{Y})-\delta}f\left(\mathcal{T}_{T_{X}}\right)\right|\label{eq:-73}\\
 & \geq e^{nH(T_{Y})}-\sum_{T_{X}:H(T_{X})<H(T_{Y})-\delta}e^{nH(T_{X})}\\
 & \doteq e^{nH(T_{Y})}-\max_{T_{X}:H(T_{X})<H(T_{Y})-\delta}e^{nH(T_{X})}\\
 & \doteq e^{nH(T_{Y})}-e^{n\left(H(T_{Y})-\delta\right)}\\
 & \doteq e^{nH(T_{Y})}.\label{eq:-71}
\end{align}
Hence by Lemma \ref{lem:convexity_concavity} with the identifications
$\beta=-s$, $m=A$, $n=N$, and $b_{i}=\sum_{T_{X}:H(T_{X})\ge H(T_{Y})-\delta}\sum_{x^{n}\in\mathcal{T}_{T_{X}}}P_{X}^{n}(x^{n})1\left\{ y^{n}=f(x^{n})\right\} $,
we have \eqref{eq:-111}-\eqref{eq:-112} (given on page \pageref{eq:-111}).
\begin{figure*}
\begin{align}
 & \frac{1}{n}D_{1+s}(Q_{Y}^{n}\|P_{Y^{n}})\nonumber \\
 & \geq\frac{1}{ns}\log\sum_{T_{Y}}N\left(\frac{A}{N}\right)^{-s}e^{\left(1+s\right)n\sum_{y}T_{Y}(y)\log Q_{Y}(y)}\label{eq:-111}\\
 & \geq\frac{1}{ns}\log\sum_{T_{Y}}e^{nH(T_{Y})}\left(\sum_{T_{X}:H(T_{X})\ge H(T_{Y})-\delta}e^{-nD(T_{X}\|P_{X})-nH(T_{Y})}\right)^{-s}e^{\left(1+s\right)n\sum_{y}T_{Y}(y)\log Q_{Y}(y)}+o(1)\\
 & =\frac{1}{ns}\log\sum_{T_{Y}}e^{nH(T_{Y})}\left(\max_{T_{X}:H(T_{X})\ge H(T_{Y})-\delta}e^{-nD(T_{X}\|P_{X})-nH(T_{Y})}\right)^{-s}e^{\left(1+s\right)n\sum_{y}T_{Y}(y)\log Q_{Y}(y)}+o(1)\\
 & =\frac{1}{ns}\log\max_{T_{Y}}\min_{T_{X}:H(T_{X})\ge H(T_{Y})-\delta}e^{snD(T_{X}\|P_{X})+\left(1+s\right)nH(T_{Y})}e^{\left(1+s\right)n\sum_{y}T_{Y}(y)\log Q_{Y}(y)}+o(1)\\
 & =\max_{T_{Y}}\min_{T_{X}:H(T_{X})\ge H(T_{Y})-\delta}\left\{ D(T_{X}\|P_{X})-\frac{1+s}{s}D(T_{Y}\|Q_{Y})\right\} +o(1)\\
 & =\max_{\widetilde{P}_{Y}\in\mathcal{P}\left(\mathcal{Y}\right)}\min_{\widetilde{P}_{X}\in\mathcal{P}\left(\mathcal{X}\right):H(\widetilde{P}_{X})\ge H(\widetilde{P}_{Y})-\delta}\left\{ D(\widetilde{P}_{X}\|P_{X})-\frac{1+s}{s}D(\widetilde{P}_{Y}\|Q_{Y})\right\} +o(1)\label{eq:-112}
\end{align}

\hrulefill{} 
\end{figure*}

Since $\delta>0$ is arbitrary, letting $\delta\to0$ we have 
\begin{align}
 & \liminf_{n\to\infty}\inf_{f}\frac{1}{n}D_{1+s}(Q_{Y}^{n}\|P_{Y^{n}})\nonumber \\
 & \geq\max_{\widetilde{P}_{Y}\in\mathcal{P}\left(\mathcal{Y}\right)}\min_{\widetilde{P}_{X}\in\mathcal{P}\left(\mathcal{X}\right):H(\widetilde{P}_{X})\ge H(\widetilde{P}_{Y})}\nonumber \\
 & \qquad\left\{ D(\widetilde{P}_{X}\|P_{X})-\frac{1+s}{s}D(\widetilde{P}_{Y}\|Q_{Y})\right\} \\
 & =\max_{t\in[0,\infty]}tH_{\frac{1+s}{st+1+s}}(Q_{Y})-tH_{\frac{1}{1+t}}(P_{X}),\label{eq:-26}
\end{align}
where \eqref{eq:-26} follows from the derivation \eqref{eq:-27}-\eqref{eq:-28}.

\section{\label{sec:Proof-of-Theorem-Renyimax}Proof of Theorem \ref{thm:Renyimax} }

In the following, we only prove the case of $R=1$. In addition, we
only prove the case $\alpha=1+s>1$. Other cases can be proven by
similar proof techniques.

\emph{Achievability: }Given two type-to-type functions $g_{1}:\mathcal{P}^{\left(n\right)}\left(\mathcal{X}\right)\to\mathcal{P}^{\left(n\right)}\left(\mathcal{Y}\right),g_{2}:\mathcal{P}^{\left(n\right)}\left(\mathcal{Y}\right)\to\mathcal{P}^{\left(n\right)}\left(\mathcal{X}\right)$,
we consider a mapping $g$ that maps a set $\left\{ T_{X}\right\} $
of $n$-types on $\mathcal{X}$ to the set $g_{1}(\left\{ T_{X}\right\} )\cup g_{2}^{-1}(\left\{ T_{X}\right\} )$
of $n$-types on $\mathcal{Y}$, i.e., $g\left(\left\{ T_{X}\right\} \right)=g_{1}(\left\{ T_{X}\right\} )\cup g_{2}^{-1}(\left\{ T_{X}\right\} )$.
We design $g_{2}$ such that it satisfies $H(g_{2}(T_{Y}))\ge H(T_{Y})+\delta,\forall T_{Y}$.

For each $T_{X}$, denote $a_{T_{X}}=\left|g(\left\{ T_{X}\right\} )\right|$.
Partition $\mathcal{T}_{T_{X}}$ into $a_{T_{X}}$ subsets with size
$\left\lfloor \frac{\left|\mathcal{T}_{T_{X}}\right|}{a_{T_{X}}}\right\rfloor $
or $\left\lceil \frac{\left|\mathcal{T}_{T_{X}}\right|}{a_{T_{X}}}\right\rceil $,
and for each $T_{Y}\in g(\left\{ T_{X}\right\} )$, map the sequences
in each subset to the sequences in the set $\mathcal{T}_{T_{Y}}$
as uniformly as possible: $\left\lfloor \nicefrac{\left\lfloor \frac{\left|\mathcal{T}_{T_{X}}\right|}{a_{T_{X}}}\right\rfloor }{\left|\mathcal{T}_{T_{Y}}\right|}\right\rfloor $
or $\left\lceil \nicefrac{\left\lfloor \frac{\left|\mathcal{T}_{T_{X}}\right|}{a_{T_{X}}}\right\rfloor }{\left|\mathcal{T}_{T_{Y}}\right|}\right\rceil $
(for subsets with size $\left\lfloor \frac{\left|\mathcal{T}_{T_{X}}\right|}{a_{T_{X}}}\right\rfloor $)
or $\left\lfloor \nicefrac{\left\lceil \frac{\left|\mathcal{T}_{T_{X}}\right|}{a_{T_{X}}}\right\rceil }{\left|\mathcal{T}_{T_{Y}}\right|}\right\rfloor $
or $\left\lceil \nicefrac{\left\lceil \frac{\left|\mathcal{T}_{T_{X}}\right|}{a_{T_{X}}}\right\rceil }{\left|\mathcal{T}_{T_{Y}}\right|}\right\rceil $
(for subsets with size $\left\lceil \frac{\left|\mathcal{T}_{T_{X}}\right|}{a_{T_{X}}}\right\rceil $)
sequences in $\mathcal{T}_{T_{X}}$ are mapped to each sequence in
$\mathcal{T}_{T_{Y}}$.

For this code, and for $\alpha=1+s>1$, analogous to \eqref{eq:-50},
we can prove that 
\begin{align}
 & \frac{1}{n}D_{1+s}(P_{Y^{n}}\|Q_{Y}^{n})\nonumber \\
 & \leq\max_{T_{Y}}\max_{T_{X}\in g^{-1}(\left\{ T_{Y}\right\} )}\Bigl\{-\frac{1+s}{s}D(T_{X}\|P_{X})+D(T_{Y}\|Q_{Y})\nonumber \\
 & \:+\left(H(T_{Y})-H(T_{X})\right)1\left\{ H(T_{X})<H(T_{Y})\right\} \Bigr\}+o(1)\\
 & =\max_{T_{Y}}\max\biggl\{\max_{T_{X}\in g_{1}^{-1}(\left\{ T_{Y}\right\} )}\Bigl\{-\frac{1+s}{s}D(T_{X}\|P_{X})\nonumber \\
 & \:+D(T_{Y}\|Q_{Y})+\left(H(T_{Y})-H(T_{X})\right)1\left\{ H(T_{X})<H(T_{Y})\right\} \Bigr\},\nonumber \\
 & \:-\frac{1+s}{s}D(g_{2}(T_{Y})\|P_{X})+D(T_{Y}\|Q_{Y})\biggr\}+o(1)\label{eq:-93}
\end{align}
and analogous to \eqref{eq:-52}, we can prove that 
\begin{align}
 & \frac{1}{n}D_{1+s}(Q_{Y}^{n}\|P_{Y^{n}})\nonumber \\
 & \leq\max_{T_{Y}}\min_{T_{X}\in g^{-1}(\left\{ T_{Y}\right\} ):H(T_{X})\ge H(T_{Y})+\delta}\nonumber \\
 & \qquad\left\{ D(T_{X}\|P_{X})-\frac{1+s}{s}D(T_{Y}\|Q_{Y})\right\} +o(1)\\
 & \leq\max_{T_{Y}}D(g_{2}(T_{Y})\|P_{X})-\frac{1+s}{s}D(T_{Y}\|Q_{Y})+o(1).\label{eq:-97}
\end{align}
Therefore, 
\begin{equation}
\frac{1}{n}D_{\alpha}^{\mathsf{max}}(P_{Y^{n}},Q_{Y}^{n})\leq\max\left\{ \eqref{eq:-93},\eqref{eq:-97}\right\} .\label{eq:-98}
\end{equation}

Choose the function $g_{1}(T_{X})$ as the function $g(T_{X})$ given
in Appendix \ref{sec:Proof-of-Theorem-RenyiPQ}. Then as shown in
Appendix \ref{sec:Proof-of-Theorem-RenyiPQ}, we have 
\begin{align}
 & \max_{T_{Y}}\max_{T_{X}\in g_{1}^{-1}(T_{Y})}-\frac{1+s}{s}D(T_{X}\|P_{X})+D(T_{Y}\|Q_{Y})\nonumber \\
 & \qquad+\left(H(T_{Y})-H(T_{X})\right)1\left\{ H(T_{X})<H(T_{Y})\right\} \nonumber \\
 & \leq\max_{t\in[0,1]}\left\{ tH_{\frac{1}{1-t}}(Q_{Y})-tH_{\frac{1+s}{1+s-st}}(P_{X})\right\} +o(1).
\end{align}
For each $T_{Y}$, choose $g_{2}(T_{Y})$ as a $T_{X}$ that satisfies
$H(T_{X})\ge H(T_{Y})+\delta$ and at the same time minimizes 
\begin{align}
 & \max\biggl\{-\frac{1+s}{s}D(T_{X}\|P_{X})+D(T_{Y}\|Q_{Y}),\nonumber \\
 & \qquad D(T_{X}\|P_{X})-\frac{1+s}{s}D(T_{Y}\|Q_{Y})\biggr\}.
\end{align}
Substituting $g_{1}(T_{X})$ and $g_{2}(T_{Y})$ into \eqref{eq:-98},
we obtain \eqref{eq:-113}-\eqref{eq:-53} (given on page \pageref{eq:-113}).
\begin{figure*}
\begin{align}
 & \frac{1}{n}D_{\alpha}^{\mathsf{max}}(P_{Y^{n}},Q_{Y}^{n})\nonumber \\
 & \leq\max\biggl\{\max_{t\in[0,1]}\left\{ tH_{\frac{1}{1-t}}(Q_{Y})-tH_{\frac{1+s}{1+s-st}}(P_{X})\right\} +o(1),\nonumber \\
 & \quad\max_{T_{Y}}\min_{T_{X}:H(T_{X})\ge H(T_{Y})+\delta}\max\biggl\{-\frac{1+s}{s}D(T_{X}\|P_{X})+D(T_{Y}\|Q_{Y}),\thinspace\thinspace D(T_{X}\|P_{X})-\frac{1+s}{s}D(T_{Y}\|Q_{Y})\biggr\}\biggr\}\label{eq:-113}\\
 & =\max\biggl\{\max_{t\in[0,1]}\left\{ tH_{\frac{1}{1-t}}(Q_{Y})-tH_{\frac{1+s}{1+s-st}}(P_{X})\right\} +o(1),\nonumber \\
 & \quad\max_{\widetilde{P}_{Y}\in\mathcal{P}\left(\mathcal{Y}\right)}\min_{\widetilde{P}_{X}\in\mathcal{P}\left(\mathcal{X}\right):H(\widetilde{P}_{X})\ge H(\widetilde{P}_{Y})+\delta}\max\biggl\{-\frac{1+s}{s}D(\widetilde{P}_{X}\|P_{X})+D(\widetilde{P}_{Y}\|Q_{Y}),D(\widetilde{P}_{X}\|P_{X})-\frac{1+s}{s}D(\widetilde{P}_{Y}\|Q_{Y})\biggr\}\biggr\}+o(1)\label{eq:-53}
\end{align}

\hrulefill{} 
\end{figure*}

Define 
\begin{align}
\Gamma\left(P_{X},\widetilde{P}_{Y}\right) & :=\min_{\substack{\widetilde{P}_{X}\in\mathcal{P}\left(\mathcal{X}\right):\\
H(\widetilde{P}_{X})\ge H(\widetilde{P}_{Y})
}
}D(\widetilde{P}_{X}\|P_{X})\label{eq:-101}\\
 & =\max_{t\in[0,\infty]}t\left(H(\widetilde{P}_{Y})-H_{\frac{1}{1+t}}(P_{X})\right)\\
\widehat{\Gamma}\left(P_{X},\widetilde{P}_{Y}\right) & :=\max_{\substack{\widetilde{P}_{X}\in\mathcal{P}\left(\mathcal{X}\right):\\
H(\widetilde{P}_{X})\ge H(\widetilde{P}_{Y})
}
}D(\widetilde{P}_{X}\|P_{X})\label{eq:-102}\\
 & =-\min_{\substack{\widetilde{P}_{X}\in\mathcal{P}\left(\mathcal{X}\right):\\
H(\widetilde{P}_{X})\ge H(\widetilde{P}_{Y})
}
}\sum_{x}\widetilde{P}_{X}(x)\log P_{X}(x)\nonumber \\
 & \qquad-H(\widetilde{P}_{Y})\label{eq:-99}\\
 & =\min_{t\in[0,\infty]}\left(1+t\right)\left(H_{\frac{-1}{t}}(P_{X})-H(\widetilde{P}_{Y})\right),\label{eq:-100}
\end{align}
where \eqref{eq:-99} and \eqref{eq:-100} follow since, on one hand,
$\widehat{\Gamma}\left(P_{X},\widetilde{P}_{Y}\right)\leq\eqref{eq:-99}=\eqref{eq:-100}$
due to the constraint $H(\widetilde{P}_{X})\ge H(\widetilde{P}_{Y})$;
and on the other hand, by setting $\widetilde{P}_{X}=\nicefrac{P_{X}^{\frac{-1}{t}}\left(\cdot\right)}{\sum_{x}P_{X}^{\frac{-1}{t}}\left(x\right)}$
with $t\in[0,\infty]$ satisfying $H(\widetilde{P}_{X})=H(\widetilde{P}_{Y})$,
we have $\widehat{\Gamma}\left(P_{X},\widetilde{P}_{Y}\right)\geq\eqref{eq:-100}$.

Since $\delta>0$ is arbitrary and all the functions in \eqref{eq:-53}
are continuous, we have \eqref{eq:-114}-\eqref{eq:-115} (given on
page \pageref{eq:-114}). 
\begin{figure*}
\begin{align}
 & \limsup_{n\to\infty}\frac{1}{n}D_{\alpha}^{\mathsf{max}}(P_{Y^{n}},Q_{Y}^{n})\nonumber \\
 & \leq\max\biggl\{\max_{t\in[0,1]}t\left(H_{\frac{1}{1-t}}(Q_{Y})-H_{\frac{1+s}{1+s-st}}(P_{X})\right),\nonumber \\
 & \qquad\max_{\widetilde{P}_{Y}\in\mathcal{P}\left(\mathcal{Y}\right)}\min_{\widetilde{P}_{X}\in\mathcal{P}\left(\mathcal{X}\right):H(\widetilde{P}_{X})\ge H(\widetilde{P}_{Y})}\max\biggl\{-\frac{1+s}{s}D(\widetilde{P}_{X}\|P_{X})+D(\widetilde{P}_{Y}\|Q_{Y}),D(\widetilde{P}_{X}\|P_{X})-\frac{1+s}{s}D(\widetilde{P}_{Y}\|Q_{Y})\biggr\}\biggr\}\label{eq:-114}\\
 & =\max\biggl\{\max_{t\in[0,1]}t\left(H_{\frac{1}{1-t}}(Q_{Y})-H_{\frac{1+s}{1+s-st}}(P_{X})\right),\nonumber \\
 & \qquad\max_{\widetilde{P}_{Y}\in\mathcal{P}\left(\mathcal{Y}\right)}\min_{r:\Gamma\left(P_{X},\widetilde{P}_{Y}\right)\leq r\leq\widehat{\Gamma}\left(P_{X},\widetilde{P}_{Y}\right)}\max\left\{ D(\widetilde{P}_{Y}\|Q_{Y})-\frac{1+s}{s}r,\:r-\frac{1+s}{s}D(\widetilde{P}_{Y}\|Q_{Y})\right\} \biggr\}\\
 & =\max\biggl\{\max_{t\in[0,1]}t\left(H_{\frac{1}{1-t}}(Q_{Y})-H_{\frac{1+s}{1+s-st}}(P_{X})\right),\nonumber \\
 & \qquad\max_{\widetilde{P}_{Y}\in\mathcal{P}\left(\mathcal{Y}\right)}\biggl\{\max\biggl\{-\frac{1}{s}D(\widetilde{P}_{Y}\|Q_{Y}),\:D(\widetilde{P}_{Y}\|Q_{Y})-\frac{1+s}{s}\widehat{\Gamma}\left(P_{X},\widetilde{P}_{Y}\right),\:\Gamma\left(P_{X},\widetilde{P}_{Y}\right)-\frac{1+s}{s}D(\widetilde{P}_{Y}\|Q_{Y})\biggr\}\biggr\}\\
 & =\max\biggl\{\max_{t\in[0,1]}t\left(H_{\frac{1}{1-t}}(Q_{Y})-H_{\frac{1+s}{1+s-st}}(P_{X})\right),\nonumber \\
 & \qquad\max\left\{ 0,\:\max_{t\in[0,\infty]}\frac{1+s}{s}\left(1+t\right)\left(H_{\frac{1}{1-\frac{1+s}{s}\left(1+t\right)}}(Q_{Y})-H_{\frac{-1}{t}}(P_{X})\right),\:\max_{t\in[0,\infty]}t\left(H_{\frac{1+s}{1+s+st}}(Q_{Y})-H_{\frac{1}{1+t}}(P_{X})\right)\right\} \biggr\}\\
 & =\max\left\{ \max_{t\in[0,1]\cup[\frac{1+s}{s},\infty]}t\left(H_{\frac{1}{1-t}}(Q_{Y})-H_{\frac{1+s}{1+s-st}}(P_{X})\right),\max_{t\in[0,\infty]}t\left(H_{\frac{1+s}{1+s+st}}(Q_{Y})-H_{\frac{1}{1+t}}(P_{X})\right)\right\} \\
 & =\max\left\{ \max_{t\in[0,1]\cup[\frac{\alpha}{\alpha-1},\infty]}t\left(H_{\frac{1}{1-t}}(Q_{Y})-H_{\frac{1}{1-\frac{\alpha-1}{\alpha}t}}(P_{X})\right),\max_{t\in[0,\infty]}t\left(H_{\frac{1}{1+\frac{\alpha-1}{\alpha}t}}(Q_{Y})-tH_{\frac{1}{1+t}}(P_{X})\right)\right\} \label{eq:-115}
\end{align}

\hrulefill{} 
\end{figure*}

\emph{Converse:} By the converse part of Theorem \ref{thm:RenyiPQ},
we have 
\begin{align}
 & \liminf_{n\to\infty}\frac{1}{n}D_{\alpha}^{\mathsf{max}}(P_{Y^{n}},Q_{Y}^{n})\nonumber \\
 & \geq\max_{t\in[0,1]}\left\{ tH_{\frac{1}{1-t}}(Q_{Y})-tH_{\frac{1+s}{1+s-st}}(P_{X})\right\} 
\end{align}
Next we prove 
\begin{align}
 & \liminf_{n\to\infty}\frac{1}{n}D_{\alpha}^{\mathsf{max}}(P_{Y^{n}},Q_{Y}^{n})\nonumber \\
 & \geq\max\biggl\{\max_{t\in[\frac{\alpha}{\alpha-1},\infty]}\left\{ tH_{\frac{1}{1-t}}(Q_{Y})-tH_{\frac{1}{1-\frac{\alpha-1}{\alpha}t}}(P_{X})\right\} ,\nonumber \\
 & \qquad\max_{t\in[0,\infty]}\left\{ tRH_{\frac{1}{1+\frac{\alpha-1}{\alpha}t}}(Q_{Y})-tH_{\frac{1}{1+t}}(P_{X})\right\} \biggr\}.
\end{align}
\emph{ }

For $s>0$, we have \eqref{eq:-116}-\eqref{eq:-117} (given on page
\pageref{eq:-116}). 
\begin{figure*}
\begin{align}
 & \frac{1}{n}D_{1+s}(Q_{Y}^{n}\|P_{Y^{n}})\nonumber \\
 & =\frac{1}{ns}\log\sum_{y^{n}}Q_{Y}^{n}(y^{n})^{1+s}P_{Y^{n}}(y^{n})^{-s}\label{eq:-116}\\
 & =\frac{1}{ns}\log\sum_{T_{Y}}\sum_{y^{n}\in\mathcal{T}_{T_{Y}}}\left(\sum_{T_{X}}\sum_{x^{n}\in\mathcal{T}_{T_{X}}}P_{X}^{n}(x^{n})1\left\{ y^{n}=f(x^{n})\right\} \right)^{-s}e^{\left(1+s\right)n\sum_{y}T_{Y}(y)\log Q_{Y}(y)}\\
 & \geq\frac{1}{ns}\log\sum_{T_{Y}}\sum_{y^{n}\in\mathcal{T}_{T_{Y}}\backslash\bigcup_{T_{X}:H(T_{X})<H(T_{Y})-\delta}f\left(\mathcal{T}_{T_{X}}\right)}\left(\sum_{T_{X}}\sum_{x^{n}\in\mathcal{T}_{T_{X}}}P_{X}^{n}(x^{n})1\left\{ y^{n}=f(x^{n})\right\} \right)^{-s}e^{\left(1+s\right)n\sum_{y}T_{Y}(y)\log Q_{Y}(y)}\\
 & =\frac{1}{ns}\log\sum_{T_{Y}}\sum_{y^{n}\in\mathcal{T}_{T_{Y}}\backslash\bigcup_{T_{X}:H(T_{X})<H(T_{Y})-\delta}f\left(\mathcal{T}_{T_{X}}\right)}\left(\sum_{T_{X}:H(T_{X})\ge H(T_{Y})-\delta}\sum_{x^{n}\in\mathcal{T}_{T_{X}}}P_{X}^{n}(x^{n})1\left\{ y^{n}=f(x^{n})\right\} \right)^{-s}\nonumber \\
 & \qquad e^{\left(1+s\right)n\sum_{y}T_{Y}(y)\log Q_{Y}(y)}\label{eq:-117}
\end{align}

\hrulefill{} 
\end{figure*}

Same as \eqref{eq:-70} and \eqref{eq:-71}, we have 
\begin{align}
N & :=\left|\mathcal{T}_{T_{Y}}\backslash\bigcup_{T_{X}:H(T_{X})<H(T_{Y})-\delta}f\left(\mathcal{T}_{T_{X}}\right)\right|\\
 & \dotge e^{nH(T_{Y})},
\end{align}
and 
\begin{align}
A & :=\sum_{y^{n}\in\mathcal{T}_{T_{Y}}\backslash\bigcup_{T_{X}:H(T_{X})<H(T_{Y})-\delta}f\left(\mathcal{T}_{T_{X}}\right)}\sum_{T_{X}:H(T_{X})\ge H(T_{Y})-\delta}\nonumber \\
 & \qquad\sum_{x^{n}\in\mathcal{T}_{T_{X}}}P_{X}^{n}(x^{n})1\left\{ y^{n}=f(x^{n})\right\} \\
 & \dotle\max_{T_{X}:H(T_{X})\ge H(T_{Y})-\delta}e^{-nD(T_{X}\|P_{X})}.
\end{align}
Furthermore, $A$ can be lower bounded as follows. 
\begin{align}
A & \geq N\min_{T_{X}:H(T_{X})\ge H(T_{Y})-\delta}e^{n\sum_{x}T_{X}(x)\log P_{X}(x)}\\
 & \doteq e^{nH(T_{Y})}\min_{T_{X}:H(T_{X})\ge H(T_{Y})-\delta}e^{n\sum_{x}T_{X}(x)\log P_{X}(x)}.
\end{align}
Define $r:=-\frac{1}{n}\log A$. Then 
\begin{align}
 & \min_{T_{X}:H(T_{X})\ge H(T_{Y})-\delta}D(T_{X}\|P_{X})\nonumber \\
 & \leq r\\
 & \leq-H(T_{Y})-\min_{T_{X}:H(T_{X})\ge H(T_{Y})-\delta}\sum_{x}T_{X}(x)\log P_{X}(x).
\end{align}

Hence by Lemma \ref{lem:convexity_concavity}, we have 
\begin{align}
 & \frac{1}{n}D_{1+s}(Q_{Y}^{n}\|P_{Y^{n}})\nonumber \\
 & \geq\frac{1}{ns}\log\sum_{T_{Y}}N\left(\frac{A}{N}\right)^{-s}e^{\left(1+s\right)n\sum_{y}T_{Y}(y)\log Q_{Y}(y)}\\
 & =\frac{1}{ns}\log\sum_{T_{Y}}e^{\left(1+s\right)nH(T_{Y})}A^{-s}e^{\left(1+s\right)n\sum_{y}T_{Y}(y)\log Q_{Y}(y)}\nonumber \\
 & \qquad+o(1)\\
 & =\frac{1}{ns}\log\sum_{T_{Y}}A^{-s}e^{-n\left(1+s\right)D(T_{Y}\|Q_{Y})}+o(1)\\
 & =\max_{T_{Y}}\left\{ r-\frac{1+s}{s}D(T_{Y}\|Q_{Y})\right\} +o(1).\label{eq:-45}
\end{align}

On the other hand,

\begin{align}
 & \frac{1}{n}D_{1+s}(P_{Y^{n}}\|Q_{Y}^{n})\nonumber \\
 & =\frac{1}{ns}\log\sum_{T_{Y}}\sum_{y^{n}\in\mathcal{T}_{T_{Y}}}Q_{Y}^{n}(y^{n})^{-s}\nonumber \\
 & \quad\times\left(\sum_{T_{X}}\sum_{x^{n}\in\mathcal{T}_{T_{X}}}P_{X}^{n}(x^{n})1\left\{ y^{n}=f(x^{n})\right\} \right)^{1+s}\\
 & \geq\frac{1}{ns}\log\sum_{T_{Y}}\sum_{y^{n}\in\mathcal{T}_{T_{Y}}\backslash\bigcup_{T_{X}:H(T_{X})<H(T_{Y})-\delta}f\left(\mathcal{T}_{T_{X}}\right)}Q_{Y}^{n}(y^{n})^{-s}\nonumber \\
 & \quad\times\left(\sum_{T_{X}}\sum_{x^{n}\in\mathcal{T}_{T_{X}}}P_{X}^{n}(x^{n})1\left\{ y^{n}=f(x^{n})\right\} \right)^{1+s}\\
 & =\frac{1}{ns}\log\sum_{T_{Y}}\sum_{y^{n}\in\mathcal{T}_{T_{Y}}\backslash\bigcup_{T_{X}:H(T_{X})<H(T_{Y})-\delta}f\left(\mathcal{T}_{T_{X}}\right)}Q_{Y}^{n}(y^{n})^{-s}\nonumber \\
 & \quad\times\left(\sum_{\substack{T_{X}:\\
H(T_{X})\ge H(T_{Y})-\delta
}
}\sum_{x^{n}\in\mathcal{T}_{T_{X}}}P_{X}^{n}(x^{n})1\left\{ y^{n}=f(x^{n})\right\} \right)^{1+s}\\
 & \geq\frac{1}{ns}\log\sum_{T_{Y}}N\left(\frac{A}{N}\right)^{1+s}Q_{Y}^{n}(y^{n})^{-s}\\
 & =\frac{1}{ns}\log\sum_{T_{Y}}e^{-snH(T_{Y})}A^{1+s}e^{-sn\sum_{y}T_{Y}(y)\log Q_{Y}(y)}+o(1)\\
 & =\frac{1}{ns}\log\sum_{T_{Y}}A^{1+s}e^{nsD(T_{Y}\|Q_{Y})}+o(1)\\
 & =\max_{T_{Y}}\left\{ D(T_{Y}\|Q_{Y})-\frac{1+s}{s}r\right\} +o(1).\label{eq:-46}
\end{align}

Define 
\begin{align}
\Gamma_{\delta}^{(n)}\left(P_{X},T_{Y}\right) & :=\min_{\substack{T_{X}\in\mathcal{P}^{\left(n\right)}\left(\mathcal{X}\right):\\
H(T_{X})\ge H(T_{Y})-\delta
}
}D(T_{X}\|P_{X})\\
\widehat{\Gamma}_{\delta}^{(n)}\left(P_{X},T_{Y}\right) & :=-\min_{\substack{T_{X}\in\mathcal{P}^{\left(n\right)}\left(\mathcal{X}\right):\\
H(T_{X})\ge H(T_{Y})-\delta
}
}\sum_{x}T_{X}(x)\log P_{X}(x)\nonumber \\
 & \qquad-H(T_{Y})
\end{align}
Combining \eqref{eq:-45} and \eqref{eq:-46}, we have 
\begin{align}
 & \frac{1}{n}D_{\alpha}^{\mathsf{max}}(P_{Y^{n}},Q_{Y}^{n})\nonumber \\
 & \geq\max_{T_{Y}}\biggl\{\max\Bigl\{ D(T_{Y}\|Q_{Y})-\frac{1+s}{s}r,\nonumber \\
 & \qquad r-\frac{1+s}{s}D(T_{Y}\|Q_{Y})\Bigr\}\biggr\}+o(1)\\
 & \geq\max_{T_{Y}}\biggl\{\min_{r:\Gamma_{\delta}^{(n)}\left(P_{X},T_{Y}\right)\leq r\leq\widehat{\Gamma}_{\delta}^{(n)}\left(P_{X},T_{Y}\right)}\max\Bigl\{ D(T_{Y}\|Q_{Y})\nonumber \\
 & \qquad-\frac{1+s}{s}r,\:r-\frac{1+s}{s}D(T_{Y}\|Q_{Y})\Bigr\}\biggr\}+o(1).\label{eq:-47}
\end{align}
Since $\delta>0$ is arbitrary and all the functions involved in \eqref{eq:-47}
are continuous, letting $n\to\infty$ and $\delta\to0$, we have \eqref{eq:-103}-\eqref{eq:-104}
(given on page \pageref{eq:-103}), where $\Gamma\left(P_{X},\widetilde{P}_{Y}\right)$
and $\widehat{\Gamma}\left(P_{X},\widetilde{P}_{Y}\right)$ are respectively
defined in \eqref{eq:-101} and \eqref{eq:-102} (recall the equation
\eqref{eq:-99}).

\begin{figure*}
\begin{align}
 & \liminf_{n\to\infty}\frac{1}{n}D_{\alpha}^{\mathsf{max}}(P_{Y^{n}},Q_{Y}^{n})\nonumber \\
 & \geq\max_{\widetilde{P}_{Y}\in\mathcal{P}\left(\mathcal{Y}\right)}\biggl\{\min_{r:\Gamma\left(P_{X},\widetilde{P}_{Y}\right)\leq r\leq\widehat{\Gamma}\left(P_{X},\widetilde{P}_{Y}\right)}\max\left\{ D(\widetilde{P}_{Y}\|Q_{Y})-\frac{1+s}{s}r,\:r-\frac{1+s}{s}D(\widetilde{P}_{Y}\|Q_{Y})\right\} \biggr\}\label{eq:-103}\\
 & =\max_{\widetilde{P}_{Y}\in\mathcal{P}\left(\mathcal{Y}\right)}\biggl\{\max\biggl\{-\frac{1}{s}D(\widetilde{P}_{Y}\|Q_{Y}),\:D(\widetilde{P}_{Y}\|Q_{Y})-\frac{1+s}{s}\widehat{\Gamma}\left(P_{X},\widetilde{P}_{Y}\right),\:\Gamma\left(P_{X},\widetilde{P}_{Y}\right)-\frac{1+s}{s}D(\widetilde{P}_{Y}\|Q_{Y})\biggr\}\biggr\}\\
 & =\max\left\{ 0,\:\max_{t\in[0,\infty]}\frac{1+s}{s}\left(1+t\right)\left(H_{\frac{1}{1-\frac{1+s}{s}\left(1+t\right)}}(Q_{Y})-H_{\frac{-1}{t}}(P_{X})\right),\:\max_{t\in[0,\infty]}t\left(H_{\frac{\frac{1+s}{s}}{\frac{1+s}{s}+t}}(Q_{Y})-H_{\frac{1}{1+t}}(P_{X})\right)\right\} \\
 & =\max\left\{ \max_{t\in[\frac{1+s}{s},\infty]}t\left(H_{\frac{1}{1-t}}(Q_{Y})-H_{\frac{1}{1-\frac{s}{1+s}t}}(P_{X})\right),\:\max_{t\in[0,\infty]}t\left(H_{\frac{1}{1+\frac{s}{1+s}t}}(Q_{Y})-H_{\frac{1}{1+t}}(P_{X})\right)\right\} \label{eq:-104}
\end{align}

\hrulefill{} 
\end{figure*}

\section{\label{sec:Proof-of-Theorem-Renyi1Con}Proof of Theorem \ref{thm:Renyi1Con} }

The equality in \eqref{eq:} follows from Theorem \ref{thm:RenyiPQ}.
For \eqref{eq:-58}, the case $\alpha=0$ can be proven easily. The
converse parts for the cases $\alpha\in(0,1]\cup\left\{ \infty\right\} $
follow from \eqref{eq:}. The achievability parts for $\alpha\in\left\{ 1,\infty\right\} $
follow from \eqref{eq:-57}. The achievability parts for $\alpha\in(0,1)$
are implied by the achievability part for $\alpha=1$, since the conversion
rates for these cases are all equal to $\frac{H(P_{X})}{H(Q_{Y})}$.
Hence here we only need to prove \eqref{eq:-57}.

\begin{figure}[t]
\centering\includegraphics[width=1\columnwidth]{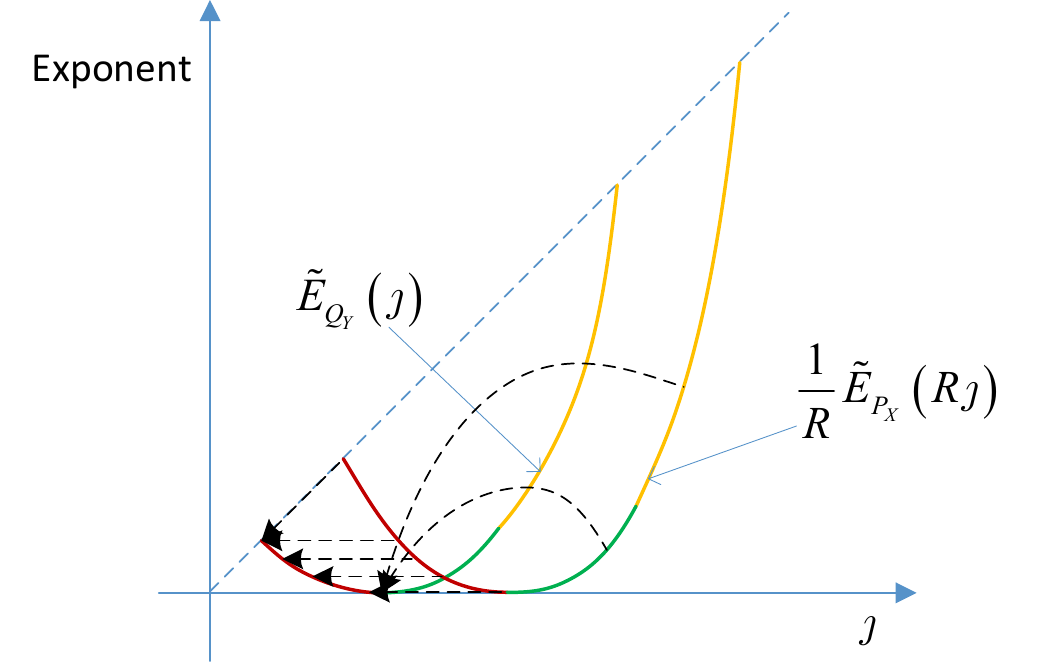}

\caption{Illustration of the code used to prove the achievability for $\alpha\in[1,\infty]$
in Theorem \ref{thm:Renyi1Con} by using information spectrum exponents.}
\label{fig:DPQ} 
\end{figure}

Define $\mathcal{A}:=\left\{ y^{n}:Q_{Y}^{n}(y^{n})\geq e^{-n\left(H(Q_{Y})+\delta\right)}\right\} $
for $\delta>0$. Define $\widetilde{Q}_{Y^{n}}(y^{n}):=\frac{Q_{Y}^{n}(y^{n})}{Q_{Y}^{n}(\mathcal{A})}1\left\{ y^{n}\in\mathcal{A}\right\} $.
Use Mapping 1 given in Appendix \ref{subsec:Mappings} to map the
sequences in $\mathcal{X}^{k}$ to the sequences in $\mathcal{A}$,
where the distributions $P_{X}$ and $Q_{Y}$ are respectively replaced
by $P_{X}^{k}$ and $\widetilde{Q}_{Y^{n}}$. That is, for each $i\in[1:|\mathcal{X}|^{k}]$,
$x_{i}^{k}$ is mapped to $y_{j}^{n}$ where $j=G_{Y^{n}}^{-1}(G_{X^{k}}(i))$.
This code is illustrated in Fig. \ref{fig:DPQ}. Hence the following
properties hold: 
\begin{enumerate}
\item If $P_{X}^{k}(x_{i}^{k})\geq\widetilde{Q}_{Y^{n}}(y_{j}^{n})$ where
$i:=G_{X^{k}}^{-1}(G_{Y^{n}}(j))$, then $|\left\{ i:G_{Y^{n}}^{-1}(G_{X^{k}}(i))=j\right\} |\leq1$.
Hence $P_{Y^{n}}(y_{j}^{n})\le P_{X}^{k}(x_{i}^{k})$. 
\item If $P_{X}^{k}(x_{i}^{k})<\widetilde{Q}_{Y^{n}}(y_{j}^{n})$ where
$i:=G_{X^{k}}^{-1}(G_{Y^{n}}(j))$, then $|\left\{ i:G_{Y^{n}}^{-1}(G_{X^{k}}(i))=j\right\} |\geq1$
and 
\begin{equation}
\frac{1}{2}\widetilde{Q}_{Y^{n}}(y_{j}^{n})\leq P_{Y^{n}}(y_{j}^{n})\le\widetilde{Q}_{Y^{n}}(y_{j}^{n})+P_{X}^{k}(x_{i}^{k}).
\end{equation}
\item $P_{Y^{n}}(y^{n})=0$ for $y^{n}\notin\mathcal{A}$. 
\end{enumerate}
For brevity, we denote $i\left(y^{n}\right):=G_{X^{k}}^{-1}(G_{Y^{n}}(j))$
where $j$ is the index of $y^{n}$, and denote $j\left(x^{k}\right):=G_{Y^{n}}^{-1}(G_{X^{k}}(i))$
where $i$ is the index of $x^{k}$.

For this code, and for $0\le s\le1$, we have \eqref{eq:-118}-\eqref{eq:-42}
(given on page \pageref{eq:-118}), 
\begin{figure*}
\begin{align}
D_{1+s}(P_{Y^{n}}||Q_{Y}^{n}) & =\frac{1}{s}\log\sum_{y^{n}}P_{Y^{n}}(y^{n})^{1+s}Q_{Y}^{n}(y^{n})^{-s}\label{eq:-118}\\
 & \leq\frac{1}{s}\log\sum_{y^{n}}P_{Y^{n}}(y^{n})\Biggl[P_{X}^{k}(x_{i\left(y^{n}\right)}^{k})1\{P_{X}^{k}(x_{i\left(y^{n}\right)}^{k})\geq\widetilde{Q}_{Y^{n}}(y^{n})\}\nonumber \\
 & \qquad+\left(\widetilde{Q}_{Y^{n}}(y^{n})+P_{X}^{k}(x_{i\left(y^{n}\right)}^{k})\right)1\{P_{X}^{k}(x_{i\left(y^{n}\right)}^{k})<\widetilde{Q}_{Y^{n}}(y^{n})\}\Biggr]^{s}Q_{Y}^{n}(y^{n})^{-s}\\
 & =\frac{1}{s}\log\sum_{y^{n}}P_{Y^{n}}(y^{n})\Biggl[\left(\frac{P_{X}^{k}(x_{i\left(y^{n}\right)}^{k})}{Q_{Y^{n}}(y^{n})}\right)^{s}1\{P_{X}^{k}(x_{i\left(y^{n}\right)}^{k})\geq\widetilde{Q}_{Y^{n}}(y^{n})\}\nonumber \\
 & \qquad+\left(\frac{\widetilde{Q}_{Y^{n}}(y^{n})}{Q_{Y^{n}}(y^{n})}\right)^{s}\left(1+\frac{P_{X}^{k}(x_{i\left(y^{n}\right)}^{k})}{\widetilde{Q}_{Y^{n}}(y^{n})}\right)^{s}1\{P_{X}^{k}(x_{i\left(y^{n}\right)}^{k})<\widetilde{Q}_{Y^{n}}(y^{n})\}\Biggr]\\
 & \leq\frac{1}{s}\log\sum_{y^{n}}P_{Y^{n}}(y^{n})\Biggl[\left(\frac{P_{X}^{k}(x_{i\left(y^{n}\right)}^{k})}{Q_{Y^{n}}(y^{n})}\right)^{s}1\{P_{X}^{k}(x_{i\left(y^{n}\right)}^{k})\geq\widetilde{Q}_{Y^{n}}(y^{n})\}\nonumber \\
 & \qquad+\left(\frac{\widetilde{Q}_{Y^{n}}(y^{n})}{Q_{Y^{n}}(y^{n})}\right)^{s}\left(1+\left(\frac{P_{X}^{k}(x_{i\left(y^{n}\right)}^{k})}{\widetilde{Q}_{Y^{n}}(y^{n})}\right)^{s}\right)1\{P_{X}^{k}(x_{i\left(y^{n}\right)}^{k})<\widetilde{Q}_{Y^{n}}(y^{n})\}\Biggr]\label{eq:-41}\\
 & =\frac{1}{s}\log Q_{Y}^{n}(\mathcal{A})^{-s}\sum_{y^{n}}P_{Y^{n}}(y^{n})\left(\left(\frac{P_{X}^{k}(x_{i\left(y^{n}\right)}^{k})}{\widetilde{Q}_{Y^{n}}(y^{n})}\right)^{s}+1\{P_{X}^{k}(x_{i\left(y^{n}\right)}^{k})<\widetilde{Q}_{Y^{n}}(y^{n})\}\right)\\
 & \leq\frac{1}{s}\log Q_{Y}^{n}(\mathcal{A})^{-s}\sum_{y^{n}}P_{Y^{n}}(y^{n})\left(\left(\frac{P_{X}^{k}(x_{i\left(y^{n}\right)}^{k})}{\widetilde{Q}_{Y^{n}}(y^{n})}\right)^{s}+1\right)\\
 & \leq-\log Q_{Y}^{n}(\mathcal{A})+\frac{1}{s}Q_{Y}^{n}(\mathcal{A})^{-s}\sum_{y^{n}}P_{Y^{n}}(y^{n})\left(\frac{P_{X}^{k}(x_{i\left(y^{n}\right)}^{k})}{\widetilde{Q}_{Y^{n}}(y^{n})}\right)^{s}\label{eq:-42}
\end{align}

\hrulefill{} 
\end{figure*}

where \eqref{eq:-41} follows from Lemma \ref{lem:1+x}. To show $D_{1+s}(P_{Y^{n}}||Q_{Y}^{n})\to0$,
we only need to show both terms in \eqref{eq:-42} converge to zero.
Obviously, the first term converges to zero since $Q_{Y}^{n}(\mathcal{A})\to1$.
Next we focus on the second term. We have \eqref{eq:-119}-\eqref{eq:-120}
(given on page \pageref{eq:-119}), 
\begin{figure*}
\begin{align}
 & \sum_{y^{n}}P_{Y^{n}}(y^{n})\left(\frac{P_{X}^{k}(x_{i\left(y^{n}\right)}^{k})}{\widetilde{Q}_{Y^{n}}(y^{n})}\right)^{s}\nonumber \\
 & \leq\sum_{y^{n}}\left(P_{Y^{n}}(y^{n})1\{P_{X}^{k}(x_{i\left(y^{n}\right)}^{k})\geq\widetilde{Q}_{Y^{n}}(y^{n})\}+\left(\widetilde{Q}_{Y^{n}}(y^{n})+P_{X}^{k}(x_{i\left(y^{n}\right)}^{k})\right)1\{P_{X}^{k}(x_{i\left(y^{n}\right)}^{k})<\widetilde{Q}_{Y^{n}}(y^{n})\}\right)\left(\frac{P_{X}^{k}(x_{i\left(y^{n}\right)}^{k})}{\widetilde{Q}_{Y^{n}}(y^{n})}\right)^{s}\label{eq:-119}\\
 & \leq\sum_{y^{n}}\left(P_{Y^{n}}(y^{n})1\{P_{X}^{k}(x_{i\left(y^{n}\right)}^{k})\geq\widetilde{Q}_{Y^{n}}(y^{n})\}+2\widetilde{Q}_{Y^{n}}(y^{n})1\{P_{X}^{k}(x_{i\left(y^{n}\right)}^{k})<\widetilde{Q}_{Y^{n}}(y^{n})\}\right)\left(\frac{P_{X}^{k}(x_{i\left(y^{n}\right)}^{k})}{\widetilde{Q}_{Y^{n}}(y^{n})}\right)^{s}\\
 & \leq\sum_{x^{k}}P_{X}^{k}(x^{k})\left(\frac{P_{X}^{k}(x^{k})}{\widetilde{Q}_{Y^{n}}(y_{j(x^{k})}^{n})}\right)^{s}1\{P_{X}^{k}(x^{k})\geq\widetilde{Q}_{Y^{n}}(y_{j(x^{k})}^{n})\}+2\sum_{j=1}^{|\mathcal{A}|}\sum_{x^{k}\in\mathcal{B}_{j}}\frac{P_{X}^{k}(x^{k})}{\sum_{x^{n}\in\mathcal{B}_{j}}P_{X}^{k}(x^{k})}\widetilde{Q}_{Y^{n}}(y_{j}^{n})\left(\frac{P_{X}^{k}(x^{k})}{\widetilde{Q}_{Y^{n}}(y_{j}^{n})}\right)^{s}\nonumber \\
 & \qquad\times1\{P_{X}^{k}(x^{k})<\widetilde{Q}_{Y^{n}}(y_{j}^{n})\}\label{eq:-78}\\
 & \leq\sum_{x^{k}}P_{X}^{k}(x^{k})\left(\frac{P_{X}^{k}(x^{k})}{\widetilde{Q}_{Y^{n}}(y_{j(x^{k})}^{n})}\right)^{s}1\{P_{X}^{k}(x^{k})\geq\widetilde{Q}_{Y^{n}}(y_{j(x^{k})}^{n})\}+4\sum_{x^{k}}P_{X}^{k}(x^{k})\left(\frac{P_{X}^{k}(x^{k})}{\widetilde{Q}_{Y^{n}}(y_{j(x^{k})}^{n})}\right)^{s}1\{P_{X}^{k}(x^{k})<\widetilde{Q}_{Y^{n}}(y_{j}^{n})\}\label{eq:-43}\\
 & \leq4\sum_{x^{k}}P_{X}^{k}(x^{k})\left(\frac{P_{X}^{k}(x^{k})}{\widetilde{Q}_{Y^{n}}(y_{j(x^{k})}^{n})}\right)^{s}\label{eq:-120}
\end{align}

\hrulefill{} 
\end{figure*}

where $\mathcal{B}_{j}$ denotes the set of $x^{n}$ that are mapped
to $y_{j}^{n}$, \eqref{eq:-78} follows since $P_{X}^{k}(x_{i\left(y^{n}\right)}^{k})\le P_{X}^{k}(x^{k})$
for all $x^{n}$ that are mapped to $y^{n}$, and \eqref{eq:-43}
follows since 
\begin{align}
 & \sum_{x^{k}\in\mathcal{B}_{j}}\frac{P_{X}^{k}(x^{k})}{\sum_{x^{k}\in\mathcal{B}_{j}}P_{X}^{k}(x^{k})}\widetilde{Q}_{Y^{n}}(y_{j}^{n})\left(\frac{P_{X}^{k}(x^{k})}{\widetilde{Q}_{Y^{n}}(y_{j}^{n})}\right)^{s}\nonumber \\
 & =\sum_{x^{k}\in\mathcal{B}_{j}}\frac{P_{X}^{k}(x^{k})}{P_{Y^{n}}(y_{j}^{n})}\widetilde{Q}_{Y^{n}}(y_{j}^{n})\left(\frac{P_{X}^{k}(x^{k})}{\widetilde{Q}_{Y^{n}}(y_{j}^{n})}\right)^{s}\\
 & \leq\sum_{x^{k}\in\mathcal{B}_{j}}\frac{P_{X}^{k}(x^{k})}{\frac{1}{2}\widetilde{Q}_{Y^{n}}(y_{j}^{n})}\widetilde{Q}_{Y^{n}}(y_{j}^{n})\left(\frac{P_{X}^{k}(x^{k})}{\widetilde{Q}_{Y^{n}}(y_{j}^{n})}\right)^{s}\\
 & =2\sum_{x^{k}\in\mathcal{B}_{j}}P_{X}^{k}(x^{k})\left(\frac{P_{X}^{k}(x^{k})}{\widetilde{Q}_{Y^{n}}(y_{j}^{n})}\right)^{s}.
\end{align}
Next we prove $\sum_{x^{k}}P_{X}^{k}(x^{k})\left(\frac{P_{X}^{k}(x^{k})}{\widetilde{Q}_{Y^{n}}(y_{j(x^{k})}^{n})}\right)^{s}\to0$.

Based on the notations defined in Appendix \ref{subsec:Exponents-of-Information},
and using Lemma \ref{lem:Exponents}, we have 
\begin{align}
 & Q_{Y}^{n}\left(y_{j(x^{k})}^{n}\right)\nonumber \\
 & =Q_{Y}^{n}\left(y_{G_{Y^{n}}^{-1}\left(G_{X^{k}}(i)\right)}^{n}\right)\label{eq:-54}\\
 & \geq F_{Q_{Y}^{n}}^{-1}\left(F_{P_{X}^{k}}\left(-\frac{1}{k}\log P_{X}^{k}(x^{k})\right)\right)\\
 & =\exp\biggl\{-nE_{Q_{Y}}^{-1}\Bigl(-\frac{1}{n}\log\left\{ e^{-k\left(E_{P_{X}}(-\frac{1}{k}\log P_{X}^{k}(x^{k}))+o(1)\right)}\right\} \nonumber \\
 & \qquad+o(1)\Bigr)\biggr\}\\
 & =\exp\left\{ -nE_{Q_{Y}}^{-1}\left(\frac{k}{n}\left(E_{P_{X}}(-\frac{1}{k}\log P_{X}^{k}(x^{k}))\right)+o(1)\right)\right\} \\
 & =\exp\Biggl\{-n\max_{t\in[0,\infty]}\biggl\{ H_{1+t}(Q_{Y})-\frac{1}{t}\nonumber \\
 & \times\left(\frac{k}{n}\max_{t'\in[0,\infty]}\left\{ t'H_{1+t'}(P_{X})+\frac{t'}{k}\log P_{X}^{k}(x^{k})\right\} +o(1)\right)\biggr\}\Biggr\}
\end{align}
where $i$ (in \eqref{eq:-54}) denotes the index of $x^{n}$ in the
sequence $x_{1}^{n},x_{2}^{n},...,x_{|\mathcal{X}|^{n}}^{n}$.

Therefore, we have \eqref{eq:-121}-\eqref{eq:-122} (given on page
\pageref{eq:-121}), 
\begin{figure*}
\begin{align}
 & \limsup_{n\to\infty}\frac{1}{n}\log\sum_{x^{k}}P_{X}^{k}(x^{k})\left(\frac{P_{X}^{k}(x^{k})}{Q_{Y}^{n}(y_{j(x^{n})}^{n})}\right)^{s}\nonumber \\
 & \leq\limsup_{n\to\infty}\frac{1}{n}\log\sum_{T_{X}}\sum_{x^{n}\in\mathcal{T}_{T_{X}}}e^{sn\max_{t\in[0,\infty]}\left\{ H_{1+t}(Q_{Y})-\frac{1}{t}\left(\frac{k}{n}\max_{t'\in[0,\infty]}\left\{ t'H_{1+t'}(P_{X})+\frac{t'}{k}\log P_{X}^{k}(x^{k})\right\} +o(1)\right)\right\} }\nonumber \\
 & \qquad\times e^{\left(1+s\right)k\sum_{x}T_{X}(x)\log P_{X}(x)}\label{eq:-121}\\
 & =\limsup_{n\to\infty}\max_{T_{X}}\frac{k}{n}\left(H(T_{X})+\left(1+s\right)\sum_{x}T_{X}(x)\log P_{X}(x)\right)\nonumber \\
 & \qquad+s\max_{t\in[0,\infty]}\left\{ H_{1+\frac{1}{t}}(Q_{Y})-t\left(\frac{k}{n}\max_{t'\in[0,\infty]}\left\{ t'H_{1+t'}(P_{X})+t'\sum_{x}T_{X}(x)\log P_{X}(x)\right\} +o(1)\right)\right\} \\
 & =\limsup_{n\to\infty}\max_{\widetilde{P}_{X}\in\mathcal{P}\left(\mathcal{X}\right)}\frac{k}{n}\left(H(\widetilde{P}_{X})+\left(1+s\right)\sum_{x}\widetilde{P}_{X}(x)\log P_{X}(x)\right)\nonumber \\
 & \qquad+s\max_{t\in[0,\infty]}\left\{ H_{1+\frac{1}{t}}(Q_{Y})-t\left(\frac{k}{n}\max_{t'\in[0,\infty]}\left\{ t'H_{1+t'}(P_{X})+t'\sum_{x}\widetilde{P}_{X}(x)\log P_{X}(x)\right\} +o(1)\right)\right\} \\
 & =\max_{\widetilde{P}_{X}\in\mathcal{P}\left(\mathcal{X}\right)}\frac{1}{R}\left(H(\widetilde{P}_{X})+\left(1+s\right)\sum_{x}\widetilde{P}_{X}(x)\log P_{X}(x)\right)\nonumber \\
 & \qquad+s\max_{t\in[0,\infty]}\left\{ H_{1+\frac{1}{t}}(Q_{Y})-\frac{t}{R}\max_{t'\in[0,\infty]}\left\{ t'H_{1+t'}(P_{X})+t'\sum_{x}\widetilde{P}_{X}(x)\log P_{X}(x)\right\} \right\} \label{eq:-66}\\
 & \leq\max_{t\in[0,\infty]}\min_{t'\in[0,\infty]}\max_{\widetilde{P}_{X}\in\mathcal{P}\left(\mathcal{X}\right)}\frac{1}{R}\left(H(\widetilde{P}_{X})+\left(1+s\right)\sum_{x}\widetilde{P}_{X}(x)\log P_{X}(x)\right)\\
 & \qquad+s\left\{ H_{1+\frac{1}{t}}(Q_{Y})-\frac{t}{R}\left\{ t'H_{1+t'}(P_{X})+t'\sum_{x}\widetilde{P}_{X}(x)\log P_{X}(x)\right\} \right\} \\
 & =\max_{t\in[0,\infty]}\min_{t'\in[0,\infty]}-\frac{s}{R}H_{1+s-stt'}(P_{X})+sH_{1+\frac{1}{t}}(Q_{Y})-\frac{stt'}{R}\left(H_{1+t'}(P_{X})-H_{1+s-stt'}(P_{X})\right)\\
 & \leq\max_{t\in[0,\infty]}-\frac{s}{R}H_{1+\frac{s}{1+st}}(P_{X})+sH_{1+\frac{1}{t}}(Q_{Y})\label{eq:-55}\\
 & =\max_{t''\in[0,1]}\left\{ sH_{\frac{1}{1-t''}}(Q_{Y})-\frac{s}{R}H_{\frac{t''+s}{t''+s-st''}}(P_{X})\right\} \label{eq:-122}
\end{align}

\hrulefill{} 
\end{figure*}

where \eqref{eq:-66} follows from Lemma \ref{lem:Exponents}, and
\eqref{eq:-55} follows by choosing $t'=\frac{s}{1+st}$.

Therefore, if 
\begin{equation}
R<\min_{t''\in[0,1]}\frac{H_{\frac{t''+s}{t''+s-st''}}(P_{X})}{H_{\frac{1}{1-t''}}(Q_{Y})}
\end{equation}
then 
\begin{equation}
\limsup_{n\to\infty}\frac{1}{n}\log\sum_{x^{k}}P_{X}^{k}(x^{k})\left(\frac{P_{X}^{k}(x^{k})}{Q_{Y}^{n}(y_{j(x^{k})}^{n})}\right)^{s}<0.
\end{equation}
Hence $\sum_{x^{k}}P_{X}^{k}(x^{k})\left(\frac{P_{X}^{k}(x^{k})}{Q_{Y}^{n}(y_{j(x^{k})}^{n})}\right)^{s}\to0$.
This completes the proof for $0\le s\le1$. For other $s$, it can
be proven similarly (by other inequalities in Lemma \ref{lem:1+x}).

\section{\label{sec:Proof-of-Theorem-Renyi2Con}Proof of Theorem \ref{thm:Renyi2Con} }

The equality in \eqref{eq:-1} follows from Theorem \ref{thm:RenyiQP}.
For \eqref{eq:-59}, the case $\alpha=0$ can be proven easily. The
cases $\alpha\in(0,1]\cup\left\{ \infty\right\} $ follow by showing
the achievability parts for $\alpha=1$ and $\alpha=\infty$. Next
we prove these.

Here we assume that both $P_{X}$ and $Q_{Y}$ are not uniform. The
cases that $P_{X}$ is uniform or $Q_{Y}$ is uniform will be proven
in Theorems \ref{thm:Renyi2rate} and \ref{thm:RenyiIR2rate}, respectively.

\emph{Achievability part for $\alpha=1$:} Define 
\begin{align}
\mathcal{A} & :=\left\{ x^{k}:e^{-k\left(H(P_{X})+\delta\right)}\leq P_{X}^{k}(x^{k})\leq e^{-k\left(H(P_{X})-\delta\right)}\right\} \\
\mathcal{B} & :=\left\{ y^{n}:e^{-n\left(H(Q_{Y})+\delta\right)}\leq Q_{Y}^{n}(y^{n})\leq e^{-n\left(H(Q_{Y})-\delta\right)}\right\} .
\end{align}
Here $\delta>0$ is a number such that $H(P_{X})+\delta<H_{0}(P_{X})$
and $\frac{1}{R}\left(H(P_{X})-\delta\right)>H(Q_{Y})+\delta$. We
consider the following mapping. 
\begin{enumerate}
\item Map the sequences in $\mathcal{A}^{c}$ to the sequences in $\mathcal{B}^{c}$
such that for each $y^{n}\in\mathcal{B}^{c}$, there exists at least
one $x^{n}\in\mathcal{A}^{c}$ mapped to it. This is feasible since
\begin{align}
 & \liminf_{n\to\infty}\frac{1}{n}\log\left|\mathcal{A}^{c}\right|\nonumber \\
 & =\liminf_{n\to\infty}\frac{1}{n}\log\left(\left|\mathcal{X}\right|^{k}-\left|\mathcal{A}\right|\right)\\
 & \geq\liminf_{n\to\infty}\frac{1}{n}\log\left(e^{kH_{0}(P_{X})}-e^{k\left(H(P_{X})+\delta\right)}\right)\\
 & =\frac{H_{0}(P_{X})}{R}\\
 & >H_{0}(Q_{Y})\\
 & \geq\limsup_{n\to\infty}\frac{1}{n}\log\left|\mathcal{B}^{c}\right|,
\end{align}
i.e., $\left|\mathcal{A}^{c}\right|>\left|\mathcal{B}^{c}\right|$
for sufficiently large $n$. 
\item Use Mapping 1 given in Appendix \ref{subsec:Mappings} to map the
sequences in $\mathcal{A}$ to the sequences in $\mathcal{B}$, where
the distributions $P_{X}$ and $Q_{Y}$ are respectively replaced
by $\frac{P_{X}^{k}(x^{k})1\left\{ x^{k}\in\mathcal{A}\right\} }{P_{X}^{k}(\mathcal{A})}$
and $\frac{Q_{Y}^{n}(y^{n})1\left\{ y^{n}\in\mathcal{B}\right\} }{Q_{Y}^{n}(\mathcal{B})}$.
Observe that $\frac{1}{R}\left(H(P_{X})-\delta\right)>H(Q_{Y})+\delta$
implies that $\frac{P_{X}^{k}(x^{k})}{P_{X}^{k}(\mathcal{A})}\le\frac{Q_{Y}^{n}(y^{n})}{Q_{Y}^{n}(\mathcal{B})}$
for $x^{k}\in\mathcal{A},y^{n}\in\mathcal{B}$ and sufficiently large
$n$. Hence by the property of Mapping 1, for $m\in[1:|\mathcal{B}|]$,
$\frac{P_{X}^{k}(\mathcal{A})Q_{Y}^{n}(y_{m}^{n})}{Q_{Y}^{n}(\mathcal{B})}-P_{X}^{k}(x_{k_{m}}^{k})\leq P_{Y^{n}}(y_{m}^{n})\leq\frac{P_{X}^{k}(\mathcal{A})Q_{Y}^{n}(y_{m}^{n})}{Q_{Y}^{n}(\mathcal{B})}+P_{X}^{k}(x_{k_{m}}^{k})$.
By the asymptotic equipartition property \cite{Cover}, we know that
this step can be roughly considered as mapping a uniform distribution
(with a larger alphabet) to another one (with a smaller alphabet). 
\end{enumerate}
For this code, and for sufficiently large $n$, we have 
\begin{align}
 & D(Q_{Y}^{n}\|P_{Y^{n}})\nonumber \\
 & =\sum_{y^{n}\in\mathcal{B}}Q_{Y}^{n}(y^{n})\log\frac{Q_{Y}^{n}(y^{n})}{P_{Y^{n}}(y^{n})}+\sum_{y^{n}\in\mathcal{B}^{c}}Q_{Y}^{n}(y^{n})\log\frac{Q_{Y}^{n}(y^{n})}{P_{Y^{n}}(y^{n})}\\
 & \leq\sum_{m\in[1:|\mathcal{B}|]}Q_{Y}^{n}(y_{m}^{n})\log\frac{Q_{Y}^{n}(y_{m}^{n})}{\frac{P_{X}^{k}(\mathcal{A})Q_{Y}^{n}(y_{m}^{n})}{Q_{Y}^{n}(\mathcal{B})}-P_{X}^{k}(x_{k_{m}}^{k})}\nonumber \\
 & \qquad+\sum_{y^{n}\in\mathcal{B}^{c}}Q_{Y}^{n}(y^{n})\log\frac{\left(\max_{y}Q_{Y}(y)\right)^{n}}{\left(\min_{x}P_{X}(x)\right)^{k}}\\
 & =-\sum_{m\in[1:|\mathcal{B}|]}Q_{Y}^{n}(y_{m}^{n})\log\left(\frac{P_{X}^{k}(\mathcal{A})}{Q_{Y}^{n}(\mathcal{B})}-\frac{P_{X}^{k}(x_{k_{m}}^{k})}{Q_{Y}^{n}(y_{m}^{n})}\right)\nonumber \\
 & \qquad+nQ_{Y}^{n}(\mathcal{B}^{c})\log\frac{\max_{y}Q_{Y}(y)}{\left(\min_{x}P_{X}(x)\right)^{\frac{1}{R}}}\\
 & \leq-Q_{Y}^{n}(\mathcal{B})\log\left(\frac{P_{X}^{k}(\mathcal{A})}{Q_{Y}^{n}(\mathcal{B})}-\max_{m\in[1:|\mathcal{B}|]}\frac{P_{X}^{k}(x_{k_{m}}^{k})}{Q_{Y}^{n}(y_{m}^{n})}\right)\nonumber \\
 & \qquad+nQ_{Y}^{n}(\mathcal{B}^{c})\log\frac{\max_{y}Q_{Y}(y)}{\left(\min_{x}P_{X}(x)\right)^{\frac{1}{R}}}\\
 & \leq-Q_{Y}^{n}(\mathcal{B})\log\left(\frac{P_{X}^{k}(\mathcal{A})}{Q_{Y}^{n}(\mathcal{B})}-e^{-n\left(\frac{1}{R}\left(H(P_{X})-\delta\right)-\left(H(Q_{Y})+\delta\right)\right)}\right)\nonumber \\
 & \qquad+nQ_{Y}^{n}(\mathcal{B}^{c})\log\frac{\max_{y}Q_{Y}(y)}{\left(\min_{x}P_{X}(x)\right)^{\frac{1}{R}}}\\
 & \rightarrow0\label{eq:-3-1}
\end{align}
where \eqref{eq:-3-1} follows from $\frac{1}{R}\left(H(P_{X})-\delta\right)>H(Q_{Y})+\delta$
and the fact $P_{X}^{n}(\mathcal{A}^{c}),Q_{Y}^{n}(\mathcal{B}^{c})\rightarrow0$
exponentially fast, as shown in the following inequalities. 
\begin{align}
Q_{Y}^{n}(\mathcal{B}^{c}) & =\sum_{y^{n}\in\mathcal{B}^{c}}Q_{Y}^{n}(y^{n})\\
 & =Q_{Y}^{n}\left\{ y^{n}:-\frac{1}{n}\log Q_{Y}^{n}(y^{n})<H(Q_{Y})+\delta\right\} \nonumber \\
 & \qquad+Q_{Y}^{n}\left\{ y^{n}:-\frac{1}{n}\log Q_{Y}^{n}(y^{n})>H(Q_{Y})-\delta\right\} \\
 & \doteq e^{-nE_{Q_{Y}}(H(Q_{Y})-\delta)}+e^{-n\widehat{E}_{Q_{Y}}(H(Q_{Y})+\delta)}\nonumber \\
 & \doteq e^{-nE},
\end{align}
where 
\begin{equation}
E:=\min\left\{ E_{Q_{Y}}(H(Q_{Y})-\delta),\widehat{E}_{Q_{Y}}(H(Q_{Y})+\delta)\right\} >0.
\end{equation}

\begin{figure}[t]
\centering\includegraphics[width=1\columnwidth]{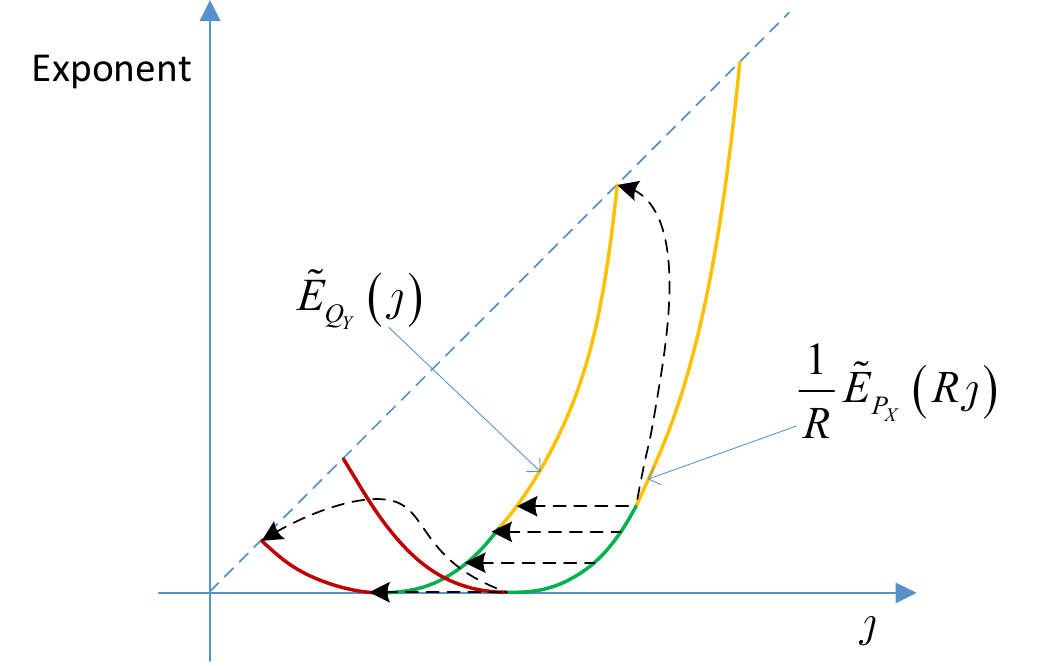}

\caption{Illustration of the code used to prove the achievability for $\alpha=\infty$
in Theorem \ref{thm:Renyi2Con} by using information spectrum exponents.}
\label{fig:DPQ-1} 
\end{figure}

\emph{Achievability part for $\alpha=\infty$: } Partition $\mathcal{X}{}^{k}$
into four parts: 
\begin{align}
 & \mathcal{A}_{1}:=\left\{ x^{k}:P_{X}^{k}(x^{k})>e^{-k\left(H(P_{X})-\delta\right)}\right\} ,\\
 & \mathcal{A}_{2}:=\left\{ x^{k}:e^{-k\left(H^{\mathrm{u}}(P_{X})-\delta\right)}<P_{X}^{k}(x^{k})\leq e^{-k\left(H(P_{X})-\delta\right)}\right\} ,\\
 & \mathcal{A}_{3}:=\left\{ x^{k}:e^{-kH^{\mathrm{u}}(P_{X})}\le P_{X}^{k}(x^{k})\leq e^{-k\left(H^{\mathrm{u}}(P_{X})-\delta\right)}\right\} ,\\
 & \mathcal{A}_{4}:=\left\{ x^{k}:P_{X}^{k}(x^{k})<e^{-kH^{\mathrm{u}}(P_{X})}\right\} .
\end{align}
Define $E^{*}:=\widehat{E}_{Q_{Y}}^{-1}\left(\frac{1}{R}\left(\widehat{E}_{P_{X}}(H^{\mathrm{u}}(P_{X}))\right)\right)$.
Partition $\mathcal{Y}^{n}$ into two parts: 
\begin{align}
 & \mathcal{B}_{1}:=\left\{ y^{n}:Q_{Y}^{n}\left(y^{n}\right)\ge e^{-nE^{*}}\right\} \\
 & \mathcal{B}_{2}:=\left\{ y^{n}:Q_{Y}^{n}\left(y^{n}\right)<e^{-nE^{*}}\right\} .
\end{align}
Consider the following code. This code is illustrated in Fig.~\ref{fig:DPQ-1}. 
\begin{enumerate}
\item Map the sequences in $\mathcal{A}_{1}\cup\mathcal{A}_{4}$ to those
in $\mathcal{Y}^{n}$ in any way. 
\item Use Mapping 1 given in Appendix \ref{subsec:Mappings} to map the
sequences in $\mathcal{A}_{2}$ to the sequences in $\mathcal{B}_{1}$. 
\item Use Mapping 2 given in Appendix \ref{subsec:Mappings} to map the
sequences in $\mathcal{A}_{3}$ to the sequences in $\mathcal{B}_{2}$. 
\end{enumerate}

Assume

\begin{equation}
R<\min_{t\in[0,\infty]}\frac{H_{\frac{1}{1+t}}(P_{X})}{H_{\frac{1}{1+t}}(Q_{Y})}.
\end{equation}
By Lemma \ref{lem:ExponentsComparison}, we have 
\begin{align}
\frac{1}{R}\widehat{E}_{P_{X}}(R\jmath) & <\widehat{E}_{Q_{Y}}(\jmath),\:\forall\jmath\in\frac{1}{R}[H(P_{X}),H^{\mathrm{u}}(P_{X})]\label{eq:-87}\\
R & <\frac{H_{0}(P_{X})}{H_{0}(Q_{Y})}.
\end{align}

We first prove $\log\max_{y^{n}\in\mathcal{B}_{1}}\frac{Q_{Y}^{n}(y^{n})}{P_{Y^{n}}(y^{n})}\rightarrow0.$
Observe that $P_{X}^{k}(\mathcal{A}_{2}),Q_{Y}^{n}\left(\mathcal{B}_{1}\right)\to1$
as $n\to\infty$. Define $\widetilde{P}_{X^{k}}\left(x^{k}\right):=\frac{P_{X}^{k}(x^{k})1\left\{ x^{k}\in\mathcal{A}_{2}\right\} }{P_{X}^{k}(\mathcal{A}_{2})}$
and $\widetilde{Q}_{Y^{n}}\left(y^{n}\right):=\frac{Q_{Y}^{n}(y^{n})1\left\{ y^{n}\in\mathcal{B}_{1}\right\} }{Q_{Y}^{n}(\mathcal{B}_{1})}$.
To prove $\log\max_{j\in\mathcal{B}_{1}}\frac{Q_{Y}^{n}(y_{j}^{n})}{P_{Y^{n}}(y_{j}^{n})}\rightarrow0$,
we only need to prove $\log\max_{y^{n}\in\mathcal{B}_{1}}\frac{Q_{Y}^{n}(y^{n})}{P_{Y^{n}}(y^{n})}\rightarrow0$,
where $\widetilde{P}_{Y^{n}}(y^{n}):=\frac{P_{Y^{n}}(y^{n})}{P_{X}^{k}(\mathcal{A}_{2})}$.
Define $\mathcal{J}_{1}:=\frac{1}{R}[H(P_{X})-\delta,H(P_{X}))$ and
$\mathcal{J}_{2}:=\frac{1}{R}[H(P_{X}),H^{\mathrm{u}}(P_{X})-\delta)$.
Then for $\jmath\in\mathcal{J}_{2}$, we have that 
\begin{align}
 & \lim_{k\to\infty}-\frac{1}{k}\log\left(1-F_{\widetilde{P}_{X^{k}}}(\jmath)\right)\nonumber \\
 & =\lim_{k\to\infty}-\frac{1}{k}\log\widetilde{P}_{X^{k}}\left(x^{k}:-\frac{1}{k}\log\widetilde{P}_{X^{k}}(x^{k})\geq\jmath\right)\\
 & =\lim_{k\to\infty}-\frac{1}{k}\log\frac{P_{X}^{k}\left(x^{k}\in\mathcal{A}_{2}:-\frac{1}{k}\log\frac{P_{X}^{k}(x^{k})}{P_{X}^{k}(\mathcal{A}_{2})}\geq\jmath\right)}{P_{X}^{k}(\mathcal{A}_{2})}\\
 & =\lim_{k\to\infty}-\frac{1}{k}\log P_{X}^{k}\left(x^{k}\in\mathcal{A}_{2}:-\frac{1}{k}\log P_{X}^{k}(x^{k})\geq\jmath+o(1)\right)\\
 & =\widehat{E}_{P_{X}}(\jmath),\label{eq:-44}
\end{align}
where \eqref{eq:-44} follows from Lemma \ref{lem:Exponents2}. Similarly,
for $\jmath\in[H_{\infty}(Q_{Y}),E^{*})$, 
\begin{align}
 & \lim_{n\to\infty}-\frac{1}{n}\log\left(1-F_{\widetilde{Q}_{Y^{n}}}(\jmath)\right)=\widehat{E}_{Q_{Y}}(\jmath),
\end{align}

Observe that by Lemma \ref{lem:Exponents}, $\widehat{E}_{Q_{Y}}(\jmath)$
is continuous. Hence \eqref{eq:-87} implies that there exists some
$\epsilon>0$ such that for any $\jmath\in\mathcal{J}_{2}$, 
\begin{equation}
\frac{1}{R}\widehat{E}_{P_{X}}(R\jmath)\leq\widehat{E}_{Q_{Y}}(\jmath-\epsilon)-\epsilon.\label{eq:-3-2-1-1-1-1}
\end{equation}
i.e., 
\begin{align}
\limsup_{n\to\infty}\frac{1}{n}\log\sup_{\jmath\in\mathcal{J}_{2}}\frac{1-F_{\widetilde{Q}_{Y^{n}}}(\jmath-\epsilon)}{1-F_{\widetilde{P}_{X^{k}}}(R\jmath)} & \leq-\epsilon.\label{eq:-40}
\end{align}
or equivalently, 
\begin{align}
 & \liminf_{n\to\infty}\inf_{\theta\in F_{\widetilde{P}_{X^{k}}}(R\mathcal{J}_{2})}\left\{ \frac{1}{R}F_{\widetilde{P}_{X^{k}}}^{-1}(\theta)-F_{\widetilde{Q}_{Y^{n}}}^{-1}(1-(1-\theta)e^{-n\epsilon})\right\} \nonumber \\
 & \geq\epsilon.\label{eq:-36-1-1}
\end{align}
Since $F_{\widetilde{Q}_{Y^{n}}}^{-1}(\theta)$ is nonincreasing in
$\theta$, \eqref{eq:-36-1-1} implies 
\begin{align}
\liminf_{n\to\infty}\inf_{\theta\in F_{\widetilde{P}_{X^{k}}}(R\mathcal{J}_{2})}\left\{ \frac{1}{R}F_{\widetilde{P}_{X^{k}}}^{-1}(\theta)-F_{\widetilde{Q}_{Y^{n}}}^{-1}(\theta)\right\}  & \geq\epsilon.\label{eq:-37-1-1}
\end{align}

On the other hand, by choosing $\delta>0$ small enough, we have $H(Q_{Y})<\frac{1}{R}(H(P_{X})-\delta)$.
This implies that for some $\epsilon>0$, 
\begin{align}
\liminf_{n\to\infty}\inf_{\theta\in F_{\widetilde{P}_{X^{k}}}(R\mathcal{J}_{1})}\left\{ \frac{1}{R}F_{\widetilde{P}_{X^{k}}}^{-1}(\theta)-F_{\widetilde{Q}_{Y^{n}}}^{-1}(\theta)\right\}  & \geq\epsilon.\label{eq:-37-1-1-1}
\end{align}

Combining \eqref{eq:-37-1-1} and \eqref{eq:-37-1-1-1} gives us that
for some $\epsilon>0$, 
\begin{align}
\liminf_{n\to\infty}\inf_{\theta\in F_{\widetilde{P}_{X^{k}}}(R\left(\mathcal{J}_{1}\cup\mathcal{J}_{2}\right))}\left\{ \frac{1}{R}F_{\widetilde{P}_{X^{k}}}^{-1}(\theta)-F_{\widetilde{Q}_{Y^{n}}}^{-1}(\theta)\right\}  & \geq\epsilon.\label{eq:-37-1-1-1-1}
\end{align}

Observe that $F_{\widetilde{P}_{X^{k}}}^{-1}(\theta)$ is finite,
hence \eqref{eq:-37-1-1} also holds if $R$ is replaced with $\frac{n}{k}$.
Furthermore, similarly in Subsection \ref{subsec:Mappings}, we sort
the elements in $\mathcal{A}_{2}$ as $x_{1}^{k},x_{2}^{k},...,x_{|\mathcal{A}_{2}|}^{k}$
such that $\widetilde{P}_{X^{k}}(x_{1}^{k})\geq\widetilde{P}_{X^{k}}(x_{2}^{k})\geq...\geq\widetilde{P}_{X^{k}}(x_{|\mathcal{A}_{2}|}^{k})$.
Define $\widetilde{G}_{X^{k}}(i):=\widetilde{P}_{X^{k}}\left(x_{l}^{k}:l\leq i\right)$
and $\widetilde{G}_{X^{k}}^{-1}(\theta):=\max\left\{ i\in\mathbb{N}:\widetilde{G}_{X^{k}}(i)\leq\theta\right\} $.
Similarly, for $\widetilde{Q}_{Y^{n}}$, we define $\widetilde{G}_{Y^{n}}(j):=\widetilde{Q}_{Y^{n}}\left(y_{l}^{n}:l\leq j\right)$
and $\widetilde{G}_{Y^{n}}^{-1}(\theta):=\min\left\{ j\in\mathbb{N}:\widetilde{G}_{Y^{n}}(j)\geq\theta\right\} $.
Hence the mapping used here is $j=\widetilde{G}_{Y^{n}}^{-1}(\widetilde{G}_{X^{k}}(i))$.
For each $i\in[1:|\mathcal{A}_{2}|]$, $\widetilde{G}_{X^{k}}(i)\in F_{\widetilde{P}_{X^{k}}}(\mathcal{J})$.
Hence we have 
\begin{align}
 & \liminf_{n\to\infty}\min_{i\in[1:|\mathcal{A}_{2}|]}\frac{1}{n}\log\frac{\widetilde{Q}_{Y^{n}}(y_{j}^{n})}{\widetilde{P}_{X^{k}}(x_{i}^{k})}\nonumber \\
 & =\liminf_{n\to\infty}\min_{i\in[1:|\mathcal{A}_{2}|]}\left\{ \frac{k}{n}F_{\widetilde{P}_{X^{k}}}^{-1}(\widetilde{G}_{X^{k}}(i))-F_{\widetilde{Q}_{Y^{n}}}^{-1}(\widetilde{G}_{X^{k}}(i))\right\} \\
 & \geq\liminf_{n\to\infty}\inf_{\theta\in F_{\widetilde{P}_{X^{k}}}(R\left(\mathcal{J}_{1}\cup\mathcal{J}_{2}\right))}\left\{ \frac{k}{n}F_{\widetilde{P}_{X^{k}}}^{-1}(\theta)-F_{\widetilde{Q}_{Y^{n}}}^{-1}(\theta)\right\} \\
 & \geq\epsilon,
\end{align}
where $j=\widetilde{G}_{Y^{n}}^{-1}(\widetilde{G}_{X^{k}}(i))$. Hence
$\frac{\widetilde{Q}_{Y^{n}}(y_{j}^{n})}{\widetilde{P}_{X^{k}}(x_{i}^{k})}\to0$
for any $i\in[1:|\mathcal{A}_{2}|]$. Therefore, we have 
\begin{align}
 & \log\max_{j\in[1:|\mathcal{B}_{1}|]}\frac{\widetilde{Q}_{Y^{n}}(y_{j}^{n})}{\widetilde{P}_{Y^{n}}(y_{j}^{n})}\nonumber \\
 & \leq\log\max_{j\in[1:|\mathcal{B}_{1}|]}\frac{\widetilde{Q}_{Y^{n}}(y_{j}^{n})}{\widetilde{Q}_{Y^{n}}(y_{j}^{n})-\max_{i:\widetilde{G}_{Y^{n}}^{-1}(\widetilde{G}_{X^{k}}(i))=j}\widetilde{P}_{X^{k}}(x_{i}^{k})}\\
 & \rightarrow0.
\end{align}
Hence $\log\max_{y^{n}\in\mathcal{B}_{1}}\frac{Q_{Y}^{n}(y^{n})}{P_{Y^{n}}(y^{n})}\rightarrow0.$

We next prove $\log\max_{y^{n}\in\mathcal{B}_{2}}\frac{Q_{Y}^{n}(y^{n})}{P_{Y^{n}}(y^{n})}\leq0.$
Observe that 
\begin{align}
\lim_{n\to\infty}-\frac{1}{n}\log Q_{Y}^{n}(\mathcal{B}_{2}) & =\frac{1}{R}\left(\widehat{E}_{P_{X}}(H^{\mathrm{u}}(P_{X}))\right)\\
 & =\frac{1}{R}D(\mathrm{Unif}\left(\mathcal{X}\right)\|P_{X}),
\end{align}
\begin{align}
 & \lim_{n\to\infty}-\frac{1}{n}\log\left(|\mathcal{Y}|^{n}p_{0}\right)\nonumber \\
 & =\frac{1}{R}H^{\mathrm{u}}(P_{X})-H_{0}(Q_{Y})\\
 & =\frac{1}{R}H_{0}(P_{X})+\frac{1}{R}D(\mathrm{Unif}\left(\mathcal{X}\right)\|P_{X})-H_{0}(Q_{Y})\\
 & >\frac{1}{R}D(\mathrm{Unif}\left(\mathcal{X}\right)\|P_{X}),
\end{align}
and 
\begin{align}
\lim_{n\to\infty}\frac{1}{n}\log P_{X}^{k}(\mathcal{A}_{3}) & =\frac{1}{R}\left(\widehat{E}_{P_{X}}(H^{\mathrm{u}}(P_{X})-\delta)\right)\\
 & <\frac{1}{R}D(\mathrm{Unif}\left(\mathcal{X}\right)\|P_{X}).
\end{align}
Hence for sufficiently large $n$, it holds that 
\begin{equation}
Q_{Y}^{n}(\mathcal{B}_{2})+|\mathcal{Y}|^{n}p_{0}\le P_{X}^{k}(\mathcal{A}_{3}),
\end{equation}
which implies that by Mapping 2, $Q_{Y}^{n}(y^{n})\leq P_{Y^{n}}(y^{n})$
for $y^{n}\in\mathcal{B}_{2}$. That is, $\log\max_{y^{n}\in\mathcal{B}_{2}}\frac{Q_{Y}^{n}(y^{n})}{P_{Y^{n}}(y^{n})}\leq0.$

\section{\label{sec:Proof-of-Theorem-Renyi3Con}Proof of Theorem \ref{thm:Renyi3Con} }

By the equality $D_{\alpha}(Q\|P)=\frac{\alpha}{1-\alpha}D_{1-\alpha}(P\|Q)$
for $\alpha\in(0,1)$, the case $\alpha\in(0,1)$ has been proven
in Theorem \ref{thm:Renyi1Con}. Furthermore, it is easy to verify
that the mapping used to prove for case $\alpha=0$ in Theorem \ref{thm:Renyi1Con}
also satisfies $D_{0}(Q_{Y}^{n}\|P_{Y^{n}})\rightarrow0$. So this
proves the case $\alpha=0$. The case $\alpha=1$ can be proven by
a proof similar to that in Appendix \ref{sec:Proof-of-Theorem-Renyi2Con}.
In the following, we consider the case $\alpha=\infty$.

We first prove the following bounds for the normalized and unnormalized
R\'enyi conversion rates for general simulation problem (the seed and
target distributions are not limited to product distributions). For
general distributions $P_{X^{n}}$ and $Q_{Y^{n}}$, we use $P_{X^{n}}$
to approximate $Q_{Y^{n}}$. Define $F_{P_{X^{k}}}(\jmath):=P_{X^{k}}\left(x^{k}:-\frac{1}{k}\log P_{X^{k}}(x^{k})<\jmath\right)$
and $F_{P_{X^{k}}}^{-1}(\theta):=\sup\left\{ \jmath:F_{P_{X^{k}}}(\jmath)\leq\theta\right\} $.
For $Q_{Y^{n}}$, we define $F_{Q_{Y^{n}}}$ and $F_{Q_{Y^{n}}}^{-1}$
similarly. Then we have the following bounds. 
\begin{lem}
\label{lem:Dmax_general} 
\begin{align}
 & \sup\biggl\{ R:\sup_{\epsilon>0}\limsup_{n\to\infty}\frac{1}{n}\log\sup_{\jmath\geq0}\frac{F_{P_{X^{k}}}(\frac{n}{k}(\jmath-\epsilon))}{F_{Q_{Y^{n}}}(\jmath)}\leq0,\nonumber \\
 & \qquad\sup_{\epsilon>0}\limsup_{n\to\infty}\frac{1}{n}\log\sup_{\jmath\geq0}\frac{1-F_{Q_{Y^{n}}}(\jmath)}{1-F_{P_{X^{k}}}(\frac{n}{k}(\jmath-\epsilon))}\leq0\biggr\}\nonumber \\
 & \geq\sup\left\{ R:\frac{1}{n}D_{\infty}^{\mathsf{max}}(P_{Y^{n}},Q_{Y^{n}})\rightarrow0\right\} \\
 & \geq\sup\left\{ R:D_{\infty}^{\mathsf{max}}(P_{Y^{n}},Q_{Y^{n}})\rightarrow0\right\} \\
 & \geq\sup\left\{ R:\liminf_{n\to\infty}\inf_{\theta\in[0,1)}\left\{ \frac{k}{n}F_{P_{X^{k}}}^{-1}(\theta)-F_{Q_{Y^{n}}}^{-1}(\theta)\right\} >0\right\} .
\end{align}
\end{lem}
\begin{rem}
The upper bound can be rewritten as 
\begin{align}
 & \sup\biggl\{ R:\nonumber \\
 & \inf_{\epsilon>0}\liminf_{n\to\infty}\inf_{\theta\in[0,e^{-n\epsilon})}\left\{ \frac{k}{n}F_{P_{X^{k}}}^{-1}(\theta e^{n\epsilon})-F_{Q_{Y^{n}}}^{-1}(\theta)\right\} \geq0,\nonumber \\
 & \inf_{\epsilon>0}\liminf_{n\to\infty}\inf_{\theta\in[0,1)}\left\{ \frac{k}{n}F_{P_{X^{k}}}^{-1}(1-\left(1-\theta\right)e^{-n\epsilon})-F_{Q_{Y^{n}}}^{-1}(\theta)\right\} \nonumber \\
 & \qquad\geq0\biggr\},
\end{align}
and the lower bound can be further lower bounded by 
\begin{align}
 & \sup\biggl\{ R:\inf_{\epsilon>0}\limsup_{n\to\infty}\sup_{\jmath\geq0}\left\{ F_{P_{X^{k}}}(\frac{n}{k}(\jmath+\epsilon))-F_{Q_{Y^{n}}}(\jmath)\right\} \nonumber \\
 & \qquad<0\biggr\}.
\end{align}
Similar expressions for bounds on the conversion rate under the TV
distance measure can be found in \cite{altug2012source}. 
\end{rem}
\begin{rem}
By similar proofs, one can show a better upper bound and a better
lower bound for the unnormalized R\'enyi conversion rate. 
\begin{align}
 & \sup\biggl\{ R:\sup_{\epsilon>0}\limsup_{n\to\infty}\sup_{\jmath\geq0}\left\{ F_{P_{X^{k}}}(\frac{n}{k}(\jmath-\epsilon))-F_{Q_{Y^{n}}}(\jmath)\right\} \nonumber \\
 & \qquad\leq0\biggr\}\nonumber \\
 & \geq\sup\left\{ R:D_{\infty}^{\mathsf{max}}(P_{Y^{n}},Q_{Y}^{n})\rightarrow0\right\} \\
 & \geq\sup\left\{ R:\liminf_{n\to\infty}\inf_{\theta\in[0,1)}\left\{ kF_{P_{X^{k}}}^{-1}(\theta)-nF_{Q_{Y^{n}}}^{-1}(\theta)\right\} =\infty\right\} .
\end{align}
\end{rem}
\begin{IEEEproof}
Achievability (Lower Bound): If $\liminf_{n\to\infty}\inf_{\theta\in[0,1)}\left\{ \frac{k}{n}F_{P_{X^{k}}}^{-1}(\theta)-F_{Q_{Y^{n}}}^{-1}(\theta)\right\} >0$,
then there exists a sufficiently small $\epsilon>0$ and a sufficiently
large $K$ such that $\frac{k}{n}F_{P_{X^{k}}}^{-1}(\theta)-F_{Q_{Y^{n}}}^{-1}(\theta)>0$
for any $\theta\in[0,1)$ and for any $k\ge K$. Assume $x_{1}^{k},x_{2}^{k},...,x_{|\mathcal{X}|^{k}}^{k}$
is a sequence such that $P_{X^{k}}(x_{1}^{k})\geq P_{X^{k}}(x_{2}^{k})\geq...\geq P_{X^{k}}(x_{|\mathcal{X}|^{k}}^{k})$.
Define $G_{X^{k}}(i)=P_{X^{k}}\left(x_{l}^{k}:l\leq i\right)$ and
$G_{X^{k}}^{-1}(\theta):=\max\left\{ i\in\mathbb{N}:F_{X^{k}}(i)\leq\theta\right\} $.
Similarly, for $Q_{Y^{n}}$, we define $G_{Y^{n}}(j):=Q_{Y^{n}}\left(y_{l}:l\leq j\right)$
and $G_{Y^{n}}^{-1}:=\min\left\{ j\in\mathbb{N}:G_{Y}(j)\geq\theta\right\} $.
Use Mapping 1 given in Appendix \ref{subsec:Mappings} to map the
sequences in $\mathcal{X}^{k}$ to the sequences in $\mathcal{Y}^{n}$,
where the distributions $P_{X}$ and $Q_{Y}$ are respectively replaced
by $P_{X^{k}}$ and $Q_{Y^{n}}$. That is, for each $i\in[1:|\mathcal{X}|^{k}]$,
$x_{i}^{k}$ is mapped to $y_{j}^{n}$ where $j=G_{Y^{n}}^{-1}(G_{X^{k}}(i))$.
This code is illustrated in Fig. \ref{fig:DPQ-1-1}.

\begin{figure}[t]
\centering\includegraphics[width=1\columnwidth]{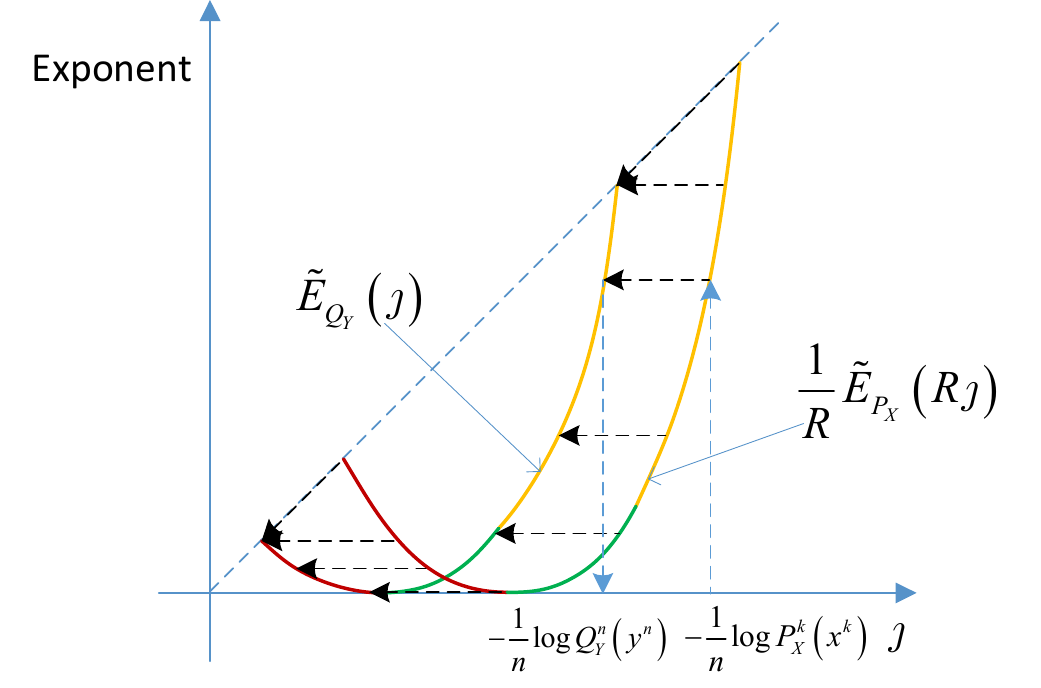}

\caption{Illustration of the code used to prove the achievability for $\alpha=\infty$
in Theorem \ref{thm:Renyi3Con} (or Lemma \ref{lem:Dmax_general})
by using information spectrum exponents.}
\label{fig:DPQ-1-1} 
\end{figure}

Hence for each $j\in[1:|\mathcal{Y}|^{n}]$,

\begin{equation}
Q_{Y^{n}}(y_{j}^{n})-P_{X^{k}}(x_{i}^{k})\leq P_{Y^{n}}(y_{j}^{n})\le Q_{Y^{n}}(y_{j}^{n})+P_{X^{k}}(x_{i}^{k}).
\end{equation}
where $i=G_{X^{n}}^{-1}(G_{Y^{k}}(j))$. By the assumption, we have
$\frac{1}{n}\log\frac{P_{X^{k}}(x_{i}^{k})}{Q_{Y^{n}}(y_{j}^{n})}=F_{Q_{Y^{n}}}^{-1}(G_{X^{k}}(i))-\frac{k}{n}F_{P_{X^{k}}}^{-1}(G_{X^{k}}(i))<0$
for $i=G_{X^{n}}^{-1}(G_{Y^{k}}(j))$. Hence $\frac{P_{X^{k}}(x_{i}^{k})}{Q_{Y^{n}}(y_{j}^{n})}\to0$.
Therefore, we have 
\begin{align}
D_{\infty}(P_{Y^{n}}\|Q_{Y^{n}}) & =\log\max_{j}\frac{P_{Y^{n}}(y_{j}^{n})}{Q_{Y^{n}}(y_{j}^{n})}\\
 & \leq\log\max_{j}\frac{Q_{Y^{n}}(y_{j}^{n})+P_{X^{k}}(x_{i}^{k})}{Q_{Y^{n}}(y_{j}^{n})}\\
 & \rightarrow0,
\end{align}
and 
\begin{align}
D_{\infty}(Q_{Y^{n}}\|P_{Y^{n}}) & =\log\max_{j}\frac{Q_{Y^{n}}(y_{j}^{n})}{P_{Y^{n}}(y_{j}^{n})}\\
 & \leq\log\max_{j}\frac{Q_{Y^{n}}(y_{j}^{n})}{Q_{Y^{n}}(y_{j}^{n})-P_{X^{k}}(x_{i}^{k})}\\
 & \rightarrow0.
\end{align}

Converse (Upper Bound): By Lemma \ref{lem:The-following-properties},
$\frac{1}{n}D_{\infty}(P_{Y^{n}}\|Q_{Y^{n}})\leq\epsilon$ implies
\begin{align}
 & \frac{1}{n}\log\sup_{\jmath\geq0}\frac{P_{Y^{n}}\left(y^{n}:-\frac{1}{n}\log Q_{Y^{n}}(y^{n})<\jmath\right)}{F_{Q_{Y^{n}}}(\jmath)}\nonumber \\
 & \leq\frac{1}{n}\log\sup_{y^{n}}\frac{P_{Y^{n}}(y^{n})}{Q_{Y^{n}}(y^{n})}\leq\epsilon.\label{eq:-67}
\end{align}
Therefore, 
\begin{align}
 & P_{Y^{n}}\left(y^{n}:-\frac{1}{n}\log Q_{Y^{n}}(y^{n})<\jmath\right)\nonumber \\
 & \geq P_{Y^{n}}\left(y^{n}:-\frac{1}{n}\log P_{Y^{n}}(y^{n})<\jmath-\epsilon\right)\\
 & =F_{P_{Y^{n}}}(\jmath-\epsilon).
\end{align}
Observe that $Y^{n}$ is a function of $X^{n}.$ By \cite[Lemma 3.5]{altug2012source}
we have 
\begin{equation}
F_{P_{X^{k}}}(\frac{n}{k}(\jmath-\epsilon))\leq F_{P_{Y^{n}}}(\jmath-\epsilon).
\end{equation}
Therefore, combining this with \eqref{eq:-67} gives 
\begin{equation}
\limsup_{n\to\infty}\frac{1}{n}\log\sup_{\jmath\geq0}\frac{F_{P_{X^{k}}}(\frac{n}{k}(\jmath-\epsilon))}{F_{Q_{Y^{n}}}(\jmath)}\leq\epsilon.
\end{equation}

On the other hand, \eqref{eq:-67} also implies 
\begin{align}
 & P_{Y^{n}}\left(y^{n}:-\frac{1}{n}\log Q_{Y^{n}}(y^{n})\geq\jmath\right)\nonumber \\
 & \leq P_{Y^{n}}\left(y^{n}:-\frac{1}{n}\log P_{Y^{n}}(y^{n})\geq\jmath-\epsilon\right)\\
 & =1-F_{P_{Y^{n}}}(\jmath-\epsilon),\label{eq:-68}
\end{align}
and $\frac{1}{n}D_{\infty}(Q_{Y^{n}}\|P_{Y^{n}})\leq\epsilon$ implies
\begin{align}
 & \frac{1}{n}\log\sup_{\jmath\geq0}\frac{1-F_{Q_{Y^{n}}}(\jmath)}{P_{Y^{n}}\left(y^{n}:-\frac{1}{n}\log Q_{Y^{n}}(y^{n})\geq\jmath\right)}\nonumber \\
 & \leq\frac{1}{n}\log\sup_{y^{n}}\frac{Q_{Y^{n}}(y^{n})}{P_{Y^{n}}(y^{n})}\leq\epsilon.\label{eq:-69}
\end{align}
Combining \eqref{eq:-68} and \eqref{eq:-69} gives 
\begin{equation}
\limsup_{k\to\infty}\frac{1}{n}\log\sup_{\jmath\geq0}\frac{1-F_{Q_{Y^{n}}}(\jmath)}{1-F_{P_{X^{k}}}(\frac{n}{k}(\jmath-\epsilon))}\leq\epsilon.
\end{equation}

Since $\epsilon>0$ can be arbitrarily small, 
\begin{align}
\sup_{\epsilon>0}\limsup_{k\to\infty}\frac{1}{n}\log\sup_{\jmath\geq0}\frac{F_{P_{X^{k}}}(\frac{n}{k}(\jmath-\epsilon))}{F_{Q_{Y^{n}}}(\jmath)}-\epsilon & \leq0,\\
\sup_{\epsilon>0}\limsup_{k\to\infty}\frac{1}{n}\log\sup_{\jmath\geq0}\frac{1-F_{Q_{Y^{n}}}(\jmath)}{1-F_{P_{X^{k}}}(\frac{n}{k}(\jmath-\epsilon))}-\epsilon & \leq0.
\end{align}
These two inequalities are equivalent to 
\begin{align}
\sup_{\epsilon>0}\limsup_{k\to\infty}\frac{1}{n}\log\sup_{\jmath\geq0}\frac{F_{P_{X^{k}}}(\frac{n}{k}(\jmath-\epsilon))}{F_{Q_{Y^{n}}}(\jmath)} & \leq0,\\
\sup_{\epsilon>0}\limsup_{k\to\infty}\frac{1}{n}\log\sup_{\jmath\geq0}\frac{1-F_{Q_{Y^{n}}}(\jmath)}{1-F_{P_{X^{k}}}(\frac{n}{k}(\jmath-\epsilon))} & \leq0.
\end{align}
\end{IEEEproof}
Now we turn back to proving Theorem \ref{thm:Renyi3Con}. We first
focus on the converse part. Consider product distributions $P_{X}^{k}$
and $Q_{Y}^{n}$. Then $\sup_{\epsilon>0}\limsup_{n\to\infty}\frac{1}{n}\log\sup_{\jmath\geq0}\frac{F_{P_{X}^{k}}(\frac{n}{k}(\jmath-\epsilon))}{F_{Q_{Y}^{n}}(\jmath)}\leq0$
and $\sup_{\epsilon>0}\limsup_{n\to\infty}\frac{1}{n}\log\sup_{\jmath\geq0}\frac{1-F_{Q_{Y}^{n}}(\jmath)}{1-F_{P_{X}^{k}}(\frac{n}{k}(\jmath-\epsilon))}\leq0$
respectively imply 
\begin{align}
 & \frac{1}{R}E_{P_{X}}(R\jmath)\geq E_{Q_{Y}}(\jmath),\:\forall\jmath\in\frac{1}{R}[H_{\infty}(P_{X}),H(P_{X})]\label{eq:-82-1-3}\\
 & \frac{1}{R}\widehat{E}_{P_{X}}(R\jmath)\leq\widehat{E}_{Q_{Y}}(\jmath),\:\forall\jmath\in\frac{1}{R}[H(P_{X}),H_{-\infty}(P_{X})].\label{eq:-85-1-2}
\end{align}
By Lemma \ref{lem:ExponentsComparison}, $R\leq\min_{\beta\in[-\infty,\infty]}\frac{H_{\beta}(P_{X})}{H_{\beta}(Q_{Y})}$.

Now we prove the achievability part (lower bound). Assume $R<\min_{\beta\in[-\infty,\infty]}\frac{H_{\beta}(P_{X})}{H_{\beta}(Q_{Y})}$.
Then by Lemma \ref{lem:ExponentsComparison}, 
\begin{align}
 & \frac{1}{R}E_{P_{X}}(R\jmath)>E_{Q_{Y}}(\jmath),\:\forall\jmath\in\frac{1}{R}[H_{\infty}(P_{X}),H(P_{X})]\label{eq:-82-1}\\
 & \frac{1}{R}\widehat{E}_{P_{X}}(R\jmath)<\widehat{E}_{Q_{Y}}(\jmath),\:\forall\jmath\in\frac{1}{R}[H(P_{X}),H_{-\infty}(P_{X})].\label{eq:-85-1}
\end{align}
Since $E_{Q_{Y}}(\jmath)$ and $\widehat{E}_{Q_{Y}}(\jmath)$ are
continuous, there exists a value $\epsilon>0$ such that 
\begin{align}
 & \frac{1}{R}E_{P_{X}}(R\jmath)>E_{Q_{Y}}(\jmath-\epsilon)-\epsilon,\:\forall\jmath\in\frac{1}{R}[H_{\infty}(P_{X}),H(P_{X})]\label{eq:-82-1-1}\\
 & \frac{1}{R}\widehat{E}_{P_{X}}(R\jmath)<\widehat{E}_{Q_{Y}}(\jmath-\epsilon)-\epsilon,\:\forall\jmath\in\frac{1}{R}[H(P_{X}),H_{-\infty}(P_{X})].\label{eq:-85-1-1}
\end{align}
That is, 
\begin{align}
\limsup_{n\to\infty}\frac{1}{n}\log\sup_{\jmath\geq0}\frac{F_{P_{X}^{k}}(R\jmath)}{F_{Q_{Y}^{n}}(\jmath-\epsilon)} & \leq-\epsilon,\\
\limsup_{n\to\infty}\frac{1}{n}\log\sup_{\jmath\geq0}\frac{1-F_{Q_{Y}^{n}}(\jmath-\epsilon)}{1-F_{P_{X}^{k}}(R\jmath)} & \leq-\epsilon,
\end{align}
which in turn respectively imply 
\begin{align}
\liminf_{n\to\infty}\inf_{\theta\in[0,1)}\left\{ \frac{1}{R}F_{P_{X}^{k}}^{-1}(\theta e^{-n\epsilon})-F_{Q_{Y}^{n}}^{-1}(\theta)\right\}  & \geq\epsilon,\label{eq:-35-1}\\
\liminf_{n\to\infty}\inf_{\theta\in[0,1)}\left\{ \frac{1}{R}F_{P_{X}^{k}}^{-1}(1-(1-\theta)e^{n\epsilon})-F_{Q_{Y}^{n}}^{-1}(\theta)\right\}  & \geq\epsilon.\label{eq:-36-1}
\end{align}
Since $F_{P_{X}^{k}}^{-1}(\theta)$ is nondecreasing in $\theta$,
we have both \eqref{eq:-35-1} and \eqref{eq:-36-1} imply 
\begin{align}
\liminf_{k\to\infty}\inf_{\theta\in[0,1)}\left\{ \frac{1}{R}F_{P_{X}^{k}}^{-1}(\theta)-F_{Q_{Y}^{n}}^{-1}(\theta)\right\}  & \geq\epsilon.\label{eq:-37-1}
\end{align}
Therefore, \eqref{eq:-37-1} always holds. Observe that $F_{P_{X}^{k}}^{-1}(\theta)\in[H_{\infty}(P_{X}),H_{-\infty}(P_{X})]$
is bounded for any $\theta\in[0,1)$, hence \eqref{eq:-37-1} also
holds if $R$ is replaced with $\frac{n}{k}$. Combining this with
Lemma \ref{lem:Dmax_general} completes the proof for the lower bound.

\section{\label{sec:Proof-of-Theorem-Renyi1rate}Proof of Theorem \ref{thm:Renyi1rate} }

Define $\mathcal{A}:=\left\{ y^{n}:Q_{Y}^{n}(y^{n})\geq e^{-n\left(H(Q_{Y})+\delta\right)}\right\} $
for $\delta>0$. Define $P_{Y^{n}}(y^{n}):=\frac{1}{\mathsf{M}}\left\lceil \frac{Q_{Y}^{n}(y^{n})}{\frac{1}{\mathsf{M}}Q_{Y}^{n}(\mathcal{A})}\right\rceil $
or $\frac{1}{\mathsf{M}}\left\lfloor \frac{Q_{Y}^{n}(y^{n})}{\frac{1}{\mathsf{M}}Q_{Y}^{n}(\mathcal{A})}\right\rfloor $
for $y^{n}\in\mathcal{A}$; $0$ otherwise. Obviously, $P_{Y^{n}}$
is an $\mathsf{M}$-type distribution. Note that this mapping corresponds
to Mapping 1 given in Appendix \ref{subsec:Mappings}. For this mapping,
we have 
\begin{align}
 & D_{\infty}(P_{Y^{n}}\|Q_{Y}^{n})\nonumber \\
 & =\log\max_{y^{n}}\frac{P_{Y^{n}}(y^{n})}{Q_{Y}^{n}(y^{n})}\\
 & \leq\log\max_{y^{n}\in\mathcal{A}}\frac{\frac{1}{\mathsf{M}}\left\lceil \frac{Q_{Y}^{n}(y^{n})}{\frac{1}{\mathsf{M}}Q_{Y}^{n}(\mathcal{A})}\right\rceil }{Q_{Y}^{n}(y^{n})}\\
 & \leq\log\max_{y^{n}\in\mathcal{A}}\frac{\frac{1}{\mathsf{M}}\left(\frac{Q_{Y}^{n}(y^{n})}{\frac{1}{\mathsf{M}}Q_{Y}^{n}(\mathcal{A})}+1\right)}{Q_{Y}^{n}(y^{n})}\\
 & \leq\log\left(\frac{1}{Q_{Y}^{n}(\mathcal{A})}+\frac{1}{\mathsf{M}}\max_{y^{n}\in\mathcal{A}}\frac{1}{Q_{Y}^{n}(y^{n})}\right)\\
 & \leq\log\left(\frac{1}{Q_{Y}^{n}(\mathcal{A})}+e^{n\left(H(Q_{Y})+\delta-\widetilde{R}\right)}\right).
\end{align}
By the fact that $Q_{Y}^{n}(\mathcal{A})\to1$ at least exponentially
fast as $n\to\infty$, we have that for $\widetilde{R}>H(Q_{Y})+\delta$,
$D_{\infty}(P_{Y^{n}}\|Q_{Y}^{n})\to0$ at least exponentially fast
as $n\to\infty$. Since $\delta>0$ is arbitrary, we have for $\widetilde{R}>H(Q_{Y})$,
$D_{\infty}(P_{Y^{n}}\|Q_{Y}^{n})\to0$ at least exponentially fast
as $n\to\infty$.

\section{\label{sec:Proof-of-Theorem-Renyi2rate}Proof of Theorem \ref{thm:Renyi2rate} }

Define $\mathcal{A}:=\left\{ y^{n}:Q_{Y}^{n}(y^{n})\geq e^{-n\left(H(Q_{Y})+\delta\right)}\right\} $.
Set $P_{Y^{n}}(y^{n}):=\frac{1}{\mathsf{M}}\left\lceil \frac{Q_{Y}^{n}(y^{n})}{\frac{1}{\mathsf{M}}}\right\rceil $
for $y^{n}\notin\mathcal{A}$ (this mapping corresponds to Mapping
2 given in Appendix \ref{subsec:Mappings}); $P_{Y^{n}}(y^{n}):=\frac{1}{\mathsf{M}}\left\lceil \frac{pQ_{Y}^{n}(y^{n})}{\frac{1}{\mathsf{M}}Q_{Y}^{n}(\mathcal{A})}\right\rceil $
or $\frac{1}{\mathsf{M}}\left\lfloor \frac{pQ_{Y}^{n}(y^{n})}{\frac{1}{\mathsf{M}}Q_{Y}^{n}(\mathcal{A})}\right\rfloor $
for $y^{n}\in\mathcal{A}$, where $p=1-\sum_{y^{n}\notin\mathcal{A}}\frac{1}{\mathsf{M}}\left\lceil \frac{Q_{Y}^{n}(y^{n})}{\frac{1}{\mathsf{M}}}\right\rceil \geq Q_{Y}^{n}(\mathcal{A})-\frac{\left|\mathrm{supp}(Q_{Y})\right|^{n}}{\mathsf{M}}$
(this mapping corresponds to Mapping 1 given in Appendix \ref{subsec:Mappings}).
Obviously, $P_{Y^{n}}$ is an $\mathsf{M}$-type distribution. For
this mapping, we have 
\begin{align}
 & D_{\infty}(Q_{Y}^{n}\|P_{Y^{n}})\nonumber \\
 & =\log\max_{y^{n}}\frac{Q_{Y}^{n}(y^{n})}{P_{Y^{n}}(y^{n})}\\
 & \leq\log\max_{y^{n}\in\mathcal{A}}\frac{Q_{Y}^{n}(y^{n})}{\frac{1}{\mathsf{M}}\left\lfloor \frac{pQ_{Y}^{n}(y^{n})}{\frac{1}{\mathsf{M}}Q_{Y}^{n}(\mathcal{A})}\right\rfloor }\\
 & \leq\log\max_{y^{n}\in\mathcal{A}}\frac{Q_{Y}^{n}(y^{n})}{\frac{pQ_{Y}^{n}(y^{n})}{Q_{Y}^{n}(\mathcal{A})}-\frac{1}{\mathsf{M}}}\\
 & \leq-\log\left(\frac{Q_{Y}^{n}(\mathcal{A})-\frac{\left|\mathrm{supp}(Q_{Y})\right|^{n}}{\mathsf{M}}}{Q_{Y}^{n}(\mathcal{A})}-\max_{y^{n}\in\mathcal{A}}\frac{1}{\mathsf{M}Q_{Y}^{n}(y^{n})}\right)\\
 & =-\log\left(1-\frac{\left|\mathrm{supp}(Q_{Y})\right|^{n}}{\mathsf{M}Q_{Y}^{n}(\mathcal{A})}-e^{n\left(H(Q_{Y})+\delta-\widetilde{R}\right)}\right).
\end{align}
By the fact that $Q_{Y}^{n}(\mathcal{A})\to1$ at least exponentially
fast as $n\to\infty$, we have that for $\widetilde{R}>\max\left\{ H_{0}(Q_{Y}),H(Q_{Y})+\delta\right\} $,
$D_{\infty}(Q_{Y}^{n}\|P_{Y^{n}})\to0$ at least exponentially fast
as $n\to\infty$. Since $\delta>0$ is arbitrary, we have for $\widetilde{R}>H_{0}(Q_{Y})$,
$D_{\infty}(Q_{Y}^{n}\|P_{Y^{n}})\to0$ at least exponentially fast
as $n\to\infty$.

\section{\label{sec:Proof-of-Theorem-Renyi3rate}Proof of Theorem \ref{thm:Renyi3rate} }

Define $\mathcal{A}:=\left\{ y^{n}:Q_{Y}^{n}(y^{n})\geq e^{-n\left(\widetilde{R}-\delta\right)}\right\} $.
Use the same mapping as the one in Appendix \ref{sec:Proof-of-Theorem-Renyi2rate}.
That is, set $P_{Y^{n}}(y^{n}):=\frac{1}{\mathsf{M}}\left\lceil \frac{Q_{Y}^{n}(y^{n})}{\frac{1}{\mathsf{M}}}\right\rceil $
for $y^{n}\notin\mathcal{A}$; $P_{Y^{n}}(y^{n}):=\frac{1}{\mathsf{M}}\left\lceil \frac{pQ_{Y}^{n}(y^{n})}{\frac{1}{\mathsf{M}}Q_{Y}^{n}(\mathcal{A})}\right\rceil $
or $\frac{1}{\mathsf{M}}\left\lfloor \frac{pQ_{Y}^{n}(y^{n})}{\frac{1}{\mathsf{M}}Q_{Y}^{n}(\mathcal{A})}\right\rfloor $
for $y^{n}\in\mathcal{A}$. Here $p:=1-\sum_{y^{n}\notin\mathcal{A}}\frac{1}{\mathsf{M}}\left\lceil \frac{Q_{Y}^{n}(y^{n})}{\frac{1}{\mathsf{M}}}\right\rceil $.
Hence $Q_{Y}^{n}(\mathcal{A})-\frac{\left|\mathrm{supp}(Q_{Y})\right|^{n}}{\mathsf{M}}\leq p\leq Q_{Y}^{n}(\mathcal{A})$.
For $\alpha=1+s\in(1,\infty)$, 
\begin{align}
 & D_{1+s}(P_{Y^{n}}\|Q_{Y}^{n})\nonumber \\
 & =\frac{1}{s}\log\sum_{y^{n}}P_{Y^{n}}(y^{n})^{1+s}Q_{Y}^{n}(y^{n})^{-s}\\
 & \leq\frac{1}{s}\log\Biggl\{\sum_{y^{n}\in\mathcal{A}}P_{Y^{n}}(y^{n})\left(\frac{\frac{1}{\mathsf{M}}\left\lceil \frac{pQ_{Y}^{n}(y^{n})}{\frac{1}{\mathsf{M}}Q_{Y}^{n}(\mathcal{A})}\right\rceil }{Q_{Y}^{n}(y^{n})}\right)^{s}\nonumber \\
 & \qquad+\sum_{y^{n}\notin\mathcal{A}}\left(\frac{\frac{1}{\mathsf{M}}\left\lceil \frac{Q_{Y}^{n}(y^{n})}{\frac{1}{\mathsf{M}}}\right\rceil }{Q_{Y}^{n}(y^{n})}\right)^{1+s}\Biggr\}\\
 & \leq\frac{1}{s}\log\Biggl\{ P_{Y^{n}}(\mathcal{A})\left(1+\max_{y^{n}\in\mathcal{A}}\frac{1}{Q_{Y}^{n}(y^{n})\mathsf{M}}\right)^{s}\nonumber \\
 & \qquad+\sum_{y^{n}:Q_{Y}^{n}(y^{n})\leq e^{-n\left(\widetilde{R}-\delta\right)}}\left(Q_{Y}^{n}(y^{n})+\frac{1}{\mathsf{M}}\right)^{1+s}Q_{Y}^{n}(y^{n})^{-s}\Biggr\}\\
 & \leq\frac{1}{s}\log\Biggl\{ P_{Y^{n}}(\mathcal{A})\left(1+e^{-n\delta}\right)^{s}\nonumber \\
 & \qquad+\sum_{y^{n}:Q_{Y}^{n}(y^{n})\leq e^{-n\left(R-\delta\right)}}\left(2e^{-n\left(\widetilde{R}-\delta\right)}\right)^{1+s}Q_{Y}^{n}(y^{n})^{-s}\Biggr\}\\
 & \leq\frac{1}{s}\log\biggl\{\left(1+e^{-n\delta}\right)^{s}\nonumber \\
 & \qquad+2^{1+s}e^{-n\left(1+s\right)\left(\widetilde{R}-\delta\right)}\sum_{y^{n}}Q_{Y}^{n}(y^{n})^{-s}\biggr\}\\
 & =\frac{1}{s}\log\biggl\{\left(1+e^{-n\delta}\right)^{s}\nonumber \\
 & \qquad+2^{1+s}e^{-n\left(1+s\right)\left(\widetilde{R}-\delta\right)+n\left(1+s\right)H_{-s}(Q_{Y})}\biggr\}.\label{eq:-5}
\end{align}

Hence if 
\begin{equation}
\widetilde{R}-\delta>H_{-s}(Q_{Y})
\end{equation}
then \eqref{eq:-5} converges to zero.

On the other hand, 
\begin{align}
 & D_{1+s}(Q_{Y}^{n}\|P_{Y^{n}})\nonumber \\
 & =\frac{1}{s}\log\sum_{y^{n}}Q_{Y}^{n}(y^{n})^{1+s}P_{Y^{n}}(y^{n})^{-s}\\
 & \leq\frac{1}{s}\log\biggl\{\sum_{y^{n}\in\mathcal{A}}\left(\frac{1}{\mathsf{M}}\left\lfloor \frac{pQ_{Y}^{n}(y^{n})}{\frac{1}{\mathsf{M}}Q_{Y}^{n}(\mathcal{A})}\right\rfloor \right)^{-s}Q_{Y}^{n}(y^{n})^{1+s}\nonumber \\
 & \qquad+Q_{Y}^{n}(\mathcal{A}^{c})\biggr\}\\
 & \leq\frac{1}{s}\log\biggl\{\sum_{y^{n}\in\mathcal{A}}\left(\frac{pQ_{Y}^{n}(y^{n})}{Q_{Y}^{n}(\mathcal{A})}-\frac{1}{\mathsf{M}}\right)^{-s}Q_{Y}^{n}(y^{n})^{1+s}\nonumber \\
 & \qquad+Q_{Y}^{n}(\mathcal{A}^{c})\biggr\}\\
 & =\frac{1}{s}\log\biggl\{\sum_{y^{n}\in\mathcal{A}}Q_{Y}^{n}(y^{n})\left(\frac{p}{Q_{Y}^{n}(\mathcal{A})}-\frac{1}{\mathsf{M}Q_{Y}^{n}(y^{n})}\right)^{-s}\nonumber \\
 & \qquad+Q_{Y}^{n}(\mathcal{A}^{c})\biggr\}\\
 & \leq\frac{1}{s}\log\biggl\{ Q_{Y}^{n}(\mathcal{A})\left(\frac{Q_{Y}^{n}(\mathcal{A})-\frac{\left|\mathrm{supp}(Q_{Y})\right|^{n}}{\mathsf{M}}}{Q_{Y}^{n}(\mathcal{A})}-\frac{1}{\mathsf{M}e^{-n\left(\widetilde{R}-\delta\right)}}\right)^{-s}\nonumber \\
 & \qquad+Q_{Y}^{n}(\mathcal{A}^{c})\biggr\}\\
 & \rightarrow0,
\end{align}
where the last line follows since $Q_{Y}^{n}(\mathcal{A}^{c})\to0$
as $n\to\infty$.

\section{\label{sec:Proof-of-Theorem-Renyi1rate-1}Proof of Theorem \ref{thm:RenyiIR1rate} }

Sort the sequences in $|\mathcal{X}|^{n}$ as $x_{1}^{n},x_{2}^{n},...,x_{|\mathcal{X}|^{n}}^{n}$
such that $P_{X}^{n}(x_{1}^{n})\geq P_{X}^{n}(x_{2}^{n})\geq...\geq P_{X}^{n}(x_{|\mathcal{X}|^{n}}^{n})$.
Use Mapping 2 given in Appendix \ref{subsec:Mappings} to map the
sequences in $\mathcal{X}{}^{n}$ to the numbers in $\mathcal{M}$,
where the distributions $P_{X}$ and $Q_{Y}$ are respectively replaced
by $P_{X}^{n}$ and $Q_{M_{n}}$. That is, denote $k_{m},m\in[1:L]$
with $k_{L}:=|\mathcal{X}|^{n}$ as a sequence of integers such that
for $m\in[1:L-1]$, $\sum_{i=k_{m-1}+1}^{k_{m}-1}P_{X}^{n}(x_{i}^{n})<\frac{1}{\mathsf{M}}\le\sum_{i=k_{m-1}+1}^{k_{m}}P_{X}^{n}(x_{i}^{n})$,
and $\sum_{i=k_{L-1}+1}^{k_{L}}P_{X}^{n}(x_{i}^{n})\leq\frac{1}{\mathsf{M}}$
or $\sum_{i=k_{L-1}+1}^{k_{L}-1}P_{X}^{n}(x_{i}^{n})<\frac{1}{\mathsf{M}}\le\sum_{i=k_{L-1}+1}^{k_{L}}P_{X}^{n}(x_{i}^{n})$.
Map $x_{k_{m-1}+1}^{n},...,x_{k_{m}}^{n}$ to $m\in[1:L]$. Define
$T_{X,m}$ as the type of $x_{k_{m}}^{n}$. Then for $s>0$, we have
\begin{align}
 & D_{1+s}(P_{M_{n}}\|Q_{M_{n}})\nonumber \\
 & =\frac{1}{s}\log\sum_{m}P_{M_{n}}(m)^{1+s}(\frac{1}{\mathsf{M}})^{-s}\\
 & \leq\frac{1}{s}\log\biggl(\sum_{m=1}^{L}\mathsf{M}^{s}P_{X}^{n}(x_{k_{m}}^{n})^{1+s}1\left\{ P_{X}^{n}(x_{k_{m}}^{n})\geq\frac{1}{\mathsf{M}}\right\} \nonumber \\
 & +\sum_{m=1}^{L}P_{M_{n}}(m)\left(1+\mathsf{M}P_{X}^{n}(x_{k_{m}}^{n})\right)^{s}1\left\{ P_{X}^{n}(x_{k_{m}}^{n})<\frac{1}{\mathsf{M}}\right\} \biggr),\label{eq:-96}
\end{align}
where \eqref{eq:-96} follows since $P_{M_{n}}(m)=P_{X}^{n}(x_{k_{m}}^{n})$
if $P_{X}^{n}(x_{k_{m}}^{n})\geq\frac{1}{\mathsf{M}}$, and $P_{M_{n}}(m)\leq\frac{1}{\mathsf{M}}+P_{X}^{n}(x_{k_{m}}^{n})$
if $P_{X}^{n}(x_{k_{m}}^{n})<\frac{1}{\mathsf{M}}$.

By Lemma \ref{lem:1+x}, we have \eqref{eq:-6}-\eqref{eq:-123} (given
on page \pageref{eq:-6}) for $0\le s\le1$.

\begin{figure*}
\begin{align}
D_{1+s}(P_{M_{n}}\|Q_{M_{n}}) & \leq\frac{1}{s}\log\biggl(\sum_{m=1}^{L}\mathsf{M}^{s}P_{X}^{n}(x_{k_{m}}^{n})^{1+s}1\left\{ P_{X}^{n}(x_{k_{m}}^{n})\geq\frac{1}{\mathsf{M}}\right\} \nonumber \\
 & \qquad+\sum_{m=1}^{L}P_{M_{n}}(m)\left(1+\left(\mathsf{M}P_{X}^{n}(x_{k_{m}}^{n})\right)^{s}\right)1\left\{ P_{X}^{n}(x_{k_{m}}^{n})<\frac{1}{\mathsf{M}}\right\} \biggr)\label{eq:-6}\\
 & \leq\frac{1}{s}\log\biggl(1+\sum_{m=1}^{L}\mathsf{M}^{s}e^{n\left(1+s\right)\sum_{x}T_{X,m}(x)\log P_{X}(x)}1\left\{ e^{n\sum_{x}T_{X,m}(x)\log P_{X}(x)}\geq\frac{1}{\mathsf{M}}\right\} \nonumber \\
 & \qquad+\sum_{m=1}^{L}\frac{2}{\mathsf{M}}\left(\mathsf{M}e^{n\sum_{x}T_{X,m}(x)\log P_{X}(x)}\right)^{s}1\left\{ e^{n\sum_{x}T_{X,m}(x)\log P_{X}(x)}<\frac{1}{\mathsf{M}}\right\} \biggr)\\
 & \leq\frac{1}{s}\log\biggl(1+\sum_{T_{X}}\left|\mathcal{T}_{T_{X}}\right|\mathsf{M}^{s}e^{n\left(1+s\right)\sum_{x}T_{X}(x)\log P_{X}(x)}1\left\{ e^{n\sum_{x}T_{X}(x)\log P_{X}(x)}\geq\frac{1}{\mathsf{M}}\right\} \nonumber \\
 & \qquad+\sum_{T_{X}}\frac{P_{X}^{n}(\mathcal{T}_{T_{X}})}{\frac{1}{\mathsf{M}}}\frac{2}{\mathsf{M}}\left(Me^{n\sum_{x}T_{X}(x)\log P_{X}(x)}\right)^{s}1\left\{ e^{n\sum_{x}T_{X}(x)\log P_{X}(x)}<\frac{1}{\mathsf{M}}\right\} \biggr)\\
 & \leq\frac{1}{s}\log\biggl(1+\sum_{T_{X}}e^{nH(T_{X})+no\left(1\right)}\mathsf{M}^{s}e^{n\left(1+s\right)\sum_{x}T_{X}(x)\log P_{X}(x)}1\left\{ e^{n\sum_{x}T_{X}(x)\log P_{X}(x)}\geq\frac{1}{\mathsf{M}}\right\} \nonumber \\
 & \qquad+\sum_{T_{X}}\frac{e^{-nD(T_{X}\|P_{X})+no\left(1\right)}}{\frac{1}{\mathsf{M}}}\frac{2}{\mathsf{M}}\left(Me^{n\sum_{x}T_{X}(x)\log P_{X}(x)}\right)^{s}1\left\{ e^{n\sum_{x}T_{X}(x)\log P_{X}(x)}<\frac{1}{\mathsf{M}}\right\} \biggr)\\
 & \leq\frac{1}{s}\log\left(1+2\sum_{T_{X}}e^{nH(T_{X})+n\left(1+s\right)\sum_{x}T_{X}(x)\log P_{X}(x)+no\left(1\right)}\mathsf{M}^{s}\right)\\
 & \leq\frac{1}{s}\log\left(1+2\max_{T_{X}}\left(e^{ns\ensuremath{\widehat{R}}+nH(T_{X})+n\left(1+s\right)\sum_{x}T_{X}(x)\log P_{X}(x)+no\left(1\right)}\right)\right)\\
 & =\frac{1}{s}\log\left(1+2\max_{\widetilde{P}_{X}\in\mathcal{P}\left(\mathcal{X}\right)}\left(e^{ns\ensuremath{\widehat{R}}+nH(\widetilde{P}_{X})+n\left(1+s\right)\sum_{x}\widetilde{P}_{X}(x)\log P_{X}(x)+no\left(1\right)}\right)\right)\\
 & =\frac{1}{s}\log\left(1+2e^{ns\left(\ensuremath{\widehat{R}}-H_{1+s}(P_{X})+o\left(1\right)\right)}\right)\label{eq:-123}
\end{align}

\hrulefill{} 
\end{figure*}

Similarly, for $1\le s\le2$, 
\begin{align}
 & D_{1+s}(P_{M_{n}}\|Q_{M_{n}})\nonumber \\
 & \leq\frac{1}{s}\log\Bigl\{1+2e^{ns\left(\ensuremath{\widehat{R}}-H_{1+s}(P_{X})+o\left(1\right)\right)}\nonumber \\
 & \qquad+2se^{ns\left(\ensuremath{\widehat{R}}-H_{2}(P_{X})+o\left(1\right)\right)}\Bigr\}
\end{align}
and for $s\ge2$, 
\begin{align}
 & D_{1+s}(P_{M_{n}}\|Q_{M_{n}})\nonumber \\
 & \leq\frac{1}{s}\log\Bigl\{1+2e^{ns\left(\ensuremath{\widehat{R}}-H_{1+s}(P_{X})+o\left(1\right)\right)}\nonumber \\
 & \qquad+2s\left(2^{s-1}-1\right)e^{ns\left(\ensuremath{\widehat{R}}-H_{2}(P_{X})+o\left(1\right)\right)}\Bigr\}.\label{eq:-7}
\end{align}

Therefore, no matter for $0\le s\le1$, $1\le s\le2$, or $s\ge2$,
$D_{1+s}(P_{M_{n}}\|Q_{M_{n}})\to0$ if $\ensuremath{\widehat{R}}<H_{1+s}(P_{X})$.

\section{\label{sec:Proof-of-Theorem-Renyi1rate-1-1}Proof of Theorem \ref{thm:RenyiIRRateQP} }

We consider the following mapping\footnote{Although there may exist simpler mappings than the one considered
here, the mapping here will be reused in Appendix \ref{sec:Proof-of-Theorem-Renyi1rate-1-2}.}. Sort the sequences in $|\mathcal{X}|^{n}$ as $x_{1}^{n},x_{2}^{n},...,x_{|\mathcal{X}|^{n}}^{n}$
such that $P_{X}^{n}(x_{1}^{n})\geq P_{X}^{n}(x_{2}^{n})\geq...\geq P_{X}^{n}(x_{|\mathcal{X}|^{n}}^{n})$.
Assume $\delta>0$ is a number such that $\ensuremath{\widehat{R}}+\delta<H(P_{X})$.
Define $\mathcal{A}:=\left\{ x^{n}:P_{X}^{n}(x^{n})\geq\frac{e^{-n\delta}}{\mathsf{M}}\right\} $.
Denote $k_{m},m\in[1:\mathsf{M}]$ as a sequence of integers such
that for $m\in[1:L]$, $\sum_{i=k_{m-1}+1}^{k_{m}-1}P_{X}^{n}(x_{i}^{n})<\frac{1}{\mathsf{M}}\le\sum_{i=k_{m-1}+1}^{k_{m}}P_{X}^{n}(x_{i}^{n})$,
where $L$ is the maximum integer such that $P_{X}^{n}(x_{k_{L}}^{n})\geq\frac{e^{-n\delta}}{\mathsf{M}}$;
and for $m\in[L+1:\mathsf{M}]$, $\sum_{i=k_{m-1}+1}^{k_{m}}P_{X}^{n}(x_{i}^{n})\le\frac{p_{0}}{\mathsf{M}_{0}}<\sum_{i=k_{m-1}+1}^{k_{m}+1}P_{X}^{n}(x_{i}^{n})$.
Here 
\begin{equation}
p_{0}:=1-\sum_{i=1}^{k_{L}}P_{X}^{n}(x_{i}^{n})\geq P_{X}^{n}(\mathcal{A}^{c})\geq P_{X}^{n}(\mathcal{T}_{\epsilon}^{n})\to1
\end{equation}
for some $\epsilon>0$ such that $\ensuremath{\widehat{R}}+\delta<\left(1-\epsilon\right)H(P_{X})$,
and 
\begin{equation}
\mathsf{M}_{0}:=\mathsf{M}-L\geq\mathsf{M}-\frac{\sum_{i=1}^{k_{L}}P_{X}^{n}(x_{i}^{n})}{\frac{1}{\mathsf{M}}}=\mathsf{M}p_{0}.\label{eq:-19}
\end{equation}

Obviously, $\sum_{i=1}^{k_{\mathsf{M}}}P_{X}^{n}(x_{i}^{n})\le1$,
hence $k_{\mathsf{M}}\leq|\mathcal{X}|^{n}$. We consider the following
mapping.

Step 1: For each $m\in[1:\mathsf{M}]$, map $x_{k_{m-1}+1}^{n},...,x_{k_{m}}^{n}$
to $m$.

Step 2: Map $x_{k_{\mathsf{M}}+1}^{n},...,x_{|\mathcal{X}|^{n}}^{n}$
to $m\in[L+1:\mathsf{M}]$ such that the resulting $P_{M_{n}}(m),m\in[L+1:\mathsf{M}]$
satisfy $\sum_{i=k_{m-1}+1}^{k_{m}}P_{X}^{n}(x_{i}^{n})\leq P_{M_{n}}(m)\leq\sum_{i=k_{m-1}+1}^{k_{m}+1}P_{X}^{n}(x_{i}^{n})$.

Note that this mapping for $m\in[1:L]$ corresponds to Mapping 2 given
in Appendix \ref{subsec:Mappings}, and for $m\in[L+1:\mathsf{M}]$
corresponds to Mapping 1 given in Appendix \ref{subsec:Mappings}.
Hence for $m\in[1:L]$, $\frac{1}{\mathsf{M}}\le P_{M_{n}}(m)<\frac{1}{\mathsf{M}}+P_{X}^{n}(x_{k_{m}}^{n})$,
and for $m\in[L+1:\mathsf{M}]$, $\frac{p_{0}}{\mathsf{M}_{0}}-P_{X}^{n}(x_{k_{m}}^{n})\leq P_{M_{n}}(m)\leq\frac{p_{0}}{\mathsf{M}_{0}}+P_{X}^{n}(x_{k_{m}}^{n})$.
\begin{align}
 & D_{\infty}(Q_{M_{n}}\|P_{M_{n}})\nonumber \\
 & =\log\max_{m}\frac{\frac{1}{\mathsf{M}}}{P_{M_{n}}(m)}\\
 & \leq\log\max_{m\in[L+1:\mathsf{M}]}\frac{\frac{1}{\mathsf{M}}}{\frac{1}{\mathsf{M}_{0}}p_{0}-P_{X}^{n}(x_{k_{m}+1}^{n})}\\
 & =-\log\left(\frac{\mathsf{M}}{\mathsf{M}_{0}}p_{0}-\max_{m\in[L+1:\mathsf{M}]}\mathsf{M}P_{X}^{n}(x_{k_{m}+1}^{n})\right)\\
 & \leq-\log\left(\frac{\mathsf{M}}{\mathsf{M}_{0}}p_{0}-e^{-n\delta}\right)\\
 & \leq-\log\left(p_{0}-e^{-n\delta}\right)\\
 & \rightarrow0.\label{eq:-14-1}
\end{align}
By the fact that $P_{X}^{n}(\mathcal{T}_{\epsilon}^{n})\to1$ at least
exponentially fast as $n\to\infty$, we have that for $\ensuremath{\widehat{R}}+\delta<H(P_{X})$,
$D_{\infty}(Q_{M_{n}}\|P_{M_{n}})\to0$ at least exponentially fast
as $n\to\infty$. Since $\delta>0$ is arbitrary, we have for $\ensuremath{\widehat{R}}<H(P_{X})$,
$D_{\infty}(Q_{M_{n}}\|P_{M_{n}})\to0$ at least exponentially fast
as $n\to\infty$.

\section{\label{sec:Proof-of-Theorem-Renyi1rate-1-2}Proof of Theorem \ref{thm:RenyiIRmax} }

Consider the mapping given in Appendix \ref{sec:Proof-of-Theorem-Renyi1rate-1-1}.

For $\alpha\in[1,\infty)$, we have 
\begin{align}
 & D_{1+s}(P_{M_{n}}\|Q_{M_{n}})\nonumber \\
 & =\frac{1}{s}\log\sum_{m}P_{M_{n}}(m)^{1+s}(\frac{1}{\mathsf{M}})^{-s}\\
 & \leq\frac{1}{s}\log\Biggl\{\sum_{m}\mathsf{M}^{s}P_{X}^{n}(x_{k_{m}}^{n})^{1+s}1\left\{ P_{X}^{n}(x_{k_{m}}^{n})\geq\frac{1}{\mathsf{M}}\right\} \nonumber \\
 & \qquad+\sum_{m}P_{M_{n}}(m)\left(1+\mathsf{M}P_{X}^{n}(x_{k_{m}}^{n})\right)^{s}\nonumber \\
 & \qquad\times1\left\{ \frac{e^{-n\delta}}{\mathsf{M}}\le P_{X}^{n}(x_{k_{m}}^{n})<\frac{1}{M}\right\} \nonumber \\
 & \qquad+\sum_{m}P_{M_{n}}(m)\left(\frac{\mathsf{M}}{\mathsf{M}_{0}}p_{0}+\mathsf{M}P_{X}^{n}(x_{k_{m}}^{n})\right)^{s}\nonumber \\
 & \qquad\times1\left\{ P_{X}^{n}(x_{k_{m}}^{n})<\frac{e^{-n\delta}}{\mathsf{M}}\right\} \Biggr\}\label{eq:-94}\\
 & \leq\frac{1}{s}\log\Biggl\{\sum_{m}\mathsf{M}^{s}P_{X}^{n}(x_{k_{m}}^{n})^{1+s}1\left\{ P_{X}^{n}(x_{k_{m}}^{n})\geq\frac{1}{\mathsf{M}}\right\} \nonumber \\
 & +\sum_{m}P_{M_{n}}(m)\left(1+\mathsf{M}P_{X}^{n}(x_{k_{m}}^{n})\right)^{s}1\left\{ P_{X}^{n}(x_{k_{m}}^{n})<\frac{1}{\mathsf{M}}\right\} \Biggr\},\label{eq:-95}
\end{align}
where \eqref{eq:-94} follows since $P_{M_{n}}(m)=P_{X}^{n}(x_{k_{m}}^{n})$
if $P_{X}^{n}(x_{k_{m}}^{n})\geq\frac{1}{\mathsf{M}}$; $P_{M_{n}}(m)\leq\frac{1}{\mathsf{M}}+P_{X}^{n}(x_{k_{m}}^{n})$
if $\frac{e^{-n\delta}}{\mathsf{M}}\le P_{X}^{n}(x_{k_{m}}^{n})<\frac{1}{M}$;
and $P_{M_{n}}(m)\leq\frac{p_{0}}{\mathsf{M}_{0}}+P_{X}^{n}(x_{k_{m}}^{n})$
if $P_{X}^{n}(x_{k_{m}}^{n})<\frac{e^{-n\delta}}{\mathsf{M}}$, and
\eqref{eq:-95} follows from \eqref{eq:-19}.

Then following steps similar to \eqref{eq:-6}-\eqref{eq:-7}, we
have $D_{\alpha}(P_{M_{n}}\|Q_{M_{n}})\to0$ if $\ensuremath{\widehat{R}}<H_{1+s}(P_{X})$.

On the other hand, 
\begin{align}
 & D_{\infty}(Q_{M_{n}}\|P_{M_{n}})\nonumber \\
 & =\log\max_{m}\frac{\frac{1}{\mathsf{M}}}{P_{M_{n}}(m)}\\
 & \leq\log\max_{m\in[L+1:\mathsf{M}]}\frac{\frac{1}{\mathsf{M}}}{\frac{1}{\mathsf{M}_{0}}p_{0}-P_{X}^{n}(x_{k_{m}}^{n})}\\
 & =-\log\left(\frac{\mathsf{M}}{\mathsf{M}_{0}}p_{0}-\max_{m\in[L+1:\mathsf{M}]}\mathsf{M}P_{X}^{n}(x_{k_{m}}^{n})\right)\\
 & =-\log\left(\frac{\mathsf{M}}{\mathsf{M}_{0}}p_{0}-e^{-n\delta}\right)\\
 & \rightarrow0.\label{eq:-14}
\end{align}
This implies $D_{\alpha}(Q_{M_{n}}\|P_{M_{n}})\to0$.

\subsection*{Acknowledgements}

The authors would like to thank the Associate Editor Prof.\ Vinod
Prabhakaran and the two reviewers for their extensive, constructive
and helpful feedback to improve the manuscript.

 \bibliographystyle{unsrt}
\bibliography{ref}
\begin{IEEEbiographynophoto}{Lei Yu}
received the B.E. and Ph.D. degrees, both in electronic engineering,
from University of Science and Technology of China (USTC) in 2010
and 2015, respectively. From 2015 to 2017, he was a postdoctoral researcher
at the Department of Electronic Engineering and Information Science
(EEIS), USTC. Currently, he is a research fellow at the Department
of Electrical and Computer Engineering, National University of Singapore.
His research interests lie in the intersection of information theory,
probability theory, and combinatorics. 
\end{IEEEbiographynophoto}

\begin{IEEEbiographynophoto}{Vincent Y.\ F.\ Tan}
(S'07-M'11-SM'15) was born in Singapore in 1981. He is currently
a Dean's Chair Associate Professor in the Department of Electrical
and Computer Engineering and the Department of Mathematics at the
National University of Singapore (NUS). He received the B.A.\ and
M.Eng.\ degrees in Electrical and Information Sciences from Cambridge
University in 2005 and the Ph.D.\ degree in Electrical Engineering
and Computer Science (EECS) from the Massachusetts Institute of Technology
(MIT) in 2011. His research interests include information theory,
machine learning, and statistical signal processing.

Dr.\ Tan received the MIT EECS Jin-Au Kong outstanding doctoral thesis
prize in 2011, the NUS Young Investigator Award in 2014, the NUS Engineering
Young Researcher Award in 2018, and the Singapore National Research
Foundation (NRF) Fellowship (Class of 2018). He is also an IEEE Information
Theory Society Distinguished Lecturer for 2018/9. He has authored
a research monograph on {\em ``Asymptotic Estimates in Information
Theory with Non-Vanishing Error Probabilities''} in the Foundations
and Trends in Communications and Information Theory Series (NOW Publishers).
He is currently serving as an Associate Editor of the IEEE Transactions
on Signal Processing. 
\end{IEEEbiographynophoto}

\end{document}